\documentclass[amsmath,floatfix]{revtex4}
\usepackage{latexsym}
\usepackage{amssymb}
\usepackage{pdfpages}
\usepackage{graphics}

\begin{document}
\newcommand{\ja}{Jakubassa-Amundsen}

\newcommand{\bfx}{\mbox{\boldmath $x$}}
\newcommand{\bfq}{\mbox{\boldmath $q$}}
\newcommand{\bfnabla}{\mbox{\boldmath $\nabla$}}
\newcommand{\bfsigma}{\mbox{\boldmath $\sigma$}}
\newcommand{\bfSigma}{\mbox{\boldmath $\Sigma$}}
\newcommand{\bfsigmas}{\mbox{{\scriptsize \boldmath $\sigma$}}}
\newcommand{\bfGamma}{\mbox{\boldmath $\Gamma$}}
\newcommand{\bfalpha}{\mbox{\boldmath $\alpha$}}
\newcommand{\bfeps}{\mbox{\boldmath $\epsilon$}}
\newcommand{\bfA}{\mbox{\boldmath $A$}}
\newcommand{\bfP}{\mbox{\boldmath $P$}}
\newcommand{\bfF}{\mbox{\boldmath $F$}}
\newcommand{\bfe}{\mbox{\boldmath $e$}}
\newcommand{\bfd}{\mbox{\boldmath $d$}}
\newcommand{\bfes}{\mbox{{\scriptsize \boldmath $e$}}}
\newcommand{\bfn}{\mbox{\boldmath $n$}}
\newcommand{\bfW}{{\mbox{\boldmath $W$}_{\!\!rad}}}
\newcommand{\bfM}{\mbox{\boldmath $M$}}
\newcommand{\bfK}{\mbox{\boldmath $K$}}
\newcommand{\bfI}{\mbox{\boldmath $I$}}
\newcommand{\bfJ}{\mbox{\boldmath $J$}}
\newcommand{\bfQ}{\mbox{\boldmath $Q$}}
\newcommand{\bfY}{\mbox{\boldmath $Y$}}
\newcommand{\bfp}{\mbox{\boldmath $p$}}
\newcommand{\bfk}{\mbox{\boldmath $k$}}
\newcommand{\bfks}{\mbox{{\scriptsize \boldmath $k$}}}
\newcommand{\bfs}{\mbox{\boldmath $s$}_0}
\newcommand{\bfv}{\mbox{\boldmath $v$}}
\newcommand{\bfw}{\mbox{\boldmath $w$}}
\newcommand{\bfb}{\mbox{\boldmath $b$}}
\newcommand{\bfxi}{\mbox{\boldmath $\xi$}}
\newcommand{\bfzeta}{\mbox{\boldmath $\zeta$}}
\newcommand{\bfr}{\mbox{\boldmath $r$}}
\newcommand{\bfrs}{\mbox{{\scriptsize \boldmath $r$}}}

\renewcommand{\theequation}{\arabic{section}.\arabic{equation}}
\renewcommand{\thesection}{\arabic{section}}

\title{\Large\bf Relativistic positron scattering from heavy ground-state nuclei}

\author{D.~H.~Jakubassa-Amundsen\\
Mathematics Institute, University of Munich, Theresienstrasse 39,\\ 80333 Munich, Germany}

\date{}


\vspace{1cm}

\begin{abstract}  
Cross sections and spin asymmetries for elastic positron scattering and for the emission of positron bremsstrahlung in collision
with  nuclei of zero and half-integer spin are calculated within the relativistic  partial-wave theory.
For elastic scattering, comparison is made with the respective results from electron scattering by $^{208}$Pb, $^{89}$Y and $^{23}$Na nuclei at high energies (up to 400 MeV).
It is found that  the difference between electron and positron intensities diminishes with energy, although the diffraction oscillations, which differ in phase for the two species,
 prevail up to the highest energies considered.
Magnetic scattering strongly reduces the spin asymmetries for nuclei with spin, both for positrons and electrons.
For the elementary process of bremsstrahlung the cross section, which is studied for the $^{197}$Au and $^{208}$Pb targets at low energies $(1-30$ MeV),
 increases with photon frequency for large scattering angles and small
photon emission angles, in contrast to the usual behaviour  of the doubly differential cross section, or of the triply differential cross section at small scattering angles.
\end{abstract}

\maketitle

\vspace{0.5cm}

\section{Introduction}

While spin asymmetries in low-energy elastic positron scattering from atoms have been  investigated recently \cite{Ma09,Ku10,SN11,Sh14},
high-energy scattering experiments from nuclei  have concentrated on the measurement of differential cross sections.
Such experiments on positron scattering from protons were aimed at investigating the two-photon exchange contribution \cite{Mar,Arr}.
There exist also scattering experiments using heavier targets, including lead and bismuth, for which the corresponding theoretical calculations were carried out by means of the relativistic phase shift analysis \cite{EP53,MR57,RF61,HCR63,GPY63,Ba64,YBR65,Bre91,Ma06}.
In these investigations, emphasis was laid on the relative cross section difference between positron and electron impact as a function of scattering angle or momentum transfer, by using collision energies up to 450 MeV.
With this method, information on the nuclear ground-state charge density could be obtained.
In contrast, studies  of the spin asymmetries in collisions beyond 1 MeV positron energy are scarce (see, e.g. \cite{TG51,IL94}).

For elastic electron scattering, on the other hand, the resulting polarization correlations
are well known, as discussed thoroughly  in the literature \cite{Mo,Ub,Ke}. More recent theoretical work can be found in \cite{CH,Jaku12a,Jaku14}, but up to now systematic experiments on the spin asymmetry  
relating to electrons spin-polarized perpendicular to the scattering plane, the so-called Sherman function,
exist only up to 14 MeV collision energy \cite{Sro}.
Actually, a few isolated measurements can be found at much higher energies, up to 3 GeV,
in the context of two-photon effects and parity violation (see, e.g. \cite{We,Ka}).

Apart from being elastically scattered, the positron or electron will also emit bremsstrahlung in the field of the nucleus. In fact, a close relation between an elastically scattered electron and a circularly polarized bremsstrahlung photon with maximum possible energy was
established for sufficiently high collision energy \cite{JR62,Jaku12}. The equivalence of these two processes at energies where the electron's rest mass is of minor importance
manifests itself in a similarity of the respective
polarization correlations which describe the spin transfer of a beam particle to an outgoing particle in a helicity eigenstate (provided that in the bremsstrahlung process the scattered lepton is not observed).

Investigations concerning positron bremsstrahlung are
scarce, based on the expectation that for fast particles there should not be much difference between electron and positron impact.
Within the relativistic Dirac partial-wave (DW)  theory, high-energy positron bremsstrahlung was in early work investigated analytically and numerically
near the short-wavelength limit (SWL) \cite{JP63,JP64}.
There is later work on the (angle-integrated) spectral distribution of positron bremsstrahlung, using the DW theory, but only for low collision energies up to 0.5 MeV \cite{FPT}.
A full theoretical account of the photon energy and angular distribution, including the linear polarization correlations between beam positron and photon (for unobserved scattered positrons), was given only recently  in the regime $0.1-1$ MeV \cite{Y12}.

For electron scattering, again much more work has been done. An investigation
 of the linear and circular bremsstrahlung polarization correlations, both experimentally and theoretically, is provided in
\cite{TP73,STP,T02,Ta15,Ni}, the theory being described in \cite{TP71} and the polarization correlations being introduced in \cite{TP73}.
Recently, particularly for advancing to higher collision energies
in the MeV region, the numerical DW code was optimized by introducing the complex-plane rotation method \cite{VF} for performing the radial integrals \cite{YS,MYS,Jaku16}.

The present work  concentrates on high-energy elastic scattering as well as on bremsstrahlung emission by positron impact in the MeV region, in comparison with the respective
results for electron scattering.
An overview over the theories is provided, which are subsequently used in the numerical calculations.
In Section II elastic scattering from the spin-zero nucleus $^{208}$Pb is considered, for which experimental cross section data on 
electron and positron scattering are available \cite{MR57}. The influence of magnetic scattering is studied 
with the help of the distorted-wave Born approximation (DWBA)
by choosing as targets the spin-$\frac12$ nucleus $^{89}$Y and the spin-$\frac{3}{2}$ nucleus $^{23}$Na. For both nuclei,  electron scattering data
at $180^\circ$ are available which isolate the magnetic contribution \cite{Wi93,Si}.
Section III deals with positron bremsstrahlung in collision with 
 $^{197}$Au  and $^{208}$Pb nuclei.
For the bremsstrahlung polarization correlations a sum rule, known from electron scattering investigations \cite{PMS,Jaku17}, is probed for the positrons.
The equivalence between elastically scattered positrons and bremsstrahlung in the vicinity of the SWL is explored in Section IV. 
The conclusion is drawn in Section V. Atomic units $(\hbar=m=e=1)$ are used unless otherwise indicated.

\section{Elastic positron scattering} 

In order to derive the scattering states of a positron one has to apply charge conjugation. Given an electronic state of a bare nucleus, $\psi_{e^-}(\bfr,\bfzeta,Z)$, with spin polarization $\bfzeta$ and nuclear charge number $Z$,
the respective state for a positron is obtained by means of \cite{BD,Ros}
\begin{equation}\label{2.1}
\psi_{e^+} (\bfr,\bfzeta,Z)\;=\;i\,\gamma_2\;\psi_{e^-}^\ast(\bfr,\bfzeta,-Z)
\end{equation}
with the Dirac matrix $\gamma_2\;=\; \left( \begin{array}{cc} 0 & \sigma_2\\
-\sigma_2&0
\end{array}\right) $, where $\sigma_2\,=\,\left( \begin{array}{cc} 0 &-i\\
i&0 
\end{array}\right)$.

Thus, let us  consider an incoming electron which impinges along the $z$-direction, being described  within the relativistic partial-wave expansion \cite{Ros,Lan},
\begin{equation}\label{2.2}
\psi^{(+)}_{i,e^-}(\bfr,\bfzeta_i,Z)\;=\;\sum_{m_i=\pm \frac12} a_{m_i} \sum_{\kappa_i} \sqrt{\frac{2l_i+1}{4\pi}}\;(l_i 0 \frac12 m_i\,|\,j_i m_i)\; i^{l_i}\;e^{i\delta_{\kappa_i}}
 \;\left( \begin{array}{c}
g_{\kappa_i}(r)\;Y_{j_il_im_i}(\hat{\bfr})\\
i\,f_{\kappa_i}(r)\;Y_{j_il_i'm_i}(\hat{\bfr})
\end{array}\right).
\end{equation}
Then the respective positron function will be  an outgoing state which reads
\begin{equation}\label{2.3}
\psi^{(-)}_{i,e^+}(\bfr,\bfzeta_i,Z)\;=\;i\sum_{m_i=\pm \frac12} a^\ast_{-m_i} \sum_{\kappa_i} \sqrt{\frac{2l_i+1}{4\pi}}\;(-1)^{\frac12-m_i} \;(l_i 0 \frac12 m_i\,|\,j_i m_i)
\; (-i)^{l_i}\;e^{-i\delta_{\kappa_i}}\;{f_{\kappa_i}(r)\;Y_{j_il_i'm_i}(\hat{\bfr}) \choose i g_{\kappa_i}(r)\;Y_{j_il_im_i}(\hat{\bfr})},
\end{equation}
where $g_{\kappa_i}$ and $f_{\kappa_i}$ are, respectively, the large and small components of the radial Dirac function. Note that $g_{\kappa_i}$ and $f_{\kappa_i}$ interchange their role when switching from electron to positron. 
The positron phase shifts $\delta_{\kappa_i}$ as well as $g_{\kappa_i}$ and $f_{\kappa_i}$ in (\ref{2.3})
result from  solutions to the Dirac equation with negative potential, $-V(r)$.
Furthermore, $a_{m_i}$ are the coefficients  of the spinors $\chi_\frac12 = {1 \choose 0}$ and $\chi_{-\frac12} = {0 \choose 1}$,  describing the direction of the polarization vector $\bfzeta_i$, see (\ref{2.7}) below. The symbol $(\cdot\cdot | \cdot)$ is a Clebsch-Gordan coefficient, and $Y_{jlm}$ denotes a spherical harmonic spinor \cite{Ed}.
The quantum numbers $\kappa_i=\pm1,\pm2,...$ are interrelated with the angular momentum 
quantum numbers $j_i$ and $l_i$ by means of $\kappa_i=l_i$ and $l_i'=l_i-1$  for $j_i=l_i-\frac12$ as well as  $\kappa_i=-l_i-1$ and $l_i'=l_i+1$ for $j_i=l_i+\frac12$.

In a similar way, the positron function which corresponds to an outgoing final electron with momentum $\bfk_f$ can be found to be
\begin{equation}\label{2.4}
\psi^{(+)}_{f,e^+}(\bfr,\bfzeta_f,Z)\;=\;i\sum_{\kappa_f m_f}\;\sum_{m_l m_{s}} Y^\ast_{l_f m_l}(\hat{\bfk}_f)\;b^\ast_{-m_{s}}\;(-1)^{\frac12 -m_{s}}\;(l_fm_l\frac12 m_{s}\,|\,j_fm_f)
\;(-i)^{l_f}\;e^{i\delta_{\kappa_f}}\;{ f_{\kappa_f}(r)\;Y_{j_fl_f'm_f}(\hat{\bfr})\choose
i g_{\kappa_f}(r)\;Y_{j_fl_fm_f}(\hat{\bfr})},
\end{equation}
where $Y_{lm}$ is a spherical harmonic function, and the coefficients $b_{m_{s}}$ relate to the final polarization vector $\bfzeta_f$. 

\subsection{Positron scattering from spin-zero nuclei}

For near-spherical spin-zero nuclei the elastic scattering process can be expressed in terms of potential scattering. This implies that the nucleus is chacterized solely by the nuclear potential which is obtained from the ground-state charge distribution, and that recoil is neglected.
In fact, recoil effects during elastic scattering from spin-zero nuclei were investigated in \cite{FF} and were
found to be small for collision energies up to several hundred MeV.

Within the phase shift analysis, the scattering amplitude for a particle of momentum $\bfk_i$ and spin polarization $\bfzeta_i$, being deflected by an angle
$\theta$, is given by
\begin{equation}\label{2.5}
f_e(\bfzeta_i,\bfzeta_f,k_i,\theta)\;=\;\langle \chi_f |\,A\,+\,B\;\bfn \bfsigma\,|\chi_i\rangle,
\end{equation}
where $\bfzeta_f$ and $\bfk_f=k_f(\sin \theta,0,\cos \theta)$ (setting $k_f=k_i$ in the recoil-free case)  are, respectively, spin polarization and momentum of the scattered particle. Moreover, $A$ and $B$ are, respectively, the spin-conserving and spin-flip parts of the transition amplitude, $\bfn=\bfe_y$ is the normal to the scattering plane (spanned by $\bfk_i$ and $\bfk_f$), $\bfsigma$ is the vector of Pauli matrices, and $\chi_i,\;\chi_f$ denote the spin states \cite{Ros},
\begin{equation}\label{2.6}
\chi_i\;= a_\frac12 \;\chi_\frac12\, + \,a_{-\frac12} \;\chi_{-\frac12},\qquad 
\chi_f\;=\;b_\frac12 \;\chi_\frac12 \,+\, b_{-\frac12} \;\chi_{-\frac12},
\end{equation}
where for an arbitrary unit vector $\bfzeta_i$, defined by the spherical angles $\alpha_s$ and $\varphi_s$ with
respect to the scattering plane,
the coefficients are
\begin{equation}\label{2.7}
a_\frac12\;=\; \cos \frac{\alpha_s}{2}\;e^{-i\varphi_s/2},\quad a_{-\frac12} \;=\; \sin \frac{\alpha_s}{2}\;e^{i \varphi_s/2}.
\end{equation}
The coefficients $b_{\pm \frac12}$ for the final spin state are defined in the same way.
If the scattered particle is in a helicity (+) state, which henceforth will be assumed, such that $\bfzeta_f =\hat{\bfk}_f,$  one has
$\;b_\frac12^\ast = \cos \frac{\theta}{2}$ and $b_{-\frac12}^\ast = \sin \frac{\theta}{2}$.

With (\ref{2.6}), the scattering amplitude reduces to
\begin{equation}\label{2.8}
f_e(\bfzeta_i,\bfzeta_f,k_i,\theta)\;=\;A\;(b_\frac12^\ast a_\frac12 +b_{-\frac12}^\ast a_{-\frac12})\;+\;i\,B\;(b_{-\frac12}^\ast a_\frac12-b_\frac12^\ast a_{-\frac12}).
\end{equation}

The differential cross section for elastic scattering
of unpolarized particles, which implies an average over $\chi_i$ as well as a sum over $\chi_f$, 
 is calculated from
\begin{equation}\label{2.9}
\left( \frac{d\sigma}{d\Omega}\right)_0\;=\;\frac12 \sum_{\zeta_i,\zeta_f} \left| f_e(\bfzeta_i,\bfzeta_f,k_i,\theta)\right|^2\;=\;|A|^2\;+\;|B|^2.
\end{equation}
Since $A$ and $B$, and hence the cross section, depend  on the positron wavefunction only in terms of the phase shifts, the change from  electron scattering to positron scattering
is accomplished by replacing the nuclear charge number $Z$ by $-Z$ in the nuclear potential when solving the Dirac equation for the phase shifts.
This procedure also induces a sign change in $f_e$.

If, instead, the polarization of the particle is kept fixed, the cross section can be expressed in terms of the
 three spin asymmetries $L,S$, and $R$ \cite{Mo}, pertaining to the three linear independent choices of $\bfzeta_i$ \ along  the coordinate axes 
$\bfe_z=\hat{\bfk}_i,\;\,\bfe_y=\bfk_i \times \bfk_f/ |\bfk_i \times \bfk_f|$  and $\bfe_x=\bfe_y \times \hat{\bfk}_i$ \cite{Mo,Jaku12},
\begin{equation}\label{2.9a}
\frac{d\sigma}{d\Omega}(\bfzeta_i,\bfzeta_f)\;=\;|f_e(\bfzeta_i,\bfzeta_f,k_i,\theta)|^2\;
=\;\frac12 \left( \frac{d\sigma}{d\Omega}\right)_0\left[ 1\,+\;S\,(\bfzeta_i \bfe_y)\,+\;L\,(\bfzeta_i \bfe_z)\,\zeta_\pm\,-\;R\,(\bfzeta_i \bfe_x)\,\zeta_\pm\right],
\end{equation}
where $\zeta_\pm = \pm 1$ for outgoing particles in a helicity $(\pm)$ state.
Accordingly, the polarization correlations 
 can be obtained in terms of relative cross section differences when
the spin is flipped,
\begin{equation}\label{2.10}
P(\bfzeta_i)\;=\;\frac{d\sigma/d\Omega(\bfzeta_i,\bfzeta_f)-d\sigma/d\Omega(-\bfzeta_i,\bfzeta_f)}{d\sigma/d\Omega(\bfzeta_i,\bfzeta_f)+d\sigma/d\Omega (-\bfzeta_i,\bfzeta_f)},
\end{equation}
and one defines $L=P(\bfe_z),\;S=P(\bfe_y)$ and $R=P(-\bfe_x)$. The Sherman function $S$ \cite{Sh56}, requiring particles which are spin-polarized perpendicular to the scattering plane,
 is the easiest one to access experimentally since it is 
independent of the spin polarization $\bfzeta_f$ of the scattered particle \cite{Mo}.

\vspace{0.5cm}

\begin{figure}[!h]
\centering
\begin{tabular}{cc}
\hspace{-1cm} \includegraphics[width=.7\textwidth]{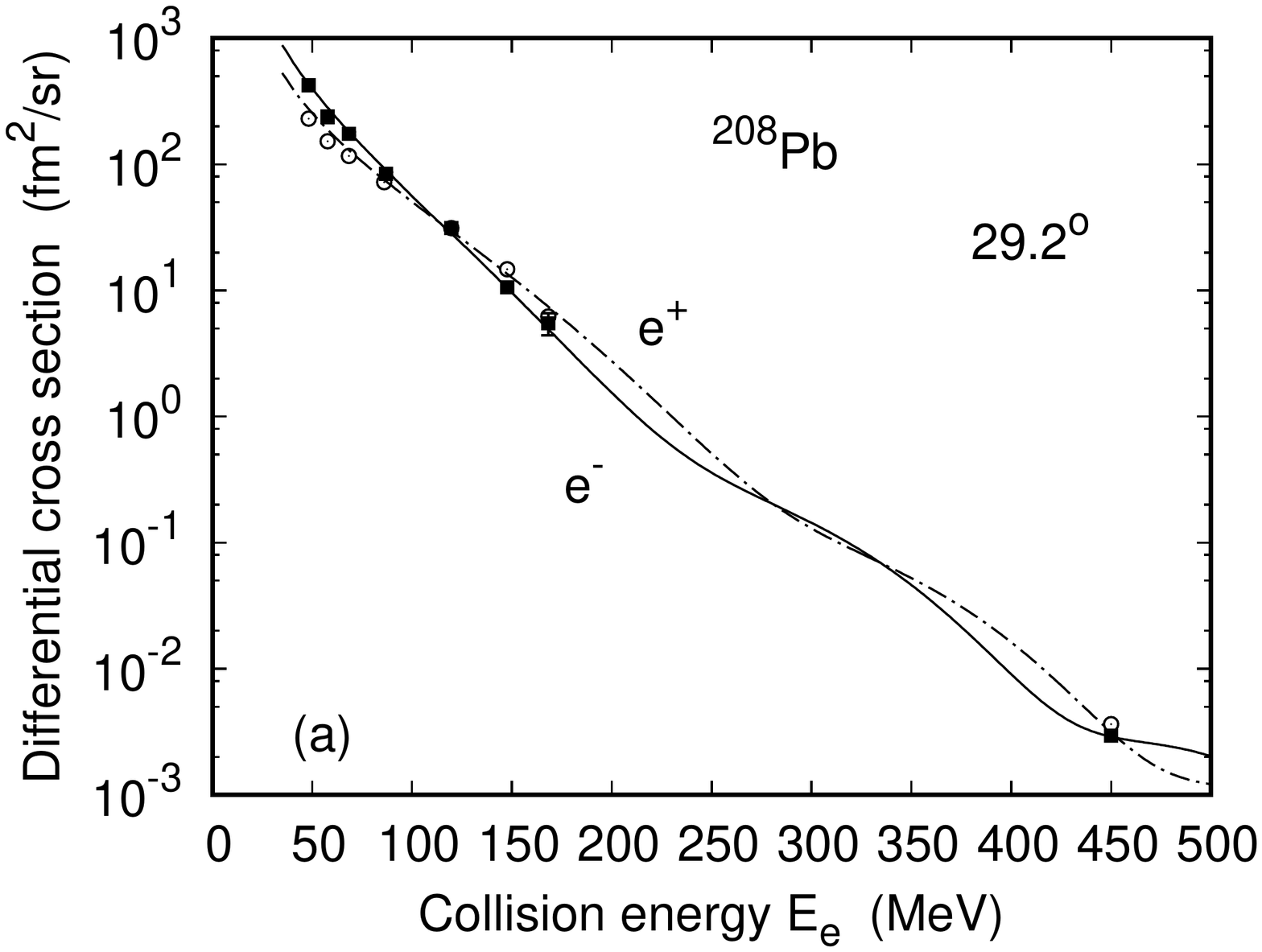}&
\hspace{-4.0cm} \includegraphics[width=.7\textwidth]{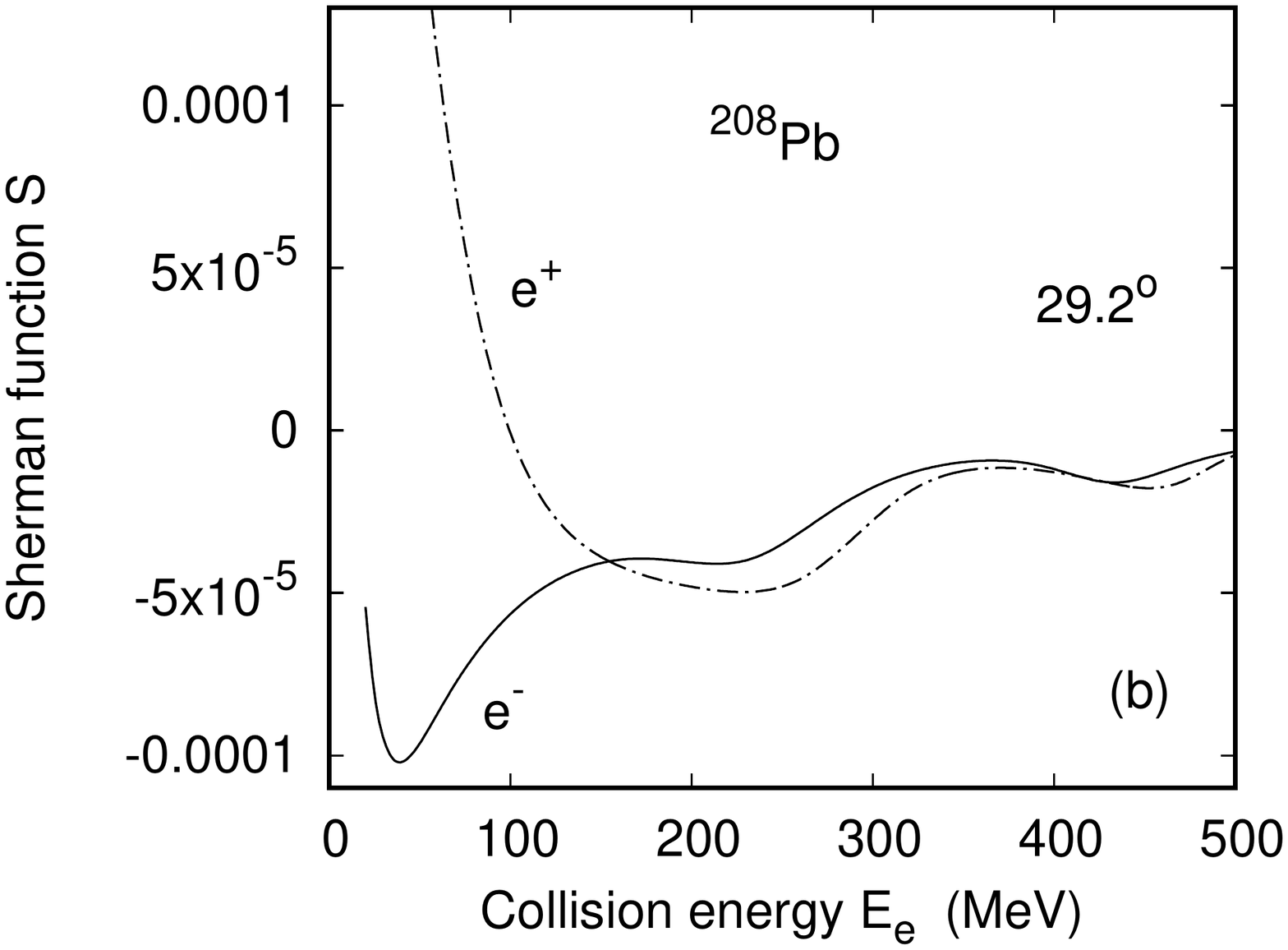}
\end{tabular}
\vspace{-1cm}
\hspace{3cm} \includegraphics[width=.7\textwidth]{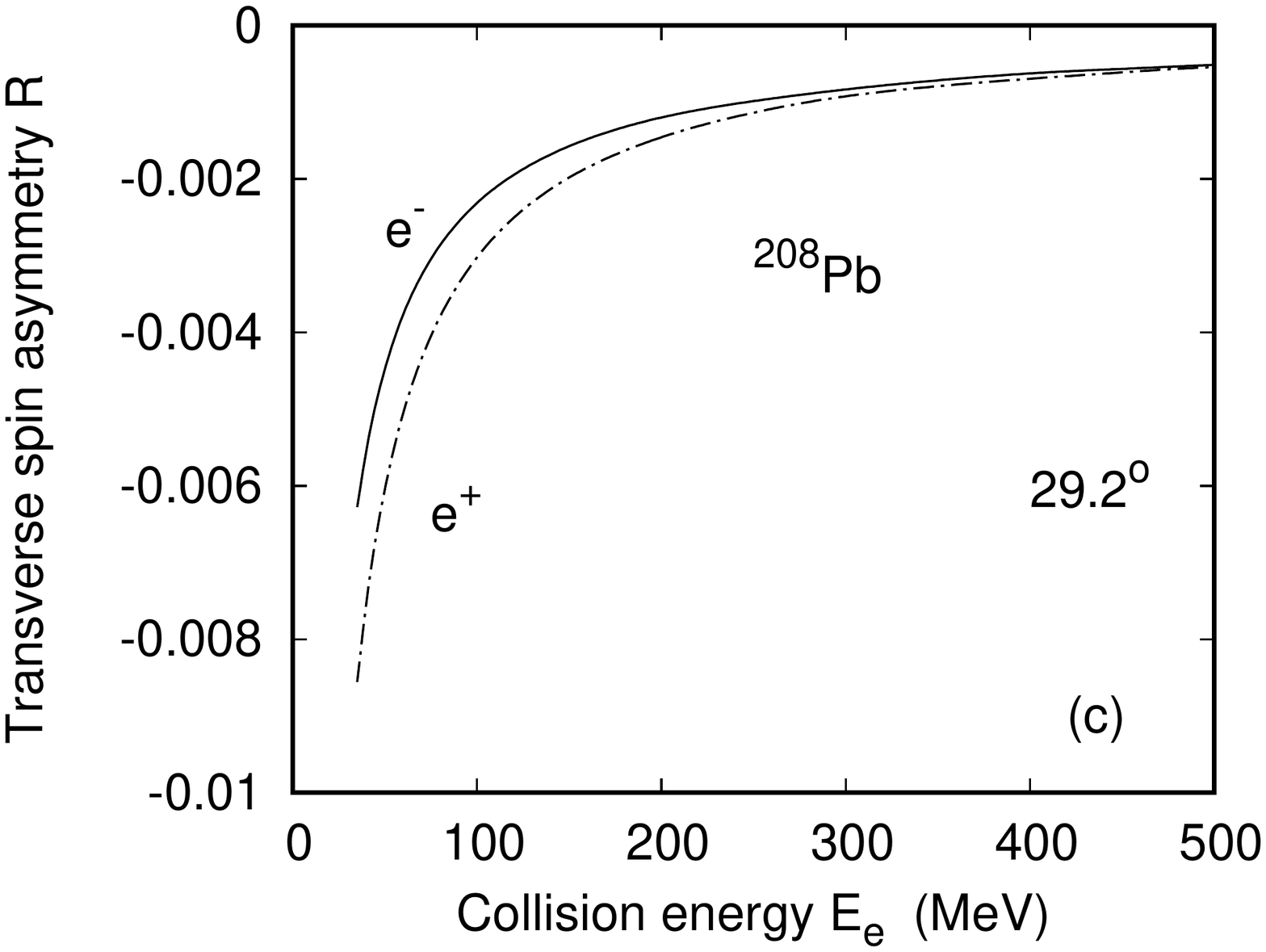}
\caption{Differential cross section $\frac{d\sigma}{d\Omega}$ (a) and spin asymmetries $S$ (b) and $R$ (c) for electrons and positrons  scattered elastically
from $^{208}$Pb ($Z=82$) as a function of collision energy $E_e$.
The scattering angle is $\theta = 29.2^\circ$. --------, electron impact; $-\cdot - \cdot -$, positron impact.
The experimental data in (a) below 200 MeV are from Miller and Robinson \cite{MR57}, those at 450 MeV are from Breton et al \cite{Bre91}:
$\blacksquare$, electron scattering; $\bigcirc$, positron scattering.}
\end{figure}

\subsection{Results for the elastic scattering from lead and from lighter spin-zero nuclei}

For the $^{208}$Pb nucleus, 
the  charge density is available in terms of a Fourier-Bessel expansion  \cite{VJ} which provides an analytical representation of the nuclear potential \cite{Jaku12a}.
The resulting Dirac equation is solved for the phase shifts with the help of the Fortran code RADIAL by Salvat et al \cite{Sal}. When performing the sum over the phase shifts in the calculation of $A$ and $B$ (see, e.g. \cite{Lan,Jaku12a}),  a three-fold convergence acceleration as introduced by Yennie at al \cite{YRW} is applied.

Fig.1a shows the dependence of the differential cross section for a lead target on the collision energy at a fixed forward scattering angle $\vartheta_f=29.2^\circ$. The electron and positron scattering data of \cite{MR57,Bre91} are well reproduced by theory.
The diffraction structures, which originate from interference effects induced by  scattering off the individual protons when the collision energy is sufficiently high such that the projectile can penetrate the nuclear surface, are clearly visible
at an energy above 50 MeV.
There is a phase shift between the corresponding oscillations for positrons relative to those for electrons.
This phase shift is interpreted in terms of an increase of the momentum $k_i$ for electrons near the nucleus, due to the mutual attraction, 
and a reduced momentum for positrons due to the repulsive potential \cite{HCR63}.
The nearly periodic diffraction pattern resembles the square of a spherical Bessel function $j_1(qR_N)$, with $q=|\bfk_i-\bfk_f| = 2k_i\sin \theta/2$ the momentum transfer to the nucleus and $R_N$ the nuclear radius, which is an exact solution 
if the nucleus is approximated by a homogeneously charged sphere \cite{Ub}.
This periodicity disappears, however, if the diffuseness of the nuclear surface is increased by at least a factor of 3.

In Figs.1b and 1c the spin asymmetries S and R are displayed.
Since they are due to purely relativistic effects, they are very small for this forward angle.
While $|R|$ decreases monotonously to zero with collision energy $E_e$, both for electrons and positrons, $S$ shows the  diffraction structures   when $E_e \gtrsim 150 $ MeV.
The longitudinal polarization correlation $L$ is always close to unity, according to the sum rule (\ref{4.2}) for potential scattering  \cite{Mo}.

\vspace{0.2cm}
\begin{figure}[!h]
\centering
\begin{tabular}{cc}
\hspace{-1cm} \includegraphics[width=.7\textwidth]{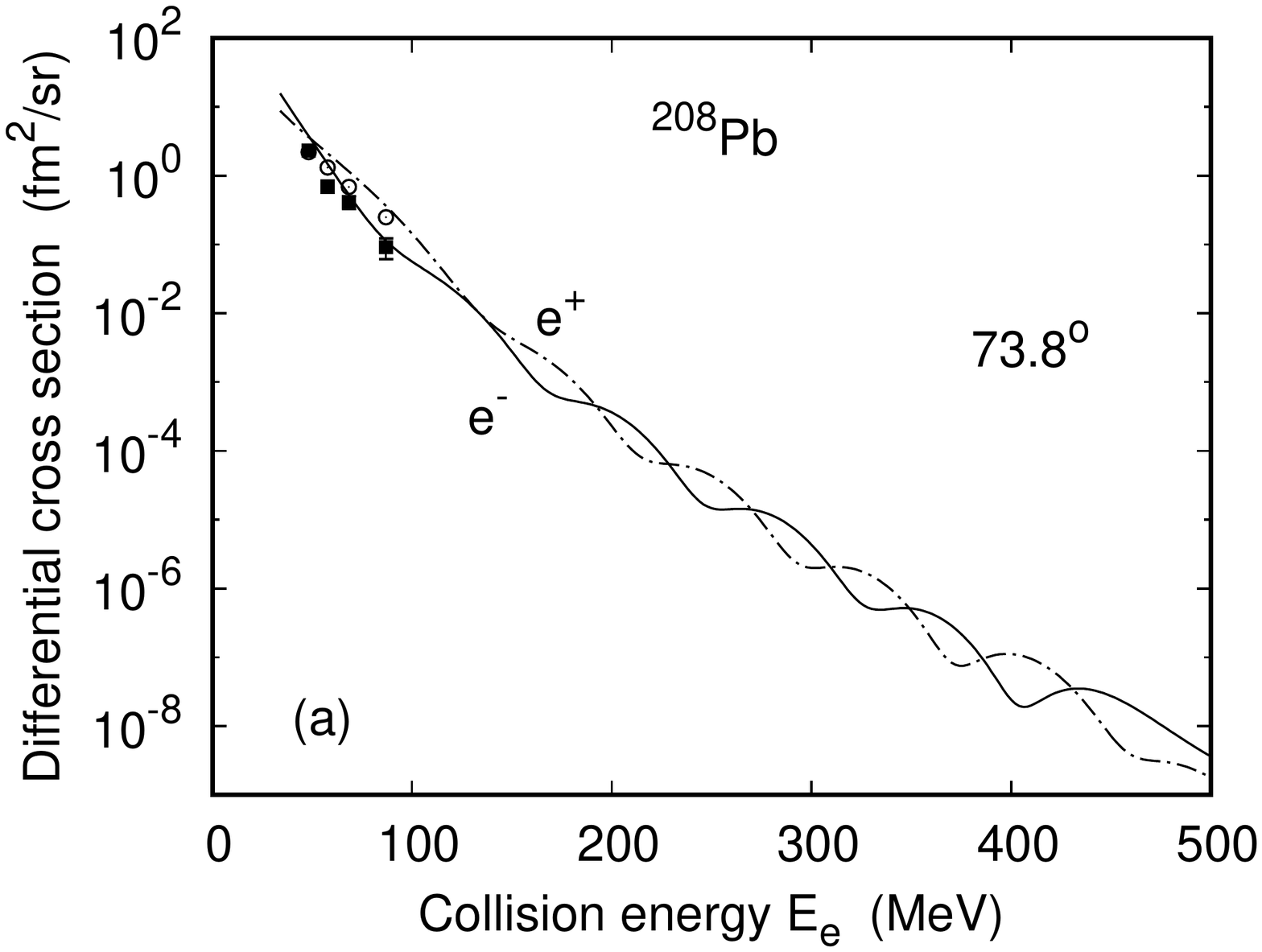}&
\hspace{-4.0cm} \includegraphics[width=.7\textwidth]{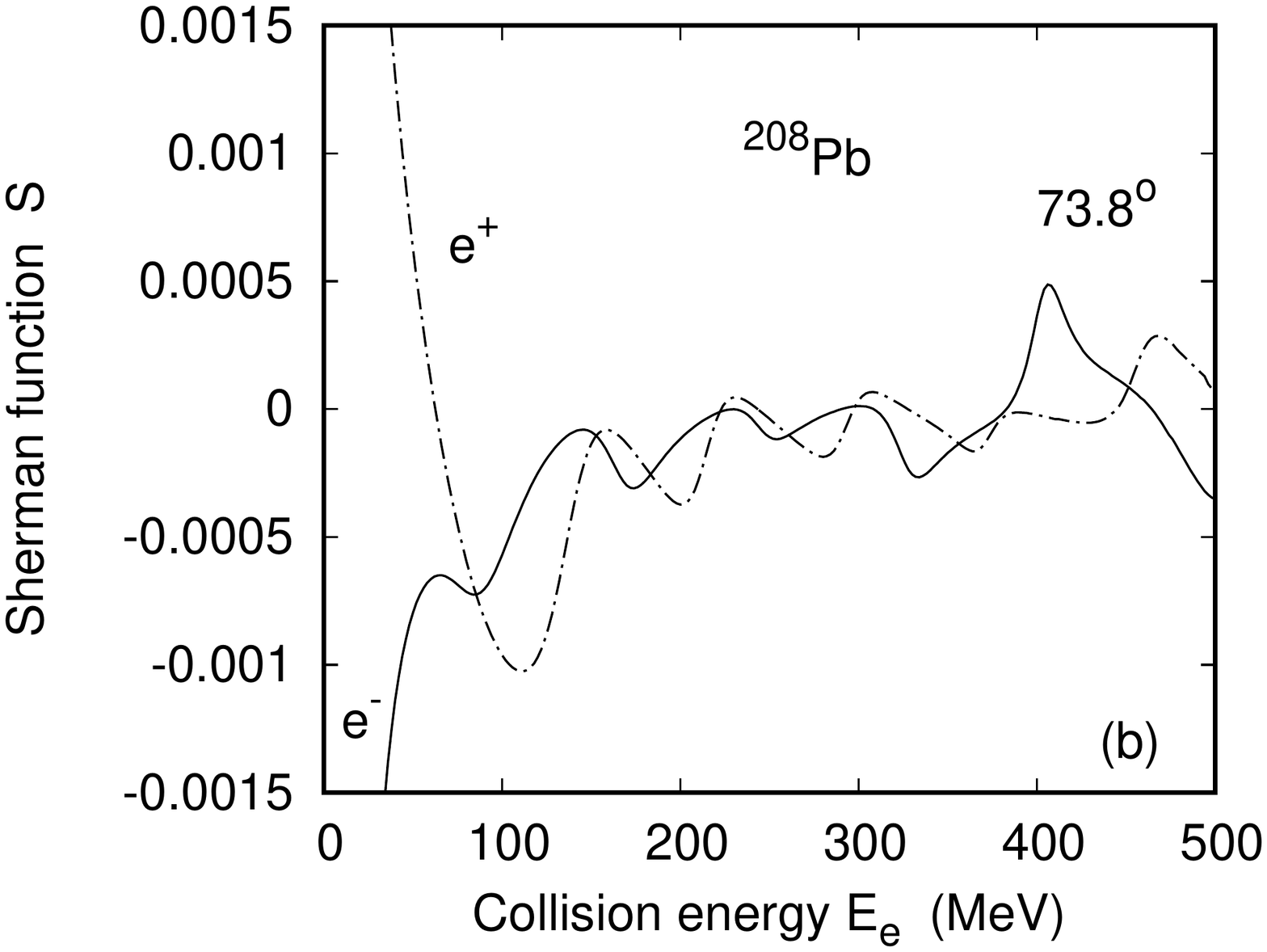}
\end{tabular}
\vspace{-1cm}
\hspace{5cm} \includegraphics[width=.7\textwidth]{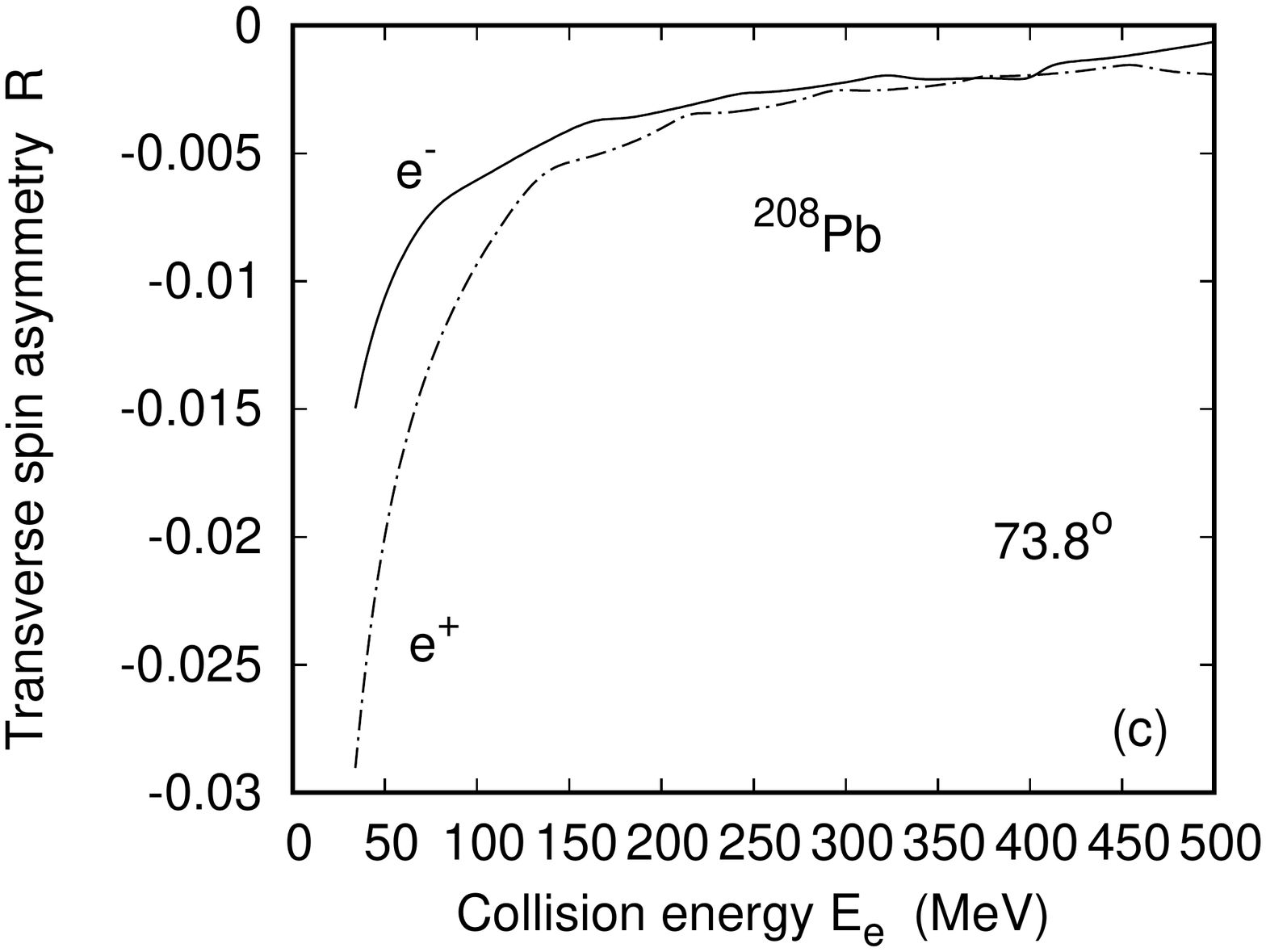}
\caption{
Same as Fig.1, but for a scattering angle of $73.8^\circ$.}
\end{figure}

Fig.2 shows the respective quantities for a larger scattering angle, $\theta=73.8^\circ$.
According to the behaviour in terms of $j_1(qR_N)$, the oscillations in the cross section have a shorter period in $k_i$ (respectively, in $E_e$) than
for the smaller angle. Even in $R$, faint structures are now visible at this angle.
The difference between electron and positron spin asymmetries decreases with energy at the smaller  scattering angle, 
but the diffraction structures veil  this behaviour at the larger angle.
In this context it was found \cite{DP} that the differential cross section approaches the Born limit (where the difference between electron and positron results tends to zero) when $k_i \to \infty$, provided that
the momentum transfer $q  \approx k_i \theta$ is kept constant.
But also when $q$ is decreased by reducing $\theta$, the distances
 between lepton and nucleus, which are relevant for the scattering process,
are getting larger. This weakens the action of the nuclear potential, leading to diminishing electron-positron differences.

In order to study the behaviour of the phase shift between the positron and electron diffraction structures we provide in Fig.3a the energy dependence of the differential cross section for the lead nucleus
when the scattering angle is varied systematically. For the sake of an easy change of parameters, when progressing to lighter nuclei,  we have chosen a Fermi-type nuclear charge density,
 $\varrho_F(r)=\varrho_0 (1+e^{(r-c)/a})^{-1}$, which is available for the isotope $^{206}$Pb \cite{VJ}. 
Significant deviations with respect to the Fourier-Bessel-type charge distribution of the $^{208}$Pb nucleus appear only above 400 MeV (for $\theta=44.1^\circ$, 
where the more accurate charge density provides better agreement with the electron experiment at 450 MeV),
but there is no change of the electron-positron phase shift at the higher $E_e$. An increase or reduction of the  parameter $c$  which describes the nuclear extension has no influence on this phase shift either.

On the other hand, as evident from Fig.3a, an increase of the scattering angle from $44.1^\circ$ to $178^\circ$
leads to a significant change of the phase shift, in addition to an increase of the number of oscillations in the given energy interval.
In particular, electrons and positrons are in phase near $\theta=140^\circ$ for the lead nucleus.

When the nuclear charge number $Z$ is reduced, the phase shift changes too. In Fig.3b the results for two lighter nuclei with a Fermi-type charge distribution, $^{88}$Sr  and $^{20}$Ne, 
are shown in addition to those for $^{206}$Pb. It is seen that the phase shift is reduced when $Z$ is decreased, i.e. when the nuclear field is weakened.
This is even the case when $Z$ is decreased while the nuclear shape is artificially kept fixed. In that latter case the number of oscillations in the given energy interval remains constant.
 
The change of phase affects also the polarization correlations. For lead, the variation of $R$ with $\theta$  follows from Fig.3c  at $178^\circ$ 
in comparison with the results for $73.8^\circ$ (Fig.2c). The $Z$-dependence of $S$ for $73.8^\circ$ is evident
when Fig.2b for $^{208}$Pb is set against Fig.3d for $^{20}$Ne.
For such a light nucleus there is a notable  symmetry between the Sherman function of electrons and positrons, with opposite signs except near the location of the minima.
Moreover, there is a marked difference in the peak shape. While for the electrons there is a slow rise to maximum, followed by a steep decrease, this is vice versa
for the positrons. This feature, although much weaker, exists also for lead. 

\vspace{0.2cm}
\begin{figure}[!h]
\centering
\begin{tabular}{cc}
\hspace{-1cm} \includegraphics[width=.7\textwidth]{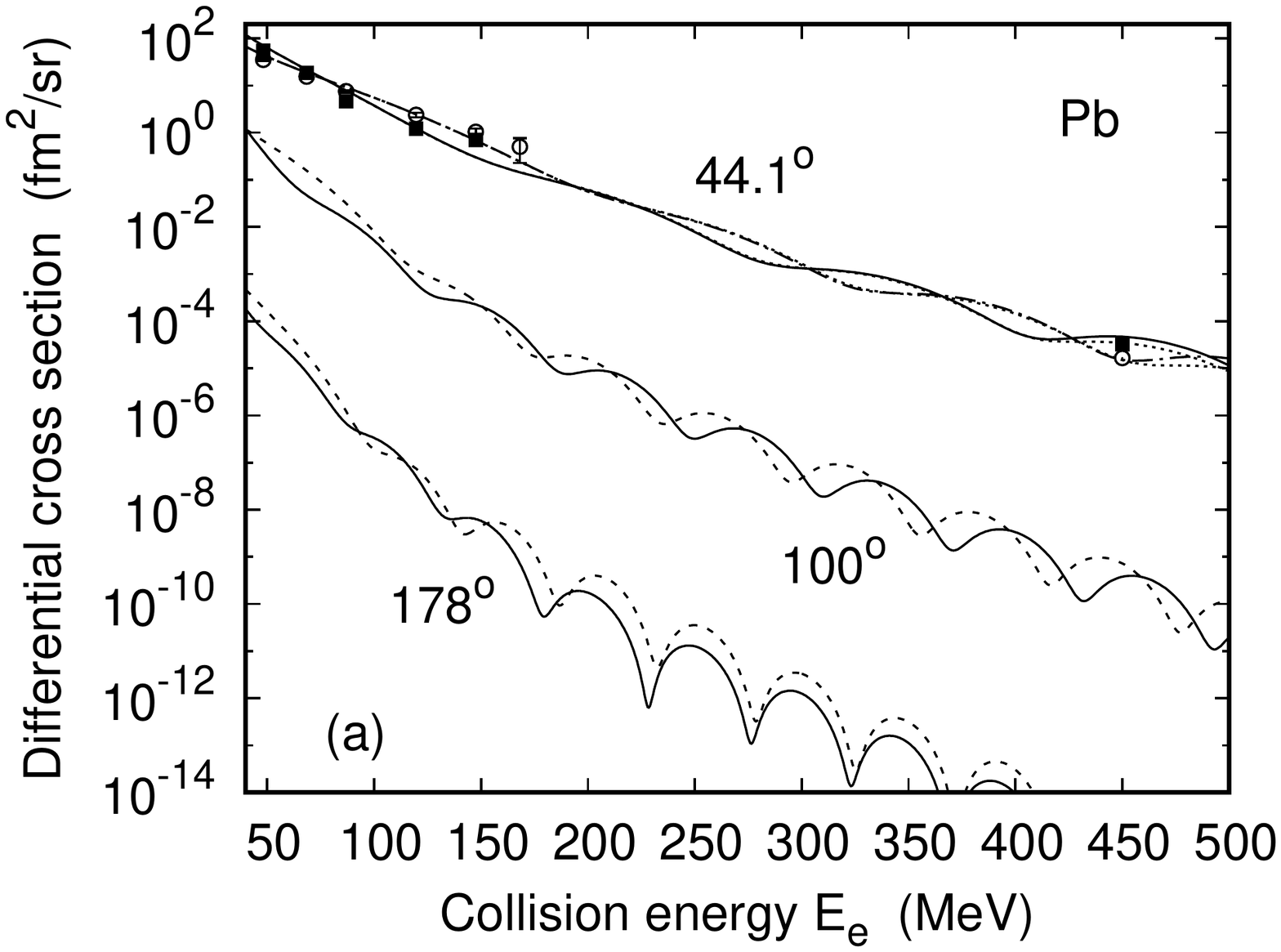}&
\hspace{-4.0cm} \includegraphics[width=.7\textwidth]{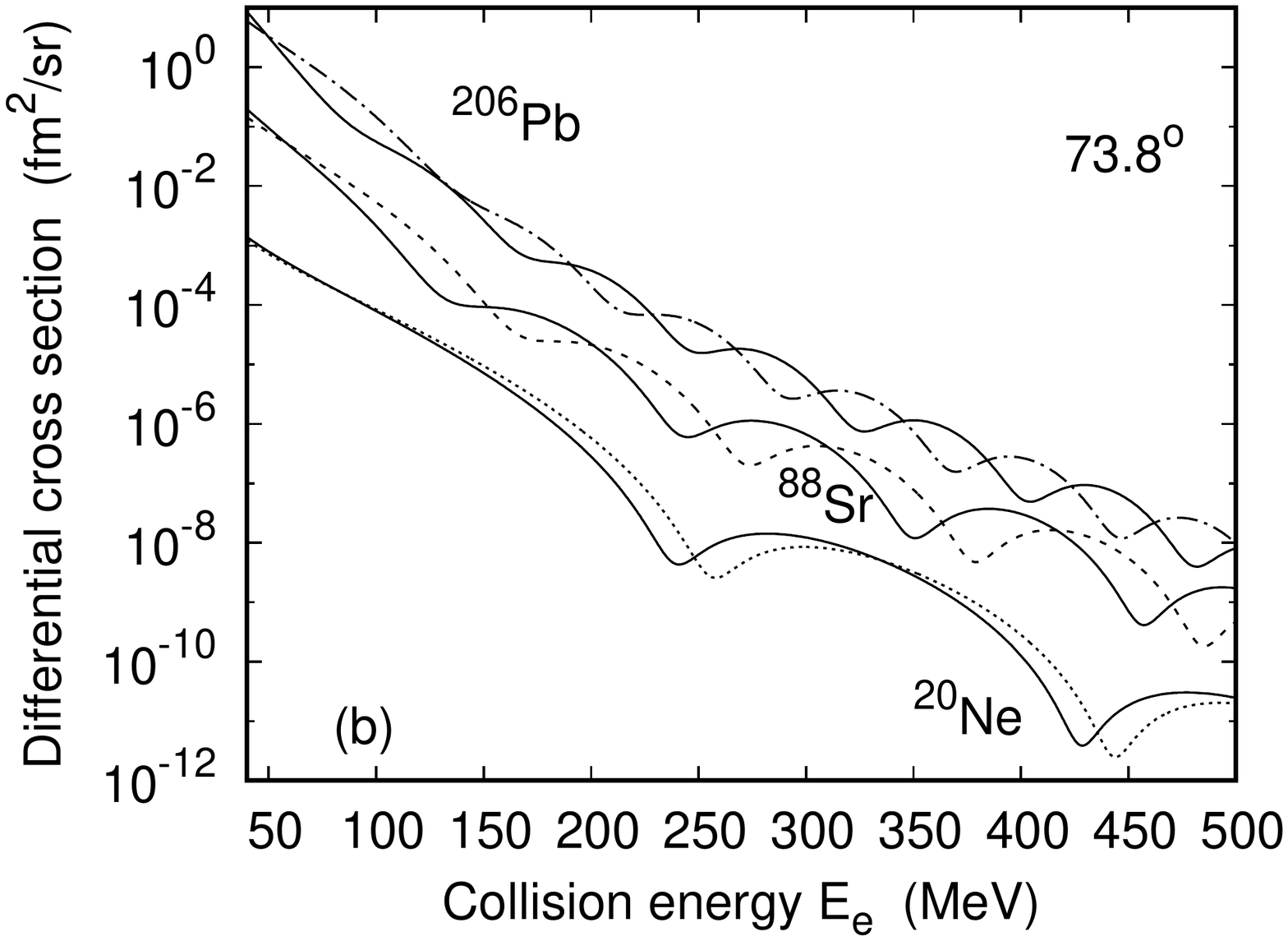}\\
\hspace{-1cm} \includegraphics[width=.7\textwidth]{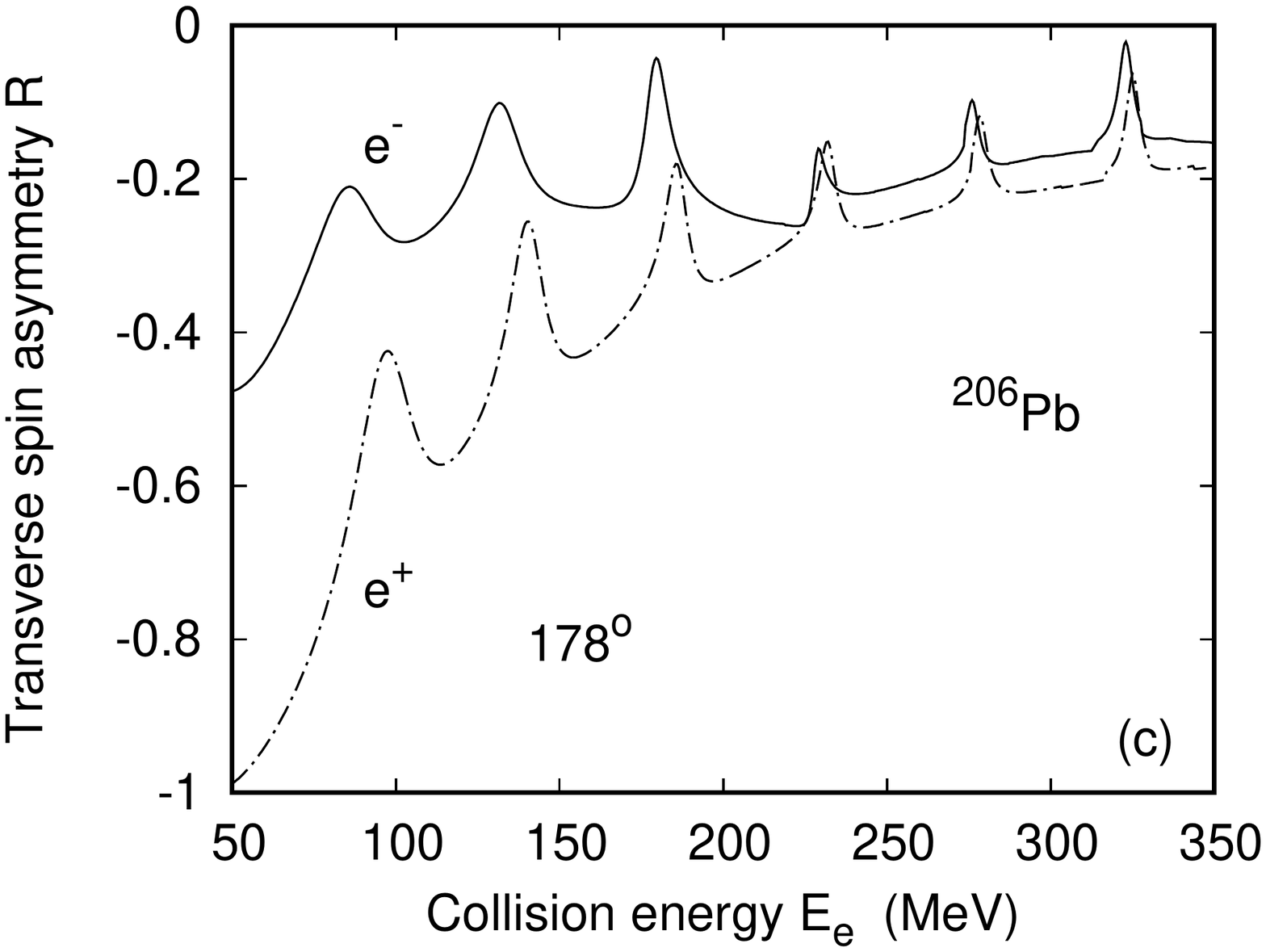}&
\hspace{-4.0cm} \includegraphics[width=.7\textwidth]{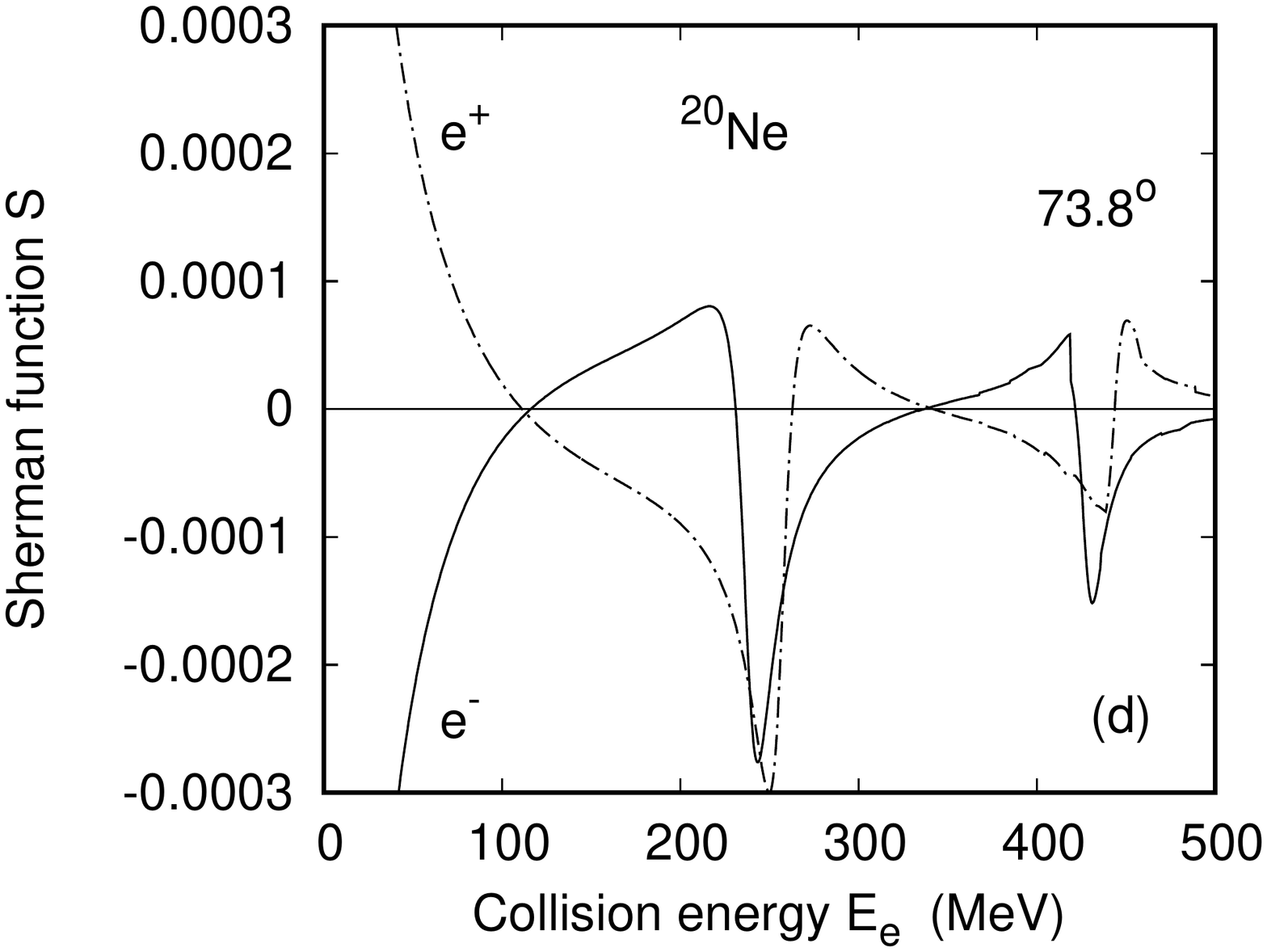}
\end{tabular}
\vspace{-1cm}
\caption{
Differential cross section $\frac{d\sigma}{d\Omega}$ for elastic electron and positron scattering as a function of collision energy $E_e$ for Pb at different scattering angles (a) and at $\theta=73.8^\circ$ for different targets (b).
-----, electron results in (a): upper curve, $44.1^\circ$; middle curve, $100^\circ$; lower curve, $178^\circ$. 
Positron results: $-\cdot - \cdot -,\;44.1^\circ;\;-----,\;$ upper curve, $100^\circ;\;$ lower curve, $\;178^\circ$.
Experimental data at $44.1^\circ$ for electrons $(\blacksquare$) and positrons ($\bigcirc$) are from \cite{MR57} for $E_e<200$ MeV, and from \cite{Bre91}
at 450 MeV interpolated to $\theta=44^\circ$.
The dotted curves at $44.1^\circ$ give the results for $^{208}$Pb using the Fourier-Bessel type charge density. \\
--------, electron results in (b): upper curve, $^{206}$Pb ($Z=82$); middle curve, $^{88}$Sr ($Z=38$); lower curve, $^{20}$Ne ($Z=10$).
Positron results: $-\cdot - \cdot -,\;^{206}$Pb; $-----,\;^{88}$Sr; $\cdots\cdots,\;^{20}$Ne.
For better visibility, the results for Sr are scaled down by a factor of 10, and those for Ne are scaled down by a factor of 100 (e.g. for Ne at 45 MeV, the true cross section is $10^{-1}$ fm$^2$/sr).\\
(c) shows the spin asymmetry $R$ for $^{206}$Pb at $\theta = 178^\circ$ and (d) shows the Sherman function $S$  for $^{20}$Ne at $\theta = 73.8^\circ$ as a function of collision energy for electron impact (----------) and positron impact $(-\cdot - \cdot -$).
}
\end{figure}

\subsection{Positron scattering from nuclei with spin}

For nuclei with spin $J_i\neq 0$ there occurs also magnetic scattering which results from the current-current interaction between the colliding particles.
In addition to the amplitude $f_e$ for potential scattering,
the elastic scattering amplitude consists therefore of the magnetic amplitude, which in DWBA reads for positrons \cite{BD}
\begin{equation}\label{2.11}
A_{fi,e^+}^{\rm mag}(M_i,M_f)\;=\;-\,\frac{1}{c}\int d\bfr_N d\bfr \left( \psi_{i,e^+}^{(-)+}(\bfr,\bfzeta_i)\;\bfalpha \;\psi_{f,e^+}^{(+)}(\bfr,\bfzeta_f)\right)\;
\frac{\stackrel{\leftrightarrow}{I}}{|\bfr-\bfr_N|}\;\left(\phi^+_{M_f}\;\bfJ_N(\bfr_N)\;\phi_{M_i}\right),
\end{equation}
where $M_i$ and $M_f$ are, respectively, the initial and final nuclear spin projections, $\bfalpha$ is a vector of Dirac matrices and $\stackrel{\leftrightarrow}{I}$ is the dyadic unit matrix.
$\bfJ_N$ denotes the nuclear current operator,  $\phi_M$ the nuclear functions, 
$\bfr_N$ is the nuclear coordinate and $\bfr$ is the positron coordinate.
Eq.(\ref{2.11}) differs from the respective  amplitude for electron scattering \cite{TW,Jaku14} not only in the lepton transition matrix element, but also in the negative sign in front of the integral.

The nuclear transition matrix element is multipole expanded \cite{TW,RW} in terms of the vector spherical harmonics $\bfY_{L\lambda}^M$ as defined in \cite{Ed},
\begin{equation}\label{2.12}
\left( \phi_{M_f}^+ \;\bfJ_N(\bfr_N)\;\phi_{M_i}\right) \equiv \;j_{fi}(\bfr_N)\;=\;-i\sum_{\lambda L M} (J_iM_iLM\,|\,J_iM_f)\;J_{L\lambda}(r_N)\;\bfY_{L\lambda}^{M}(\hat{\bfr}_N),
\end{equation}
where we have used that final and initial nuclear spins are identical, $J_f=J_i.$
Also the parity is the same for initial and final nuclear states. The multipolarity of the transition is denoted by $L$.
From parity conservation  
 it follows that for the transitions of unnatural parity, i.e.  for the transverse magnetic transitions, 
 the condition $(-1)^{L+1}=1$ with $\lambda =L$ has to be satisfied.
The transverse electric transitions (of natural parity) require $(-1)^L=1$ and $\lambda = L\pm 1$ \cite{DS,HB}.
The coefficients $J_{L \lambda} $ are the nuclear ground-state current
 densities which can either be obtained from the measured cross sections at the backmost scattering angles, or they have to be calculated from nuclear models (see, e.g., \cite{DS,HB}).

In the following we will only consider the transverse magnetic transitions, since transverse electric transitions are small for nearly spherical nuclei or for not too high momentum transfer, as compared to the contribution from potential scattering \cite{PW}.
Thus the sum in (\ref{2.12}) extends only over odd $L=\lambda$, with $M=M_f-M_i$, where only those $M_f$ are allowed which are compatible with $|M|\,\leq L.$

Upon partial-wave expanding the propagator \cite{PW},
\begin{equation}\label{2.13}
\frac{\stackrel{\leftrightarrow}{I}}{|\bfr-\bfr_N|}\;=\; \sum_{J \lambda \mu} \frac{4\pi}{2\lambda +1}\;\frac{r_<^\lambda}{r_>^{\lambda +1}}\;\bfY_{J\lambda}^\mu(\hat{\bfr})\;\bfY_{J\lambda}^{\mu \ast}(\hat{\bfr}_N),
\end{equation}
where $r_< = \min\{r,r_N\}$ and $r_>=\max\{r,r_N\}$,
inserting the multipole expansion (\ref{2.12}) and integrating over the nuclear solid angle,
the operator for the positron scattering turns into
$$\int d\Omega_N \;\bfalpha\;\frac{\stackrel{\leftrightarrow}{I}}{|\bfr-\bfr_N|}\;\left( \phi_{M_f}^+\;\bfJ_N(\bfr_N)\;\phi_{M_i}\right)$$
\begin{equation}\label{2.14}
=\;-\,i\sum_L \frac{4\pi}{2L+1}\;(J_iM_iLM\,|\,J_iM_f)\;J_{LL}(r_N)\;\frac{r_<^L}{r_>^{L+1}}\;\bfalpha\;\bfY_{LL}^M(\hat{\bfr}).
\end{equation}

With the wavefunction representations (\ref{2.3}) and (\ref{2.4})
the magnetic amplitude (\ref{2.11}) for positron scattering is eventually given by
$$A_{fi,e^+}^{\rm mag}(M_i,M_f)\;=\;-\;\frac{1}{c}\sum_L \frac{\sqrt{4\pi}}{2L+1}\;(J_iM_iLM\,|\,J_iM_f)\sum_{m_i=\pm \frac12} a_{-m_i}\,(-1)^{\frac12 - m_i}$$
\begin{equation}\label{2.15}
\times \;\sum_{m_s=\pm \frac12} b_{-m_s}^\ast\,(-1)^{\frac12 -m_s}\sum_{l_f=0}^\infty (-i)^{l_f}\;Y_{l_fm_l}^\ast (\hat{\bfk}_f) \sum_{j_f=l_f\pm \frac12}(l_fm_l\,\frac12\,m_s\,|\,j_fm_f)
\end{equation}
$$\times \sum_{\kappa_i} \sqrt{2l_i+1} \;i^{l_i}\;e^{i (\delta_{\kappa_i}+\delta_{\kappa_f})}\;(l_i\,0\,\frac12\,m_i\,|\,j_im_i)\;R_{fi}(L)\;W_{12}^{\rm mag}(l_f,l_i',L)$$
with the positron angular function
$$W_{12}^{\rm mag}(l_f,l_i',L)\;=\;\sqrt{\frac{3}{4\pi}}\;\sqrt{2L+1}\;(l_f0L\,0\,|l_i'\,0)\;\sqrt{\frac{2l_f+1}{2l_i'+1}}\;\sum_{m_{s_i}m_{s_f}}(L\mu 1 \varrho\,|\,LM)$$
\begin{equation}\label{2.16}
\times\;(l_i'\mu_i \frac12\,m_{s_i}\,|\,j_im_i)\;(l_f\mu_f\frac12\,m_{s_f}\,|\,j_fm_f)\;(\frac12\,m_{s_f} 1 \varrho\,|\,\frac12\,m_{s_i})\;(l_f\mu_f L\mu\,|\,l_i'\mu_i)
\end{equation}
and the radial integral
\begin{equation}\label{2.17}
R_{fi}(L)\;=\;\int_0^\infty r_N^2dr_N\;J_{LL}(r_N)\int_0^\infty r^2dr\;\frac{r_<^L}{r_>^{L+1}}\left[ g_{\kappa_f}(r)\,f_{\kappa_i}(r)\,+\,f_{\kappa_f}(r)\,g_{\kappa_i}(r)\right]
\end{equation}
$$
=\left( \int_0^\infty \frac{dr_N}{r_N^{L-1}}\;J_{LL}(r_N) \int_0^{r_N} dr\;r^{L+2}\;+\;\int_0^\infty dr_N r_N^{L+2}J_{LL}(r_N)\int_{r_N}^\infty \frac{dr}{r^{L-1}}\right)\left[ g_{\kappa_f}(r)\,f_{\kappa_i}(r)\;+\;f_{\kappa_f}(r)\,g_{\kappa_i}(r)\right],
$$
where use was made of $W_{12}^{\rm mag}(l_f',l_i,L)=-W_{12}^{\rm mag}(l_f,l_i',L).$
The sum over $L$ runs from $L=1$ to  $2J_i$  for half-integer $J_i$ in steps of 2.
The other variables are determined by the selection rules of the Clebsch-Gordan coefficients,
$$M\;=\;M_f-M_i,\qquad m_f+M\;=\;m_i,\qquad m_l\;=\;m_f-m_s,$$
$$\mu\;=\;\mu_i-\mu_f,\qquad \mu_f\;=\;m_f-m_{s_f},\qquad \mu_i\;=\;m_i-m_{s_i},\qquad \varrho\;=\;m_{s_i}-m_{s_f}\,=\,M-\mu,$$
\begin{equation}\label{2.18}
l_f+L+l_i'\;=\;\mbox{even}.
\end{equation}
For each $\kappa_f$, there are $2L+1$ values of $\kappa_i$, given by $\kappa_i=(-1)^n \kappa_f +n-(L+1)$ with $n=1,2,...,2L+1.$

In fact, the formal shape (\ref{2.15}) for the positron magnetic amplitude can readily be brought into the formal shape valid for electrons. To do so, one has to revert simultaneously the signs of $m_i,\;m_s,\;m_{s_i},\;m_{s_f}$ as well as of $\mu_i,\;\mu_f,\;m_f,\;m_l$.
However, $M,\;\varrho$ and $\mu$ must remain unchanged.
Using the symmetry properties of the Clebsch-Gordan coefficients and of the spherical harmonics one ends up
with $A_{fi,e^+}^{\rm mag}(M_i,M_f)$ differing from the
respective amplitude $A_{fi,e^-}^{\rm mag}(M_i,M_f)$ for electrons \cite{Jaku14} only by a global minus sign,
while the Dirac functions $f_\kappa,g_\kappa$ are now eigenfunctions to a potential with sign-reversed nuclear charge number, $-Z$.

The differential cross section for the elastic scattering of polarized positrons into the solid angle $d\Omega$,
 including recoil in the prefactor and in $A_{fi}^{\rm mag}$, is obtained from
\begin{equation}\label{2.19}
\frac{d\sigma}{d\Omega}(\bfzeta_i,\bfzeta_f)\;=\;\frac{k_f}{k_i}\;\frac{1}{f_{\rm rec}}\;\frac{1}{2J_i+1} \sum_{M_i,M_f}\left| A_{fi,e^+}^{\rm coul}\;\delta_{M_f,M_i}\;
 +\;\left(4\pi^3\;\frac{E_iE_f}{c^2}\right)^\frac12\;A_{fi,e^+}^{\rm mag}(M_i,M_f)\right|^2,
\end{equation}
where $A_{fi,e^+}^{\rm coul}$, which is only present for $M_f=M_i$, is identified with the elastic scattering amplitude (\ref{2.5}). 
Recoil effects are small, implying that the recoil factor $f_{\rm rec}$ \cite{DS,Jaku14} is  close to unity, and the final total energy $E_f$ differs only slightly from the total energy $E_i=E_e+c^2$ of the impinging particle.
When comparing with other literature, it should be noted that the factor $-i$ in the definition (\ref{2.12}) of the nuclear transition matrix element $j_{fi}(\bfr_N)$,
as well as the replacement of $\bfY_{L\lambda}^{M}$ by $\bfY_{L\lambda}^{M\ast}$ is irrelevant for the cross section.

\vspace{0.2cm}
\begin{figure}[!h]
\centering
\begin{tabular}{cc}
\hspace{-1cm} \includegraphics[width=.7\textwidth]{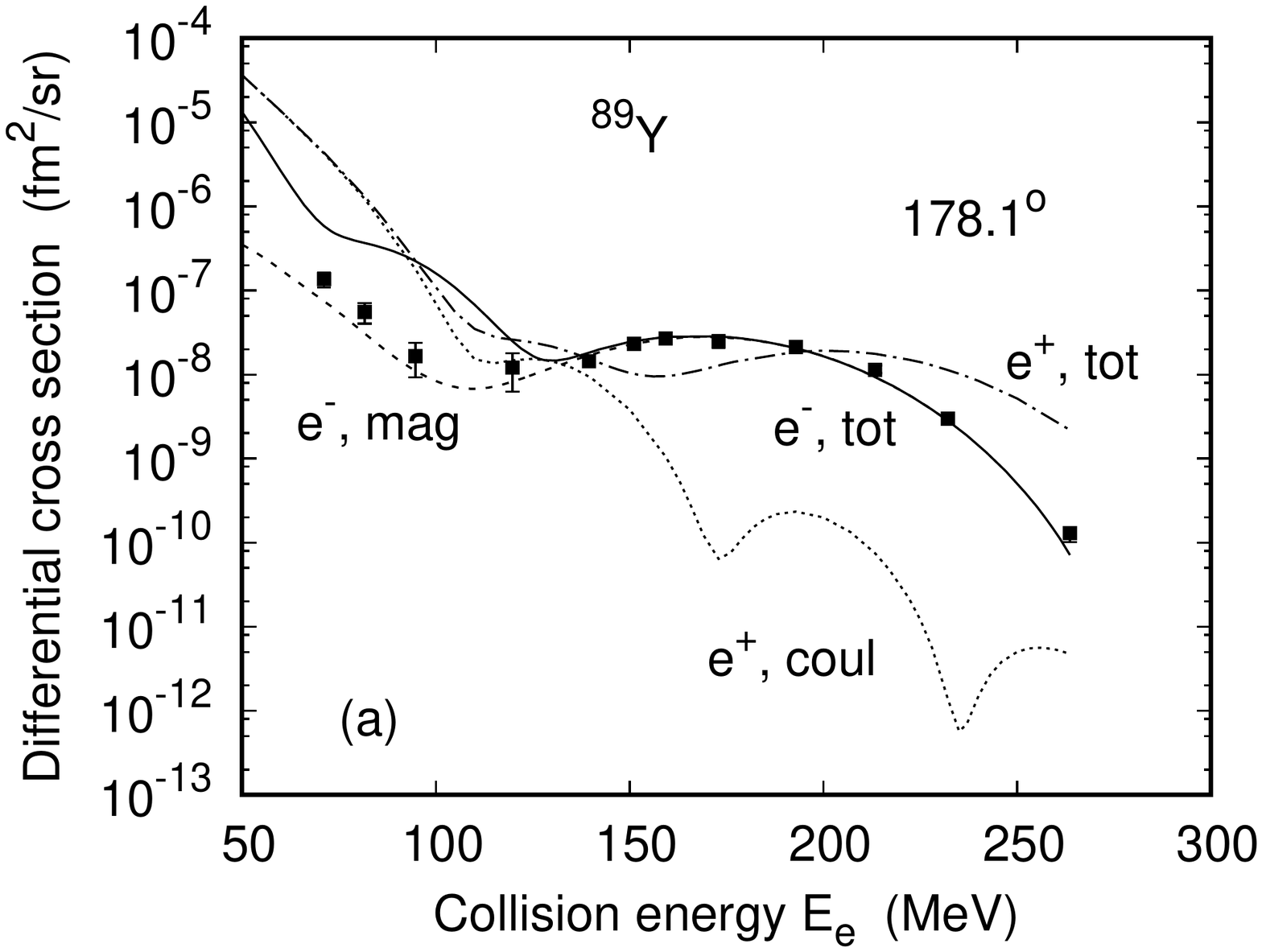}&
\hspace{-4.0cm} \includegraphics[width=.7\textwidth]{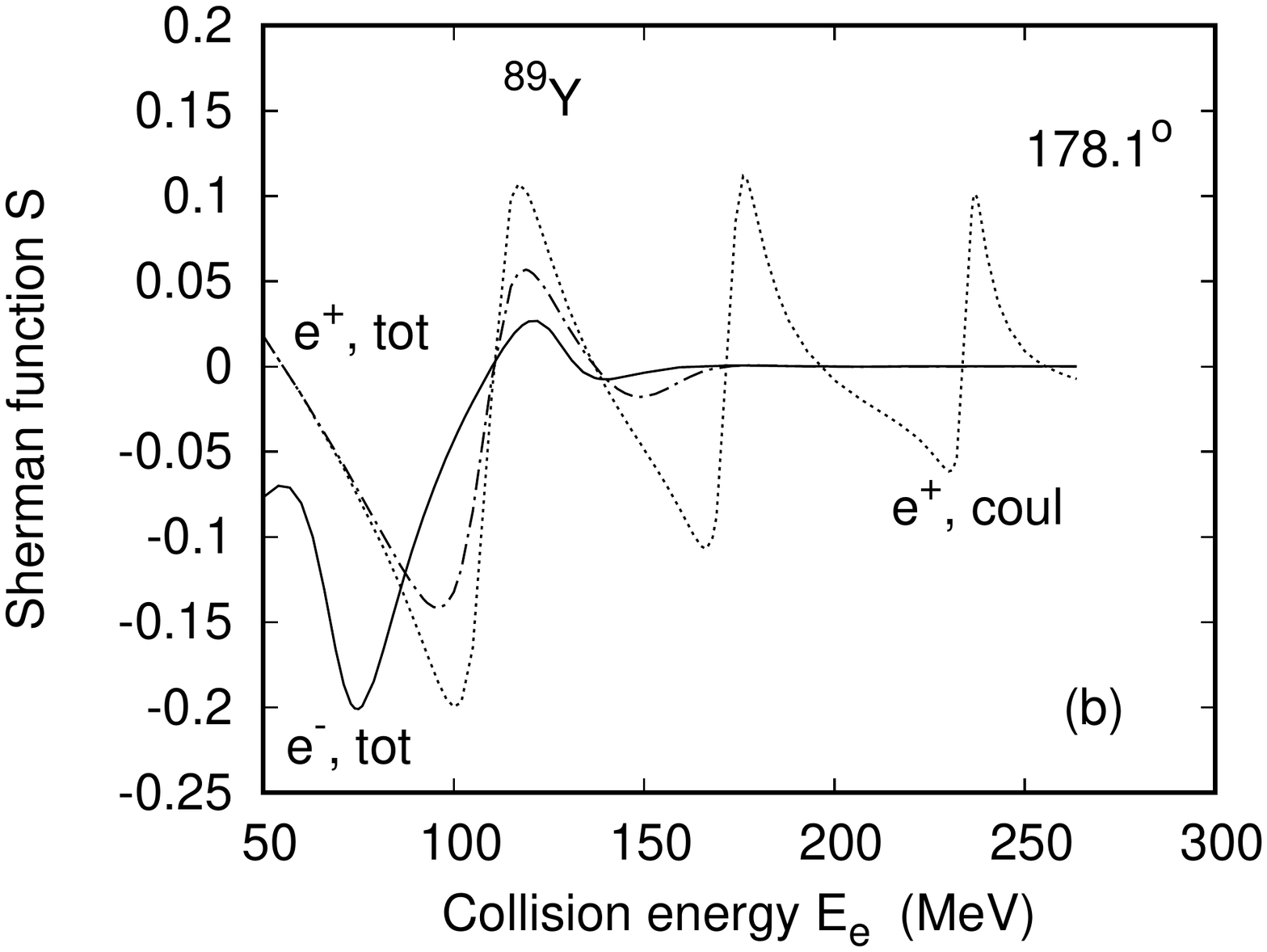}
\end{tabular}
\caption{
Differential cross section $\frac{d\sigma}{d\Omega}$ (a) and Sherman function (b) for electrons and positrons scattered elastically from $^{89}$Y ($Z=39)$ as a function of collision energy $E_e$.
For theory, the scattering angle is $\theta=178.1^\circ$.
The experimental data in (a) for magnetic electron scattering (which are corrected for the effect of potential scattering) are from Wise et al \cite{Wi93}, taken at $\theta = 178.1^\circ$ and $177.6^\circ$, respectively.
(There is no visible difference in the graph if calculated for $177.6^\circ$.)  Shown in (a) are for positron scattering the  results for the total cross section ($-\cdot - \cdot -$, including Coulombic and magnetic conributions), 
together with the separate contribution from pure Coulombic scattering ($\cdots\cdots$).
For electron scattering, ----------, total cross section; $-----$, magnetic contribution to the cross section.\\
In (b), the  results for $S$ refer to positrons ($-\cdot - \cdot -,\;S$ relating to the total cross section; $\cdots\cdots,\;S$ from Coulombic scattering only) and to electrons (----------, $S$ relating to the total cross section).}
\end{figure}

\subsection{Numerical details and results for elastic scattering from $^{89}$Y and $^{23}$Na}

In contrast to potential scattering, the DWBA theory for the magnetic interaction has to meet two difficulties, the convergence of the final-state partial-wave series and the evaluation of the infinite radial integrals.
The convergence of these integrals in terms of the upper integration limit 
 is poor 
for transitions of low multipolarity ($L\lesssim 3)$
because of the strongly oscillating integrand. Therefore the complex-plane rotation method (CRM),  applied in 
\cite{VF} for nucleon transfer, in \cite{YS} for bremsstrahlung 
and in \cite{JP} for nuclear excitation, is used.
If in (\ref{2.17}) the integral from $r_N$ to $\infty$ is split at some distance $R_m$ outside the nuclear charge distribution,
the second part of this double integral factorizes, isolating the integral
\begin{equation}\label{2.20}
\int_{R_m}^\infty dr\;\frac{1}{r^{L-1}}\;\left[ g_{\kappa_f}(r)\,f_{\kappa_i}(r)\,+\,f_{\kappa_f}(r)\,g_{\kappa_i}(r) \right].
\end{equation}
In the CRM, the real integration path
is substituted by a path which follows the positive (or negative) imaginary axis for distances $r>R_m$ and is closed to infinity by the infinitely far semicircle.
This is possible because for sufficiently large $R_m$  the solutions $f_\kappa,g_\kappa$ to the Dirac equation can be written as a superposition of regular and irregular Coulomb Dirac functions \cite{YRW,Sal}.
These functions can for sufficiently large distances readily be analytically continued into the complex plane where they can be split into a sum of two terms, one vanishing on the infinitely far semicircle of the upper half plane,
the other on the respective semicircle in the lower half plane.
With the slight inelasticity due to recoil such that $k_i>k_f$, and with $r=R_m\pm iy,\;y>0$, the resulting complex integrals converge like $\exp(-(k_i-k_f)y)$ \cite{YS,JP}.

For positron scattering, the partial-wave representation of the Coulomb Dirac functions differs in phase from the one belonging to the attractive potential for electrons. Explicitly, each partial wave, associated with some wave number $\kappa$, has to be multiplied by a factor $\delta(\kappa)$ which is given by \cite{Sal}
$$ \delta(\kappa)\;=\;\left\{ \begin{array}{rr}
-1,& \kappa <0\\
1,& \kappa >0 \end{array} \right. \quad \mbox{for electrons}$$
\begin{equation}\label{2.21}
\delta(\kappa)\;=\;1,\qquad \kappa\neq 0\qquad \qquad \mbox{for positrons}.
\end{equation}
This must explicitly be taken care of in the analytical continuation required for the CRM method.
In the finite constituents of $R_{fi}$ along real paths, the correct positron phases are automatically accounted for by the numerical solutions of the Dirac equation for negative $Z$.

 For the $^{89}$Y nucleus, the nuclear ground-state charge density can again be represented in terms of a Fourier-Bessel
expansion \cite{Wi90}.
Concerning the magnetic properties, $^{89}$Y is characterized by a $2p_\frac12$ proton outside a closed-shell core and thus has spin $J_i=\frac12.$ Hence
 there occurs only a single magnetic ground-state current density, $J_{11}$. In contrast to earlier work \cite{Jaku14} we  use here the unscaled $J_{11}$ as provided in \cite{Wi93},
since it gives a better fit to the high-energy experimental data for the magnetic electron scattering.
When including these experimental results in the figure shown below,
 their energy positions were extracted from \cite{Wi90},
where elastic scattering and nuclear excitation had been
measured simultaneously.

Fig.4a displays the differential cross section for electron and positron scattering from $^{89}$Y at a scattering angle of $178^\circ$ as a function of collision energy. 
For low energies, only the Coulombic (potential) scattering is important. However, already near
 100 MeV the magnetic scattering comes into play, and it is dominant above 150 MeV for this angle.
The diffraction structures, which are clearly visible in the potential contribution to elastic scattering, are considerably damped in the total cross section. However, the phase shift between the electron and positron results above 100 MeV is  very prominent.

In Fig.4b the energy dependence of the Sherman function is shown. While for low collision energies (below 60 MeV) the spin asymmetry for electrons and positrons is of opposite sign, this is no longer true in the regime of the diffraction structures.
These oscillations persist also above 150 MeV, but with an amplitude which is strongly damped by the magnetic interaction.
Apart from the phase shift between the electron and positron structures, also the shape of the individual peaks is different,
which is particularly obvious in the Coulombic contribution to $S$ (see also Figs.2b and 3d).

\vspace{0.2cm}
\begin{figure}[!h]
\centering
\begin{tabular}{cc}
\hspace{-1cm} \includegraphics[width=.7\textwidth]{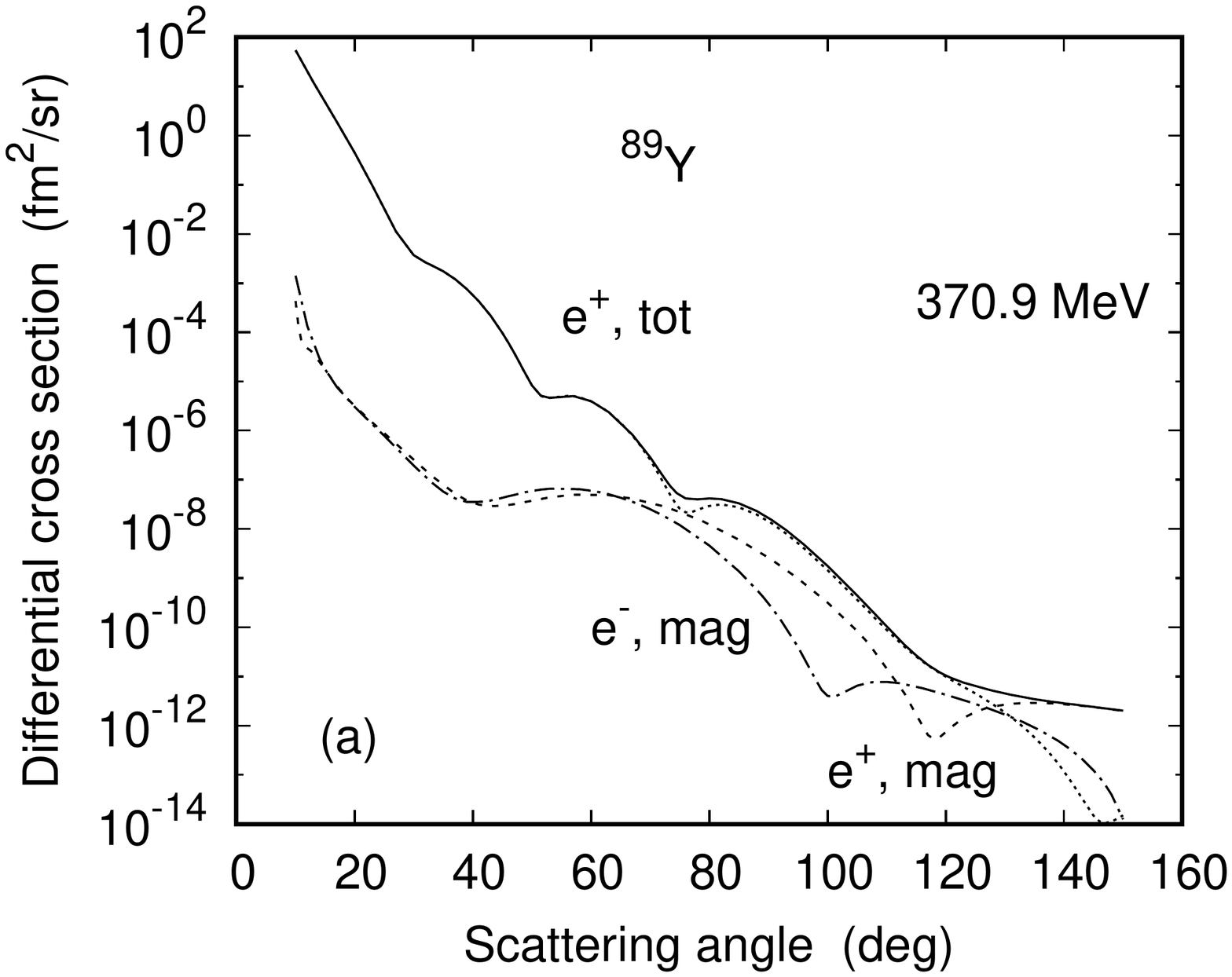}&
\hspace{-4.0cm} \includegraphics[width=.7\textwidth]{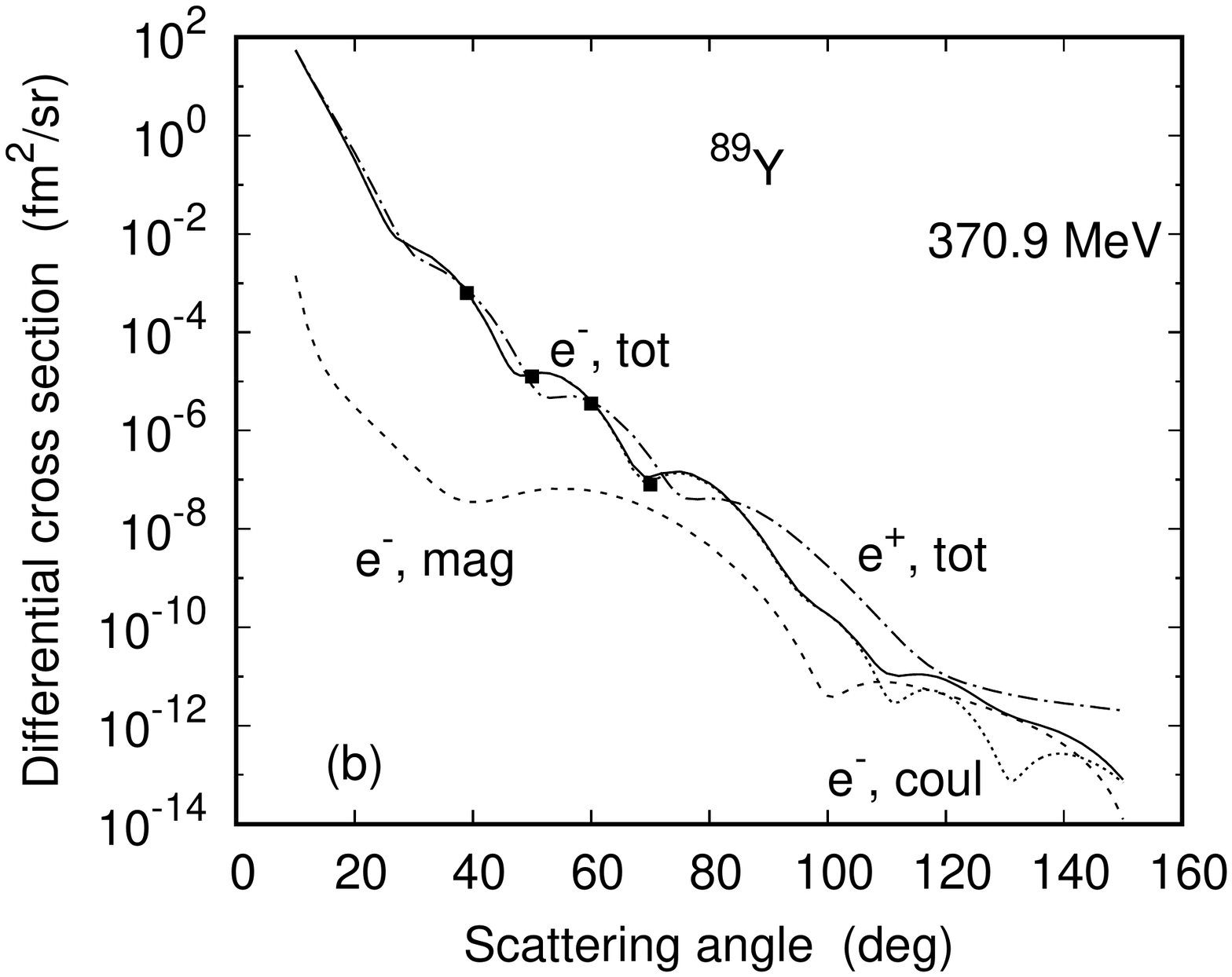}\\
\hspace{-1cm} \includegraphics[width=.7\textwidth]{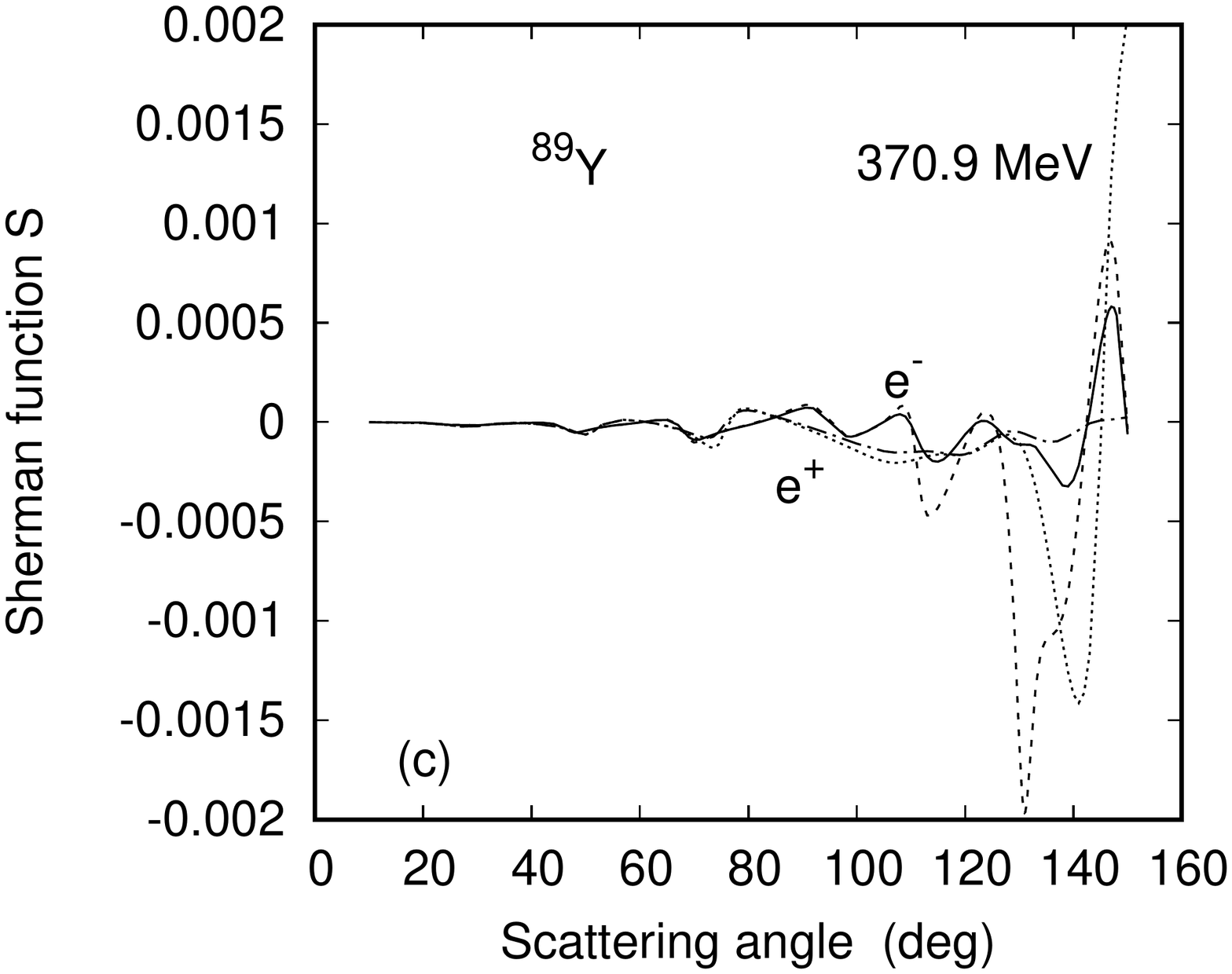}&
\hspace{-4.0cm} \includegraphics[width=.7\textwidth]{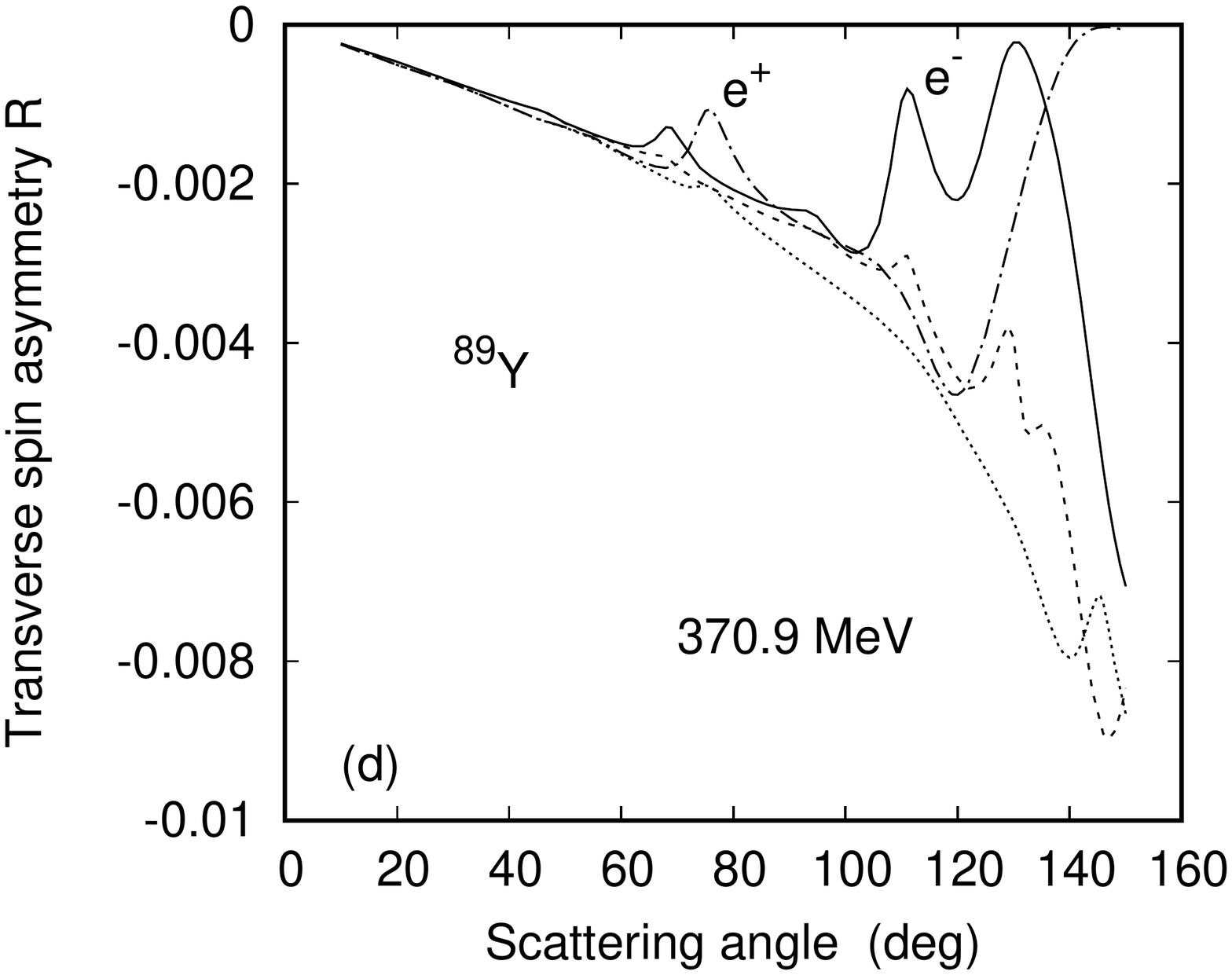}
\end{tabular}
\caption{
Differential cross section $\frac{d\sigma}{d\Omega}$ (a) and (b), Sherman function $S$ (c) and spin asymmetry $R$ (d) for 370.9 MeV positrons and electrons  scattered
elastically from $^{89}$Y as a function of scattering angle $\theta$.
Shown  in (a) are the positron  results for the total cross section (-------------), the Coulombic contribution ($\cdots\cdots$) and the magnetic contribution ($-----$) to the total cross section. For comparison, the magnetic contribution for electron scattering ($-\cdot - \cdot -$) is shown in addition.\\
In (b), electron results are shown for the total cross section (---------------), the Coulombic contribution ($\cdots\cdots$) and the magnetic contribution $(-----$).
The experimental data $(\blacksquare)$ for the total cross section are from \cite{Wi90}.
Included are the results for the total cross section for positron impact ($-\cdot - \cdot -$).\\
In (c), $S$ for positron scattering refers to spin asymmetries of the total cross section ($-\cdot - \cdot -)$ and of the Coulombic contribution ($\cdots\cdots$).
Also shown is $S$ for electron scattering, refering to the total cross section (--------------)
 and to the Coulombic contribution ($-----$).\\
In (d), $R$ is shown for positron scattering (total, $-\cdot - \cdot -$; Coulombic, $\cdots\cdots)$ and for
 electron scattering (total, ---------; Coulombic, $-----$).}
\end{figure}

The angular dependence of the differential cross section and of the polarization correlations $S$ and $R$ is displayed in Fig.5
at a collision energy of 370.9 MeV. At such a high energy the diffraction structures
start already at an angle of $30^\circ$.
Magnetic scattering gains importance at angles near $70^\circ$.
It also shows diffraction oscillations, however with a much longer period, and at backward angles there is a clear phase shift between the results for electrons and positrons (Fig.5a), 
as is the case for the total cross section at forward angles (Fig.5b).
The influence of magnetic scattering leads to a considerable damping of the diffraction structures above $100^\circ$.

Fig.5c compares the angular dependence of the Sherman function for electrons and positrons. There is a significant reduction of $|S|$  above $130^\circ$ due to the magnetic interaction, particularly for the positrons.
This corresponds to  the strong increase of the total positron cross section as compared to potential scattering.
The diffraction structures are also much less visible for positron impact.
This is related to the fact that, due to the repulsive potential, the positron cannot approach the individual protons as much as an electron,
weakening the diffraction effects at backward angles.

Finally, the polarization correlation $R$ is displayed in Fig.5d.
$R$ is even more sensitive than $S$ to the influence of magnetic scattering, causing a severe reduction of this spin asymmetry already near $100^\circ$.

\vspace{0.2cm}

The nucleus $^{23}$Na, chosen as another example of nuclei with large magnetic moments,
 has a $1d_{\frac{3}{2}}$ odd-proton configuration with a filled $1d_{\frac{5}{2}}$ proton subshell.
Its ground-state charge distribution $\varrho(r)$ was obtained from nuclear structure calculations \cite{RH} (see also \cite{HR1,HR2}). In contrast to a Fermi-type charge distribution $\varrho_F(r)$ with parameters $c=2.98$ fm (the charge radius \cite{RH}) and $a=0.49$ fm (such that the rms radius of 2.94 fm is reproduced \cite{VJ}), $\varrho(r)$ has a pronounced dip at small $r$.

Since $^{23}$Na has spin $J_i=\frac{3}{2}$, the two magnetic current densities $J_{11}$ and $J_{33}$ contribute to the magnetic amplitude.
They are obtained from the respective transverse magnetic form factors $F_L^T$, as provided in \cite{Si} and improved in \cite{Jas},
by means of the Fourier-Bessel transform
\begin{equation}\label{2.23}
J_{LL}(r)\;=\;\frac{2}{\pi}\int_0^\infty q^2dq\;F_L^T(q)\;j_L(qr),
\end{equation}
where $j_L$ is a spherical Bessel function.
In order to ensure convergence, a Gaussian tail (decaying with $q$ according to $e^{-b^2q^2/4}$, where $b$ is the oscillator length) was fitted to the numerically available $F_L^T(q)$ at large $q$.
Likewise, the power law $F_L^T(q) \sim q^L$ for small $q$ was used.
For the finite radial integrals in (\ref{2.17}) an upper cutoff at $r_N=25$ fm for $J_{11}$ and $r_N=22$ fm for $J_{33}$ was taken, ensuring that the inverse transformation to (\ref{2.23}) indeed reproduces $F_L^T(q)$.

Fig.6a displays the energy dependence of the cross section at $70^\circ$ and $178^\circ$, together with the result for potential scattering.
While at the forward angle magnetic scattering influences the cross section only in its minimum near 250 MeV,
it dominates the cross section at $178^\circ$ already for enaergies as low as 60 MeV.
Comparing with $^{89}$Y at the same angle (Fig.4a) it is seen that for $^{23}$Na the magnetic scattering is considerably
stronger in the region ranging from 50 MeV to 170 MeV.
The phase shift between the positron and the electron results is very similar for $70^\circ$ and $178^\circ$,
in contrast to the situation for Pb (Fig.3a) where there exists only potential scattering at all angles.

Included is also the result for electron scattering if the Fermi-type charge density $\varrho_F$ is used instead.
While there is hardly any difference for energies below 150 MeV,
the minimum near 230 MeV for $\theta = 70^\circ$ is slightly shifted.
The deviations increase, however, with energy or with scattering angle, where the inner region of the charge density is probed.
The dip in the nuclear-structure charge distribution leads in that region to a 
considerable decrease of intensity as compared to
the results for $\varrho_F$.
For example at $178^\circ$, or even more at $150^\circ$ where potential scattering is not yet suppressed, the Fermi-type results are well below the nuclear structure ones for $E_e >250$ MeV.

In Figs.6b, 6c and 6d the Sherman function is compared for the two leptons at the three angles $70^\circ,\;150^\circ$ and $178^\circ$.
For energies below 150 MeV there is a marked symmetry between the electron and positron results for this light nucleus, and at the smallest angle the result is similar to the one for $^{20}$Ne (Fig.3d) where only potential scattering takes place.
For $^{23}$Na the reduction by the magnetic scattering affects just the minimum of $S$ near 240 MeV at that angle.
This reduction gets stronger at $150^\circ$, and eventually leads to a nearly complete loss of spin asymmetry at the backmost angle,
as compared to $S$ from potential scattering, except for a small excursion near 210 MeV.
Nevertheless, the Sherman function increases strongly with scattering angle, which is particularly prominent at the lowest energies.
However, it is still true in the maximum of $|S|$ around 200-250 MeV.
For positrons, this maximum amounts to $2.2\times 10^{-4}$ at $70^\circ,\;8.2\times 10^{-4}$ at $150^\circ$ and $2.5\times 10^{-3}$ at $178^\circ$,
with similar results for electrons.
The influence of different types of nuclear charge distributions is also present in the Sherman function.
Against expectation, the changes in $S$ are not more prominent than those in the cross section at energies below 300 MeV.
At $150^\circ$ where the effect is largest, the Fermi-type results are shown in addition (Fig.6c).

\vspace{0.2cm}
\begin{figure}[!h]
\centering
\begin{tabular}{cc}
\hspace{-1cm} \includegraphics[width=.7\textwidth]{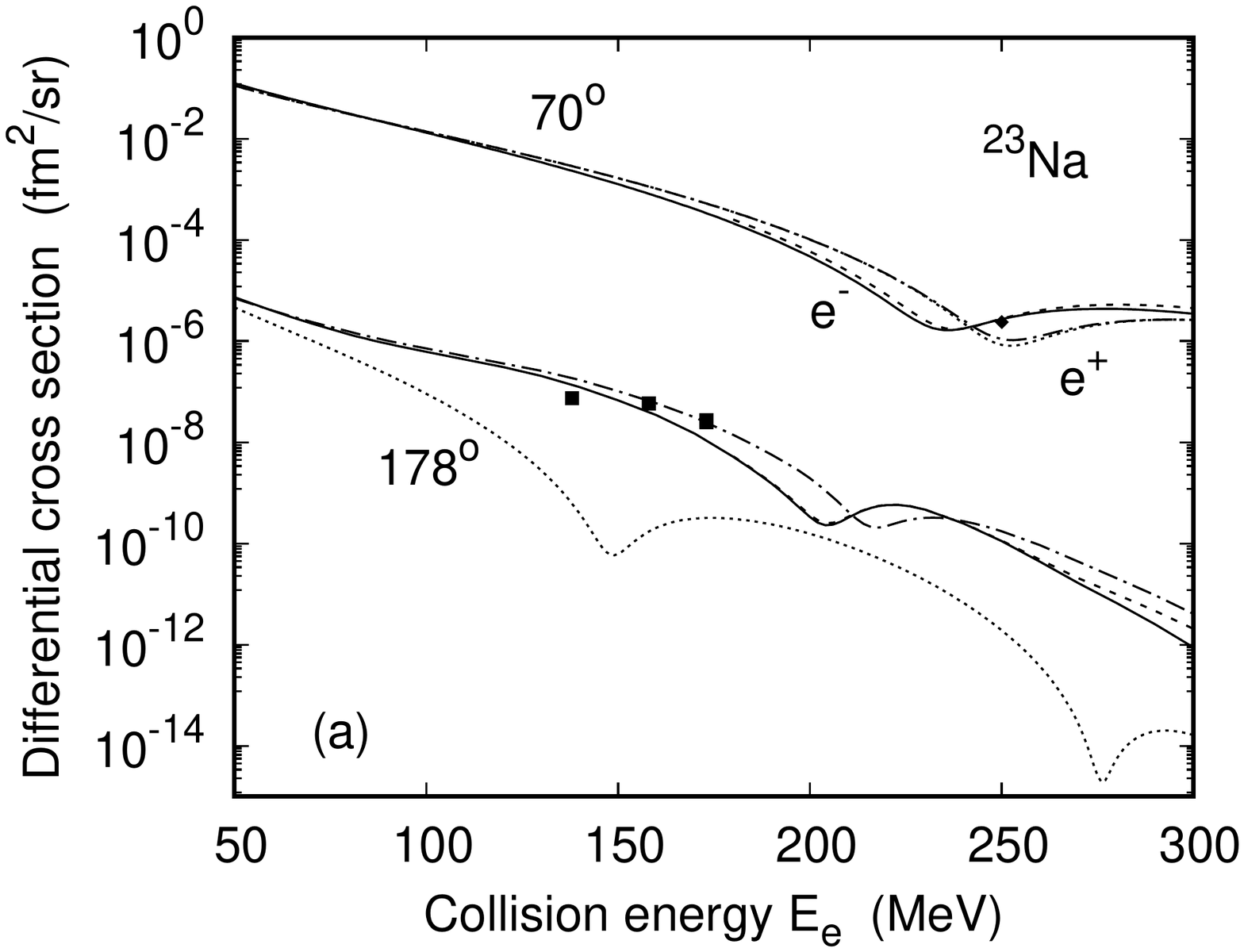}&
\hspace{-4.0cm} \includegraphics[width=.7\textwidth]{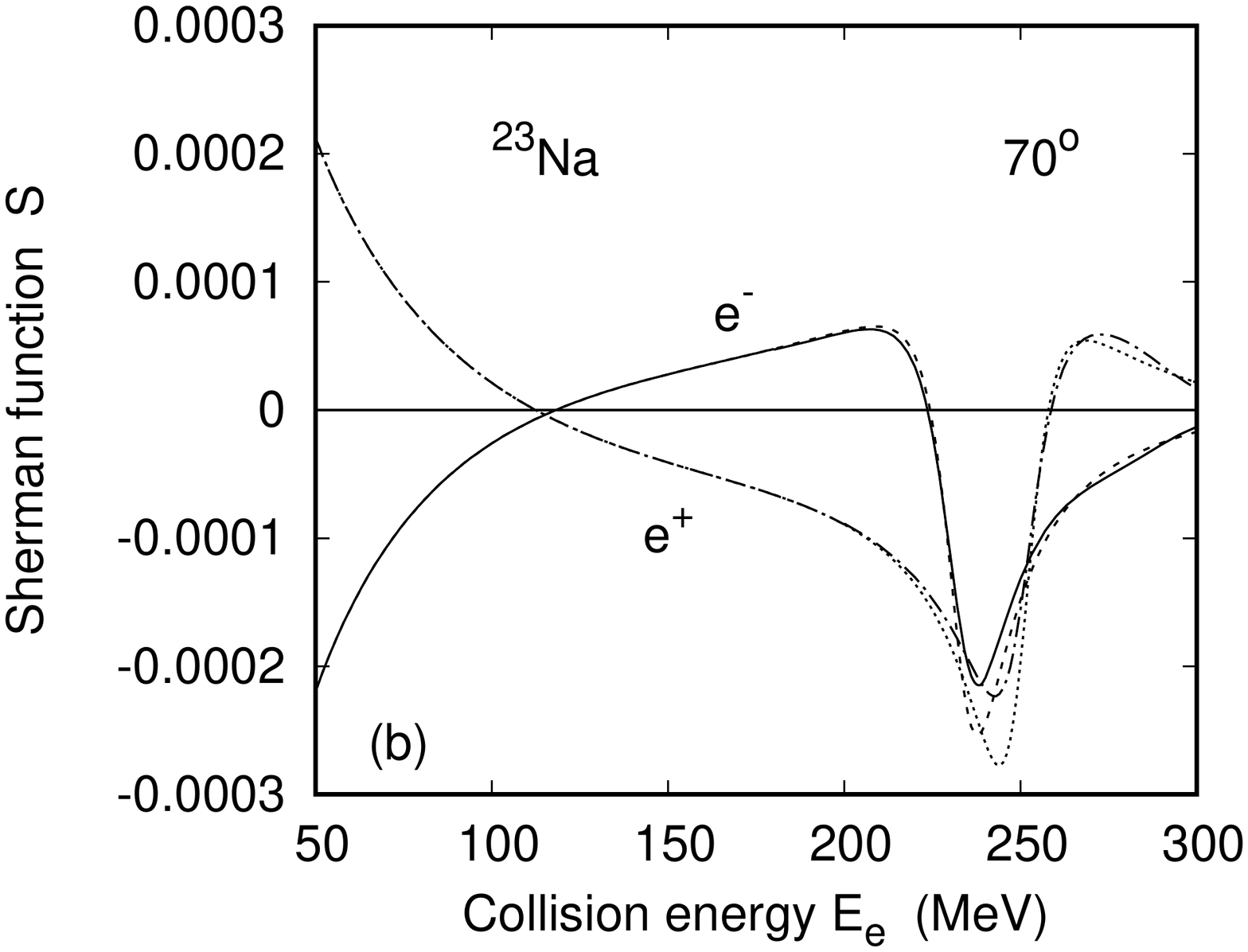}\\
\hspace{-1cm} \includegraphics[width=.7\textwidth]{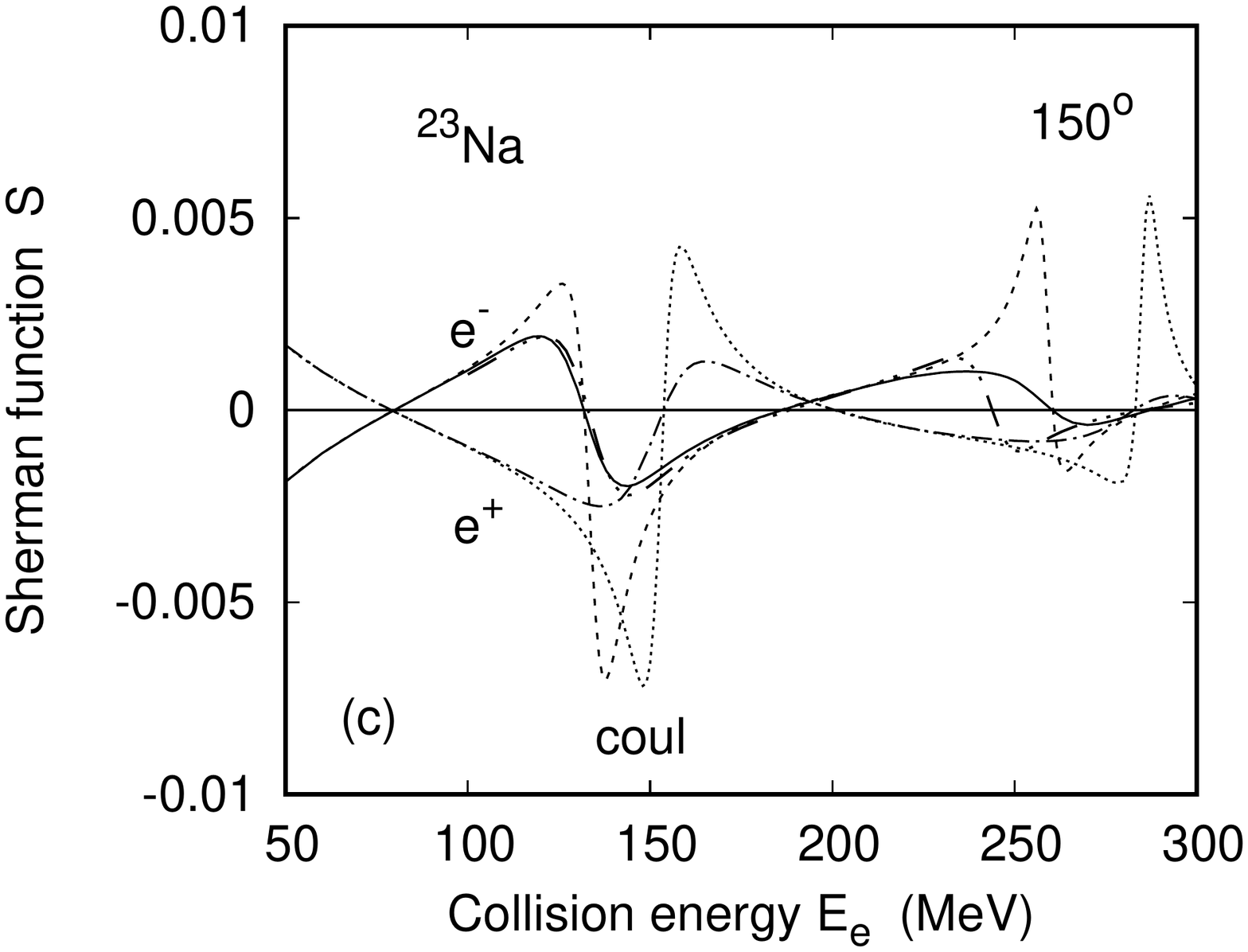}&
\hspace{-4.0cm} \includegraphics[width=.7\textwidth]{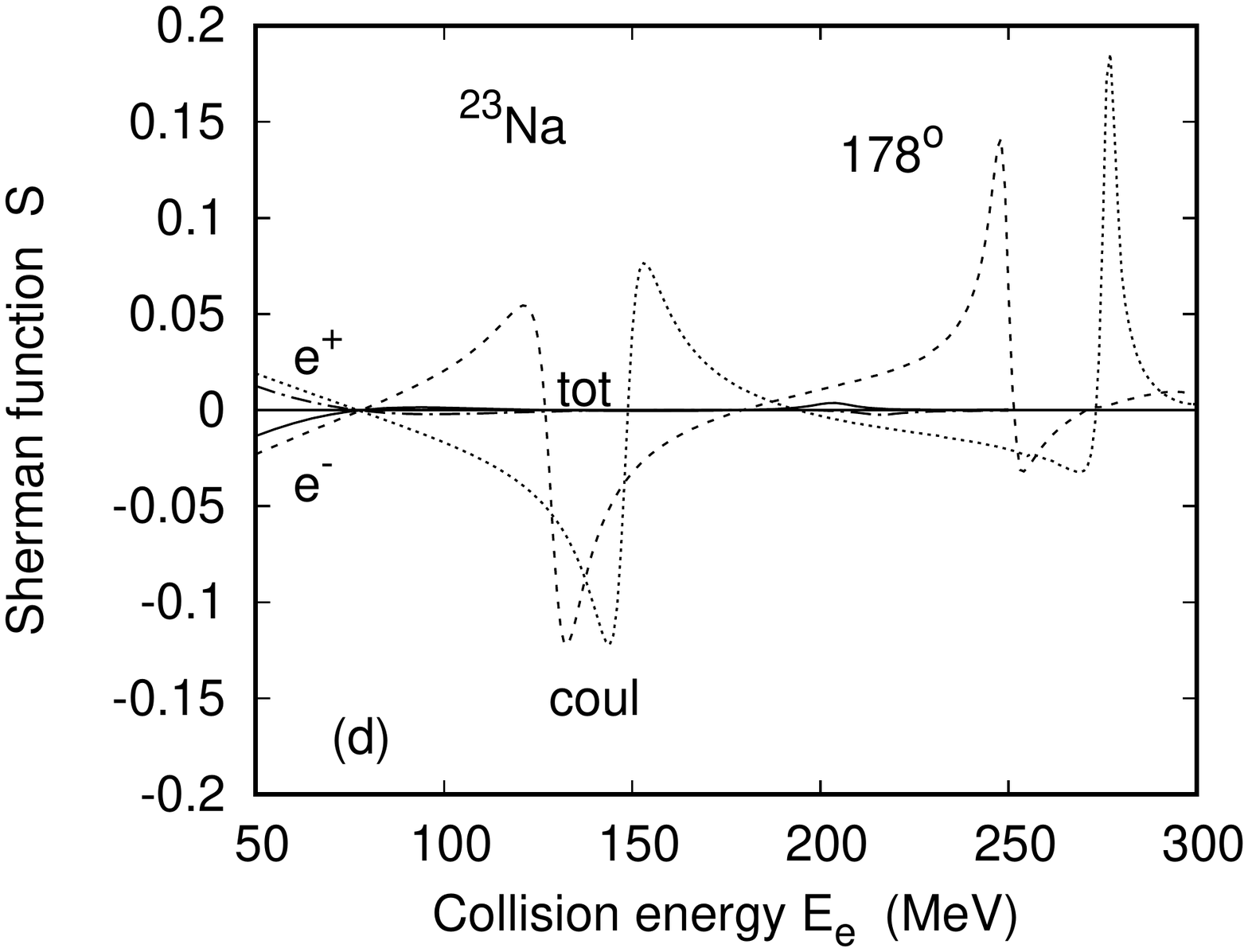}
\end{tabular}
\caption{
Differential cross section $\frac{d\sigma}{d\Omega}$ (a) and  Sherman function $S$ at scattering angles $70^\circ$ (b), $150^\circ$ (c) and $178^\circ$ (d)
 for positrons and electrons  scattered
elastically from $^{23}$Na  ($Z=11$) as a function of collision energy $E_e$.
Shown are the results including magnetic scattering for electrons (--------------) and positrons $(-\cdot - \cdot -$), and the results from positron potential scattering ($\cdots\cdots$).\\
In (a), the upper curves are for $70^\circ$ and the lower curves for $178^\circ$.
The total electron cross section results relating to  $\varrho_F$ are included ($-----$). The experimental electron data ($\blacksquare$ at $180^\circ$, corrected for the effect of potential scattering)
are from Torizuka as cited in \cite{Si},
the datum point $(\blacklozenge$) at $70^\circ$ and 250 MeV  is interpolated from Singhal et al \cite{Si}. \\
In (b), (c) and (d) the spin asymmetry from electron potential scattering ($-----$) is shown in addition. In (c) the results relating to the total electron scattering cross section, but using $\varrho_F$, are also included ($-\cdots -$).
 }
\end{figure}

\section{Positron bremsstrahlung}
\setcounter{equation}{0}

Since we use the relativistic partial-wave theory for the calculation of bremsstrahlung, which, in contrast to elastic scattering, involves multiple sums of partial waves,
we have to restrict ourselves for reasons of convergence  to collision energies below 30 MeV.
For such moderate energies it is sufficient to consider only the finite nuclear size, while  high-energy effects
like magnetic scattering or dynamical recoil for nuclei with spin need not be included. Also kinematical recoil effects are small. As a matter of fact, they are included in the triply differential cross section, while, for the sake of an analytic integration over the positron angles, they are omitted in the doubly differential cross section.

Consideration of the finite nuclear size effects is done by choosing the positron functions, like in the case of elastic scattering, as solutions to the Dirac equation with a  potential generated by the nuclear ground-state charge density. 

\subsection{Theoretical description}

The radiation matrix element for positron bremsstrahlung is calculated from \cite{BD,Y12}
\begin{equation}\label{3.1}
W_{{\rm rad},e^+}(\bfzeta_i,\bfzeta_f)\;=\;-\;\frac{ie}{c} \int d\bfr\;\psi_{i,e^+}^{(-)+}(\bfr,\bfzeta_i)\;\bfalpha\bfeps_\lambda^\ast\;e^{-i\bfks\bfrs}\;\psi_{f,e^+}^{(+)}(\bfr,\bfzeta_f),
\end{equation}
with the scattering states $\psi_{i,e^+}^{(-)}$ and $\psi_{f,e^+}^{(+)}$  given in (\ref{2.3}) and (\ref{2.4}), respectively.
The polarization of the emitted photon is denoted by $\bfeps_\lambda$,
and its momentum by $\bfk$. When comparing with the magnetic amplitude (\ref{2.11}) for elastic scattering, it is seen  that in both cases the operator $\bfalpha$
mediates the transition between the initial and final positron spinors.
We shall restrict ourselves to the case where the polarization of the scattered positron is not observed
and thus has to be summed over.
According to the formalism developed for electrons \cite{T02,Jaku16}, 
 the triply differential cross section for positrons of total energy $E_i$ emitting a photon with frequency $\omega=ck$ into the solid angle $d\Omega_k$, 
while being scattered with final total energy $E_f$ into the solid angle $d\Omega_f$, is given by 
\begin{equation}\label{3.2}
\frac{d^3\sigma}{d\omega d\Omega_k d\Omega_f}(\bfzeta_i,\bfeps_\lambda)\;=\;\frac{4\pi^2 \omega k_fE_iE_f}{c^5k_i f_{\rm re}}\sum_{m_s=\pm\frac12} \left| F_{fi,e^+}(\bfzeta_i,m_s)\right|^2,
\end{equation}
where $f_{\rm re}$ is a recoil factor close to unity, and $F_{fi,e^+}$ is defined by means of
\begin{equation}\label{3.3}
ic\;W_{{\rm rad},e^+}(\bfzeta_i,\bfzeta_f)\;=\;\sum_{m_s=\pm\frac12}b_{-m_s}^\ast\;(-1)^{\frac12-m_s}\;F_{fi,e^+}(\bfzeta_i,m_s).
\end{equation}

Upon partial-wave expanding the photon operator \cite{Ed},
\begin{equation}\label{3.4}
e^{-i\bfks\bfrs}\;=\;4\pi \sum_{l=0}^\infty (-i)^l\;j_l(kr)\;\sum_\mu Y_{l\mu}^\ast(\hat{\bfk})\;Y_{l\mu}(\hat{\bfr}),
\end{equation}
where $j_l$ is a spherical Bessel function and $Y_{l\mu}$ a spherical harmonic function, $F_{fi,e^+}$ turns into
$$F_{fi,e^+}(\bfzeta_i,m_s)\;=\;\frac{i}{\sqrt{4\pi}}\sum_{l_f=0}^\infty \sum_{m_l=-l_f}^{l_f} (-i)^{l_f}\;Y_{l_fm_l}^\ast(\hat{\bfk}_f)
\sum_{j_f=l_f\pm\frac12}\;(l_fm_l\frac12\,m_s\,|\,j_fm_f)$$
\begin{equation}\label{3.5}
\times\;\sum_{m_i=\pm\frac12} a_{-m_i}\;(-1)^{\frac12-m_i}\sum_{\kappa_i} \sqrt{2l_i+1}\;i^{l_i}\;e^{i(\delta_{\kappa_i}+\delta_{\kappa_f})}\;(l_i0\frac12\,m_i\,|\,j_im_i)\;S_{fi,e^+}.
\end{equation}
In our coordinate system the $z$-axis is taken along $\bfk_i$ as before, while the reaction $(x,z)$-plane is spanned by $\bfk_i$ and $\bfk$, such that $\bfe_y =\bfk_i \times \bfk/ |\bfk_i \times \bfk|$ and $\bfe_x=\bfe_y \times \hat{\bfk}_i$ with $\hat{\bfk}=(\sin \theta_k,0,\cos \theta_k)$ where $\theta_k$ is the emission angle of the photon.
The final positron momentum $\bfk_f$ is thus characterized by the spherical angles $\vartheta_f$ and $\varphi_f$.
The factor $S_{fi}$ includes the sum over the photon
angular momenta $l$,
\begin{equation}\label{3.6}
S_{fi,e^+}\;=\;\sum_{l=|l_i'-l_f|}^{l_i'+l_f}(-i)^l\;R_{12}(l)\;W_{12,e^+}(l_f,l_i',l)\;
-\;\sum_{l=|l_i-l_f'|}^{l_i+l_f'} (-i)^l\;R_{21}(l)\;W_{12,e^+}(l_f',l_i,l).
\end{equation}
This sum  runs in steps of 2 due to the selection rules from the Clebsch-Gordan coefficients in $W_{12,e^+}$, $l_f+l+l_i'=$ even in the first sum, and $l_f'+l+l_i=$ even
 in the second one.
The radial integrals $R_{12}$ and $R_{21}$ are given by
\begin{equation}\label{3.7}
{R_{12}(l) \choose R_{21}(l)}\;=\;\int_0^\infty r^2\,dr\;j_l(kr)\;{g_{\kappa_f}(r)\; f_{\kappa_i}(r) \choose f_{\kappa_f}(r)\;g_{\kappa_i}(r)}
\end{equation}
and agree with the ones for electron scattering (except for the negative charge number in the defining equation of the radial functions), while the functions $W_{12,e^+}$ result from the angular integration,
$$W_{12,e^+}(l_f,l_i',l)\;=\;\sqrt{3}\;(2l+1) \;(l_f0\,l\,0|
\,l_i'0)\;\sqrt{\frac{2l_f+1}{2l_i'+1}}\;\sum_{m_{s_f},m_{s_i}}\sqrt{\frac{(l-\mu)!}{(l+\mu)!}}\;P_l^\mu(\cos \theta_k)\;c_\varrho^{(\lambda)}$$
\begin{equation}\label{3.8}
\times \;(l_i'\mu_i\frac12\,m_{s_i}|\,j_im_i)\;(l_f \mu_f \frac12 m_{s_f}\,|\,j_fm_f)\;(\frac12 m_{s_f} 1 \varrho\,|\,\frac12 m_{s_i})\;(l_f\mu_f l\mu\,|\,l_i'\mu_i),
\end{equation}
where $P_l^\mu$ is a Legendre function, using the reduction of $Y_{l\mu}^\ast(\hat{\bfk})$ for  $\bfk$ in the reaction plane.
The $c_\varrho^{(\lambda)}$ are the coefficients of $\bfeps_\lambda^\ast=\sum_\varrho c_\varrho^{(\lambda)} \bfe_\varrho$ in terms of the spherical unit vectors $\bfe_\varrho$ (with $\varrho = 0,\pm 1$ \cite{Ed}).
For circularly polarized photons, defined by $\bfeps_\lambda \equiv \bfeps^{(\pm)}=\,\frac{1}{\sqrt{2}}\,(\bfeps_{\lambda_2}\mp i \bfeps_{\lambda_1})$ with $\bfeps_{\lambda_1}=(0,1,0)$ and
 $\bfeps_{\lambda_2}=(-\cos \theta_k,0,\sin \theta_k)$ the basis vectors for linear polarization, these coefficients are given by
$$c_\varrho^{(+)}\;=\left\{ \begin{array}{cc} \frac12\;(\cos \theta_k-1),& \varrho=1\\
-\,\frac12 (\cos \theta_k +1),& \varrho=-1\\
\sin \theta_k/\sqrt{2},& \varrho =0,
\end{array}\right.$$
\begin{equation}\label{3.9}
c_\varrho^{(-)}\;=\;\left\{ \begin{array}{cc}
-c_{-\varrho}^{(+)},& \varrho=\pm 1\\
c_\varrho^{(+)},& \varrho =0
\end{array} \right. \;.
\end{equation}

The selection rules for the magnetic quantum numbers imply
\begin{equation}\label{3.10}
\mu\;=\;\mu_i-\mu_f,\quad \mu_f\;=\;m_f-m_{s_f},\quad \mu_i\;=\;m_i-m_{s_i},\quad \varrho\;=\; m_{s_i}-m_{s_f},\quad m_l\;=\;m_f-m_s,
\end{equation}
which is a subset of the relations (\ref{2.18}).

\vspace{0.2cm}
\begin{figure}[!h]
\centering
\begin{tabular}{cc}
\hspace{-1cm} \includegraphics[width=.7\textwidth]{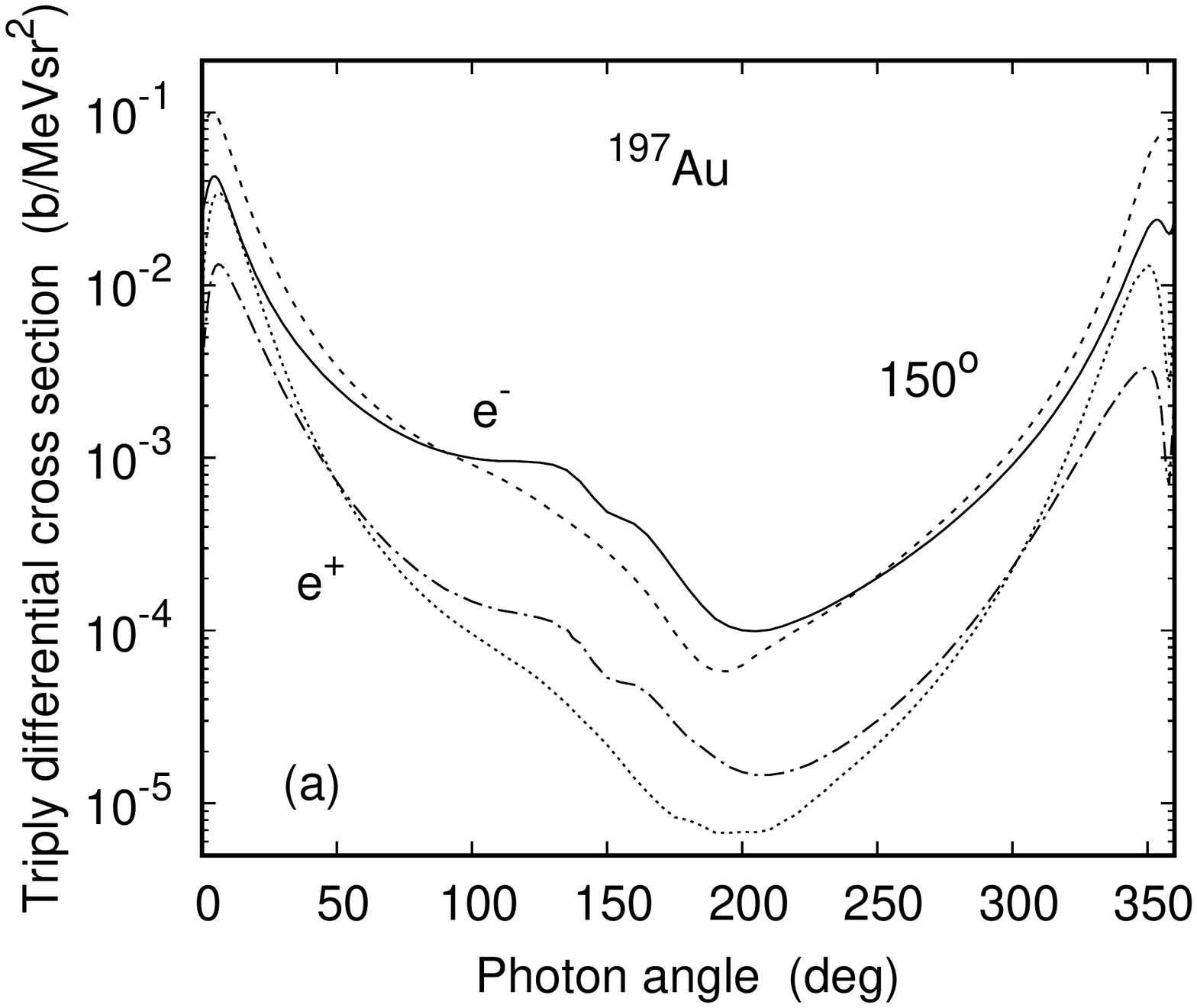}&
\hspace{-4.0cm} \includegraphics[width=.7\textwidth]{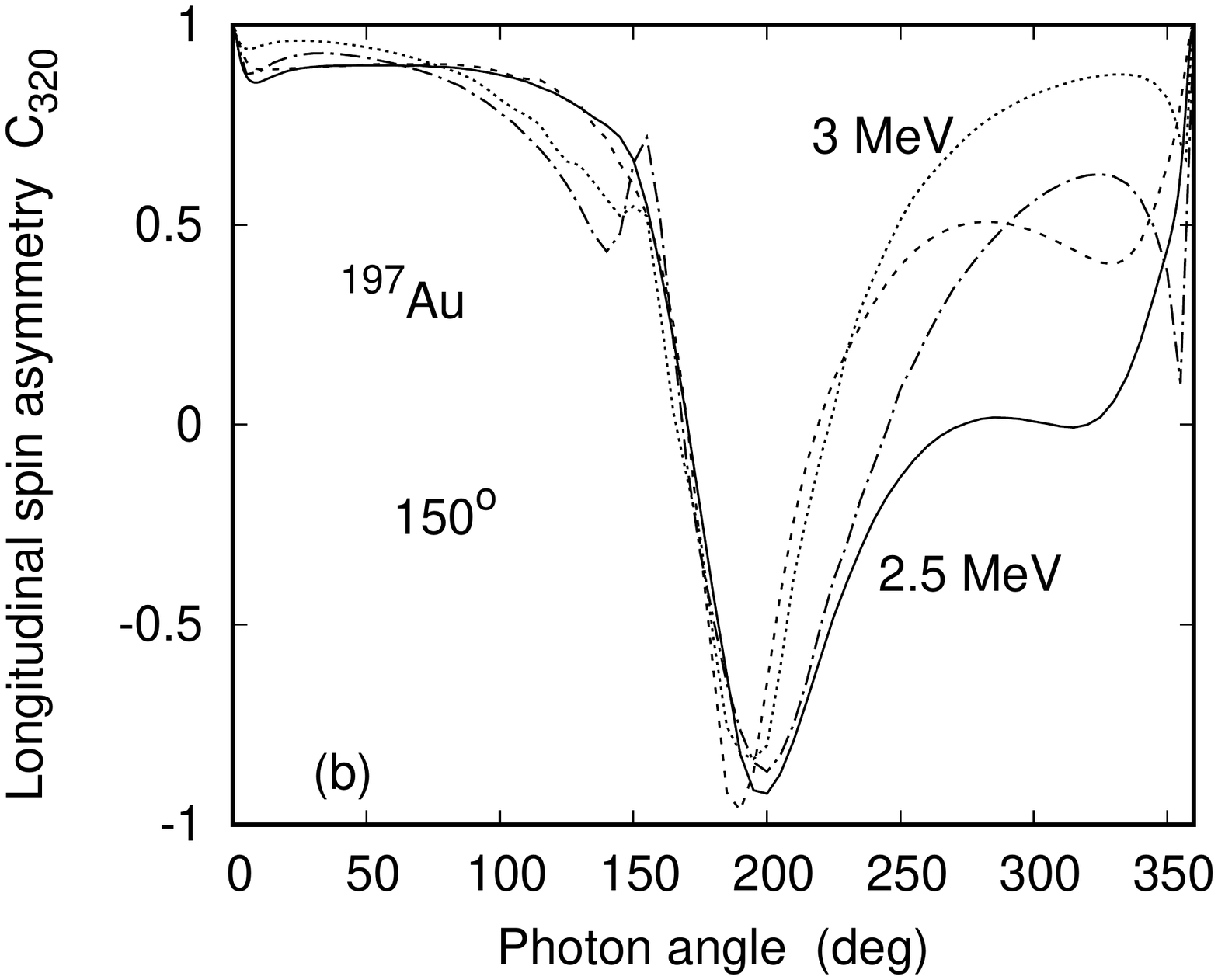}\\
\hspace{-1cm} \includegraphics[width=.7\textwidth]{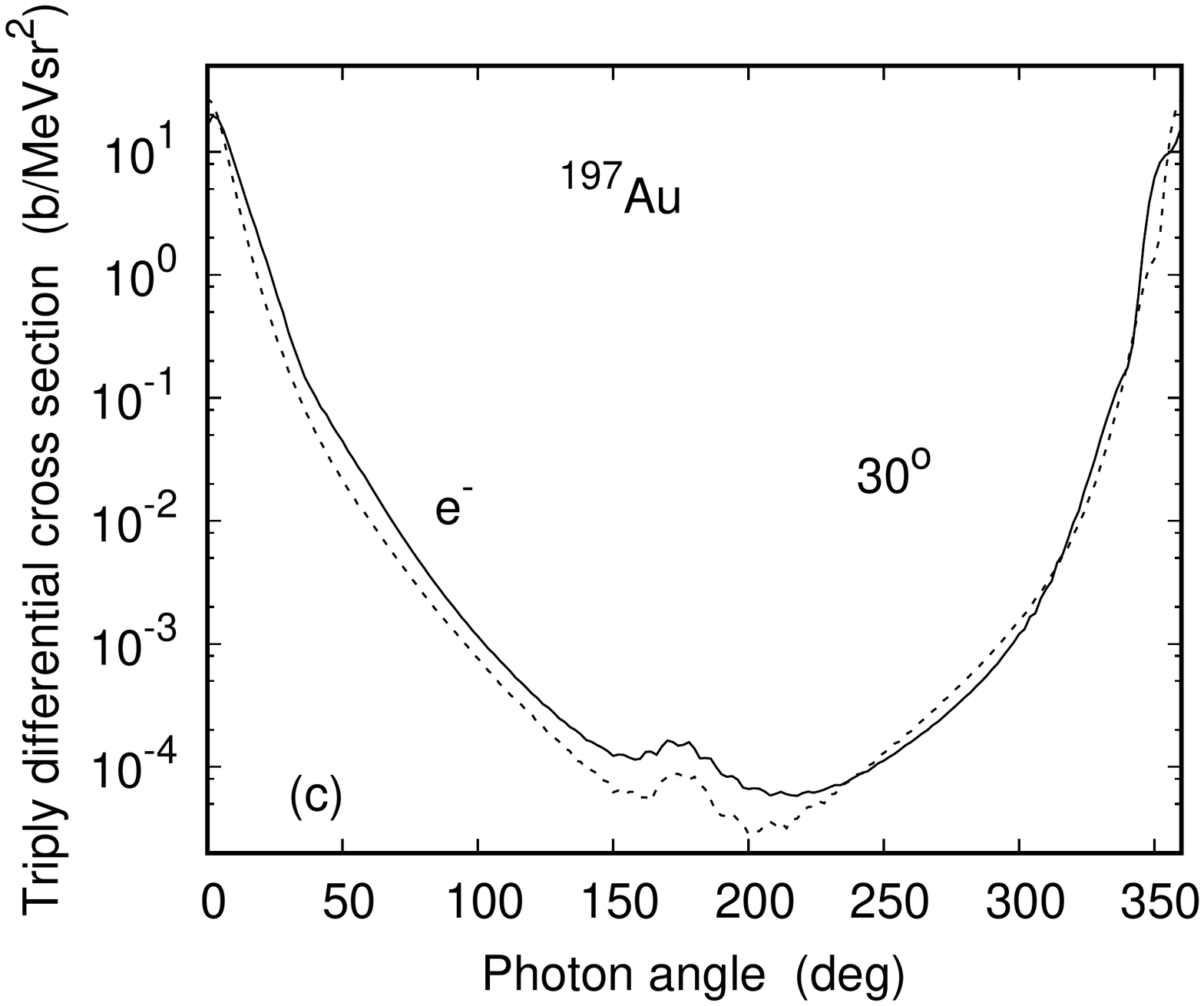}&
\hspace{-4.0cm} \includegraphics[width=.7\textwidth]{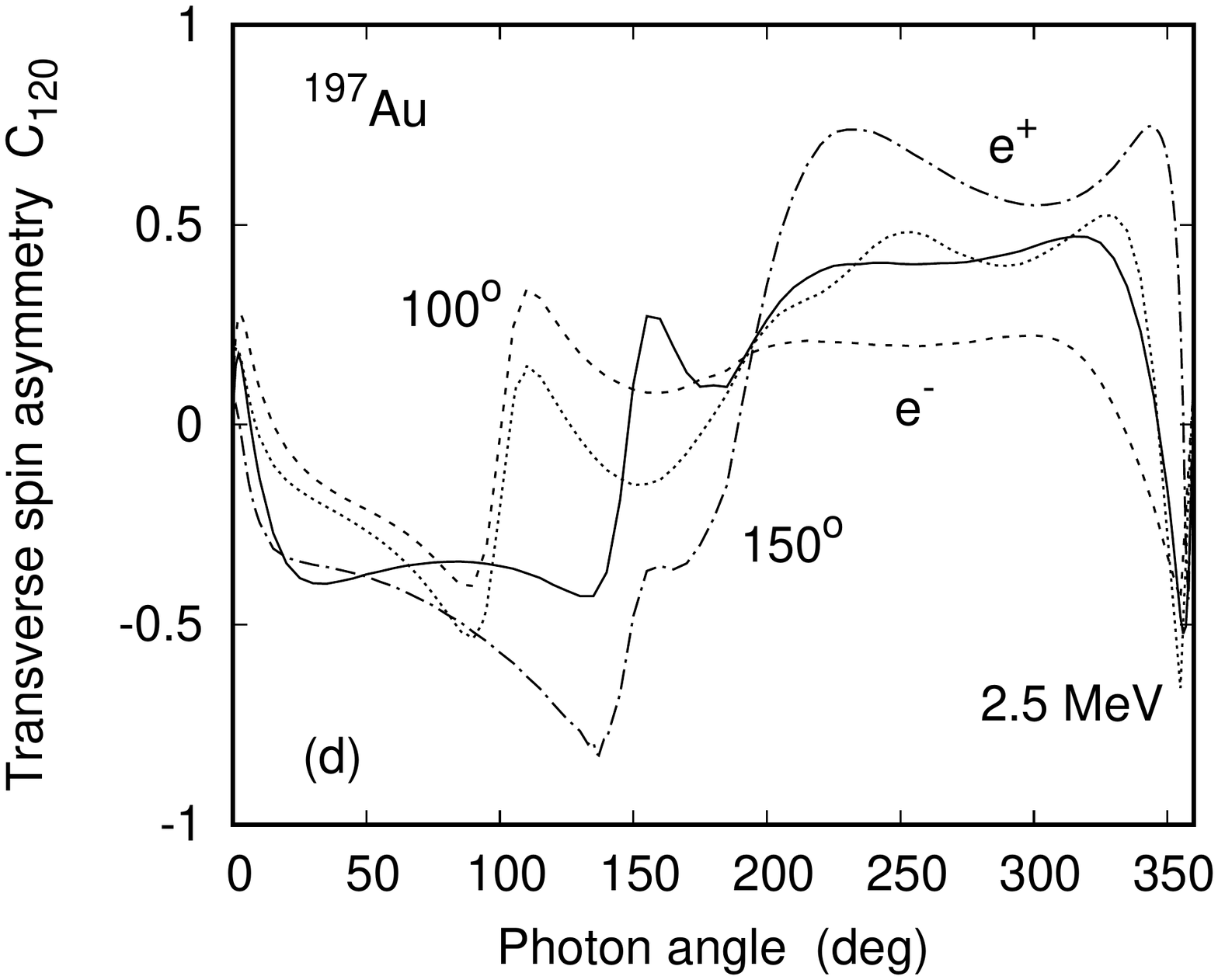}
\end{tabular}
\caption{
Triply differential bremsstrahlung cross section $\frac{d^3\sigma}{d\omega d\Omega_k d\Omega_f}$(a), (c), and polarization correlations $C_{320}$ (b) and $C_{120}$ (d) for 3.5 MeV electrons and positrons  colliding with $^{197}$Au ($Z=79$) as a function of photon angle $\theta_k$
 for coplanar geometry ($\varphi_f=0).$
Shown in (a) and (b) are the results for frequencies $\omega=2.5$ MeV
(--------------, electrons; $-\cdot - \cdot -$, positrons)
and for $\omega = 3$ MeV ($-----$, electrons; $\cdots\cdots$, positrons) at a scattering angle of $\vartheta_f=150^\circ$.
In (c), the electron cross sections for $\omega=2.5$ MeV (--------------------) and $\omega=3$ MeV ($-----$) are compared at $\vartheta_f=30^\circ$.
Large wiggles near $180^\circ$ point to numerical uncertainties. In (c) they
 are partly smoothed by means of averaging over $\theta_k$ (over a range $\Delta \theta_k  = 10^\circ$).
In (d), results are provided for $\omega=2.5 $ MeV and $\vartheta_f=150^\circ$ ( -----------------, electrons; $-\cdot - \cdot -$, positrons)
as well as for  $\vartheta_f=100^\circ\;\,( -----,$ electrons; $\cdots\cdots$, positrons).
}
\end{figure}

The interrelation between electron and positron bremsstrahlung turns out to be based on the formal identity, 
\begin{equation}\label{3.14}
\sum_{m_s}\left| F_{fi,e^+}(\bfzeta_i,m_s)\right|^2\;=\;\sum_{m_s}\left| F_{fi,e^-}(\bfzeta_i,m_s)\right|^2,
\end{equation}
taken into consideration that the respective radial integrals as well as the phase shifts differ in the sign of $Z$.

Eq.(\ref{3.14}) can be proved by changing in (\ref{3.2}) simultaneously the sign of all magnetic quantum numbers except for $\varrho$ and $\mu$, and by making use of the symmetry properties of the Clebsch-Gordan coefficients and of the spherical harmonic functions.

For the sake of completeness we provide the explicit result for electron bremsstrahlung, to be inserted into (\ref{3.2}) in place of $F_{fi,e^+}$,
$$F_{fi,e^-}(\bfzeta_i,m_s)\;=\;\frac{i}{\sqrt{4\pi}}\sum_{l_f=0}^\infty \sum_{m_l=-l_f}^{l_f}(-i)^{l_f}\;Y_{l_f m_l}(\hat{\bfk}_f)\sum_{j_f=l_f\pm \frac12}(l_fm_l\frac12\,m_s\,|\,j_fm_f)$$
$$\times\; \sum_{m_i=\pm \frac12} a_{m_i}\sum_{\kappa_i} \sqrt{2l_i+1}\;i^{l_i}\;e^{i(\delta_{\kappa_i}+\delta_{\kappa_f})}\; (l_i0\frac12\,m_i\,|\,j_im_i)\;S_{fi,e^-},$$
\begin{equation}\label{3.11}
S_{fi,e^-}\;=\;\sum_{l=|l_i'-l_f|}^{l_i'+l_f}(-i)^l\;R_{12}(l)\;W_{12,e^-}(l_f,l_i',l)
\;-\;\sum_{l=|l_i-l_f'|}^{l_i+l_f'}(-i)^l\;R_{21}(l)\;W_{12,e^-}(l_f',l_i,l),
\end{equation}
where
$$W_{12,e^-}(l_f,l_i',l)\;=\;\sqrt{3}\;(2l+1)\;(l_i'0\,l\,0|\,l_f0)\;\sqrt{\frac{2l_i'+1}{2l_f+1}}\sum_{m_{s_f},m_{s_i}}\sqrt{\frac{(l-\mu)!}{(l+\mu)!}}\;P_l^\mu(\cos \theta_k)\;c_\varrho^{(\lambda)}$$
\begin{equation}\label{3.12}
\times \;(l_i'\mu_i\frac12\,m_{s_i}|\,j_im_i)\;(l_f\mu_f\frac12\,m_{s_f}\,|\,j_fm_f)\;(\frac12\,m_{s_i}1\varrho\,|\,\frac12\,m_{s_f})\;(l_i'\mu_il\mu\,|\,l_f\mu_f)
\end{equation}
with
\begin{equation}\label{3.13}
\mu\;=\;\mu_f-\mu_i,\quad \mu_f\;=\;m_f-m_{s_f},\quad \mu_i\;=\;m_i-m_{s_i},\quad \varrho\;=\;m_{s_f}-m_{s_i},\quad m_l\;=\;m_f-m_s.
\end{equation}

If restriction is made to coplanar emission of photon and positron, such that the azimuthal angle $\varphi_f$ of the scattered positron (with respect to the reaction plane)  is either 0 or $180^\circ$,
the polarization correlations $C_{ij0}$ between the incoming positron and the emitted photon are defined in the following way \cite{T02},
$$\frac{d^3\sigma}{d\omega d\Omega_k d\Omega_f}(\bfzeta_i,\bfeps_\lambda)\;=\;\frac12\left( \frac{d^3\sigma}{d\omega d\Omega_k d\Omega_f}\right)_0 \;\left[ 1\,+\,C_{030}\xi_3\;+\;(C_{110}\xi_1-C_{120}\xi_2)\;(\bfzeta_i \bfe_x)\right.$$
\begin{equation}\label{3.15}
-\;\left. (C_{230}\xi_3+C_{200})\;(\bfzeta_i \bfe_y)\;-\;(C_{310} \xi_1-C_{320}\xi_2)\;(\bfzeta_i \bfe_z)\right],
\end{equation}
where the prefactor is the cross section for unpolarized particles.
The first index in the subscript of $C_{ij0}$
denotes the direction of $\bfzeta_i$ along the coordinate axes $\bfe_x\; \bfe_y$ and $\bfe_z$, while the index 0 refers to unpolarized scattered particles. 
The second index is associated with the photon polarization, which
 is defined in terms of the unit vector $\bfxi$ with coordinates
\begin{equation}\label{3.16}
(\xi_1,\xi_2,\xi_3)\;=\;(2\mbox{ Re}\,(\beta_1 \beta_2^\ast),\,2\mbox{ Im}\,(\beta_1\beta_2^\ast),\,|\beta_2|^2\,-\,|\beta_1|^2\,),
\end{equation}
where $\beta_1$ and $\beta_2$ are the expansion coefficients of the polarization vector $\bfeps_\lambda=\beta_1\epsilon_{\lambda_1}+\beta_2\epsilon_{\lambda_2}$ in terms of the basis vectors $\epsilon_{\lambda_1}$ and $\epsilon_{\lambda_2}$.
When circularly polarized photons are considered, for which $\xi_1=\xi_3=0$, (\ref{3.15}) reduces to
\begin{equation}\label{3.17}
\frac{d^3\sigma}{d\omega d\Omega_k d\Omega_f}(\bfzeta_i,\bfeps^{(\pm)})\;=\;\frac12\;\left( \frac{d^3\sigma}{d\omega d\Omega_k d\Omega_f}\right)_0\;\left[ 1\,-\,C_{120}\xi_2\;(\bfzeta_i \bfe_x)\;-\;C_{200}\;(\bfzeta_i \bfe_y)\;+\;C_{320}\xi_2\;(\bfzeta_i \bfe_z)\,\right],
\end{equation}
where $\xi_2=\pm 1$ for $\bfeps^{\ast(\pm)}$.

The polarization correlations $C_{ij0}$ can thus be obtained in terms of relative cross section differences, similar to (\ref{2.10}), see, e.g. \cite{Jaku10}.

\vspace{0.2cm}
\begin{figure}[!h]
\centering
\begin{tabular}{cc}
\hspace{-1cm} \includegraphics[width=.7\textwidth]{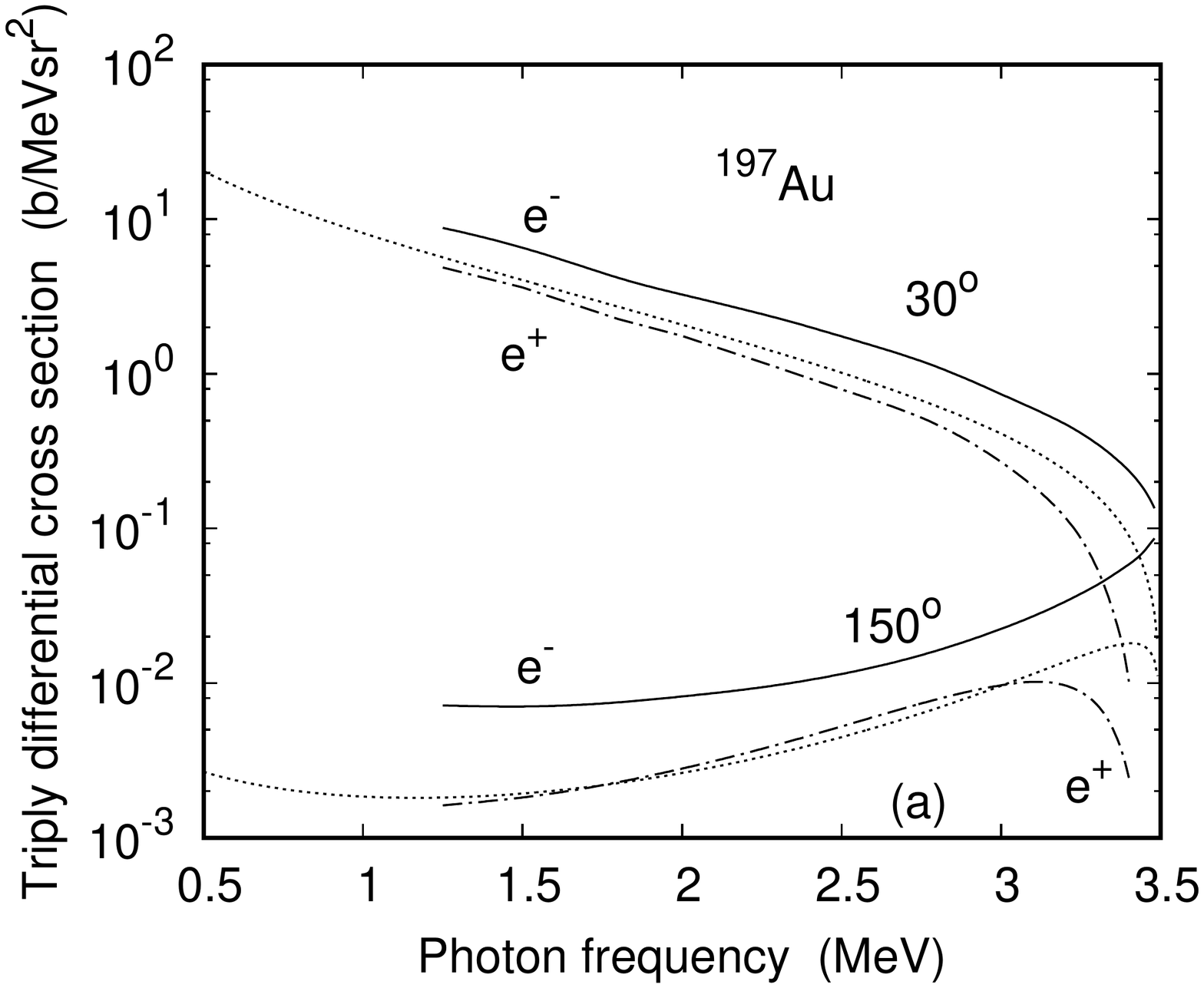}&
\hspace{-4cm} \includegraphics[width=.7\textwidth]{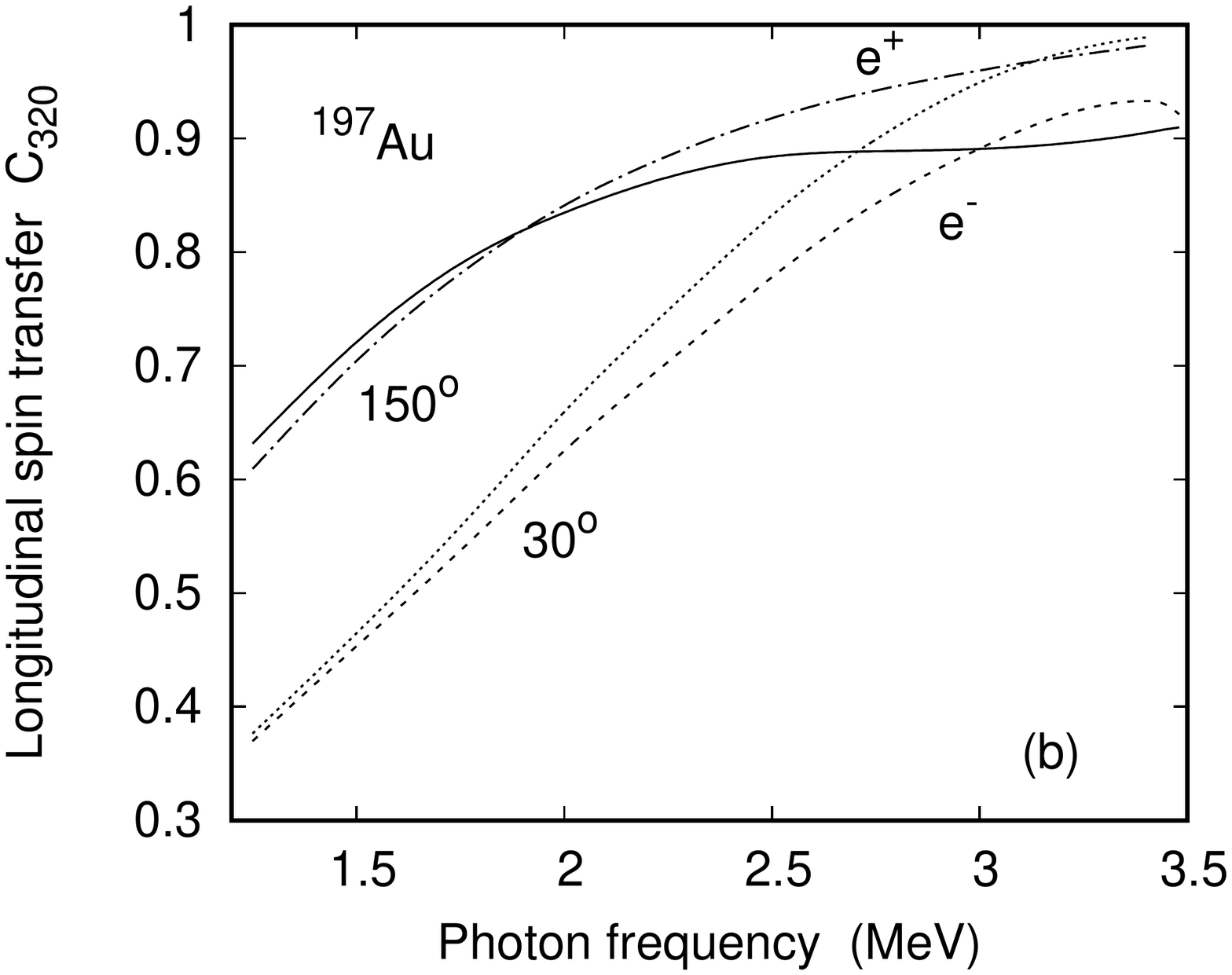}
\end{tabular}
\hspace{5cm} \includegraphics[width=.7\textwidth]{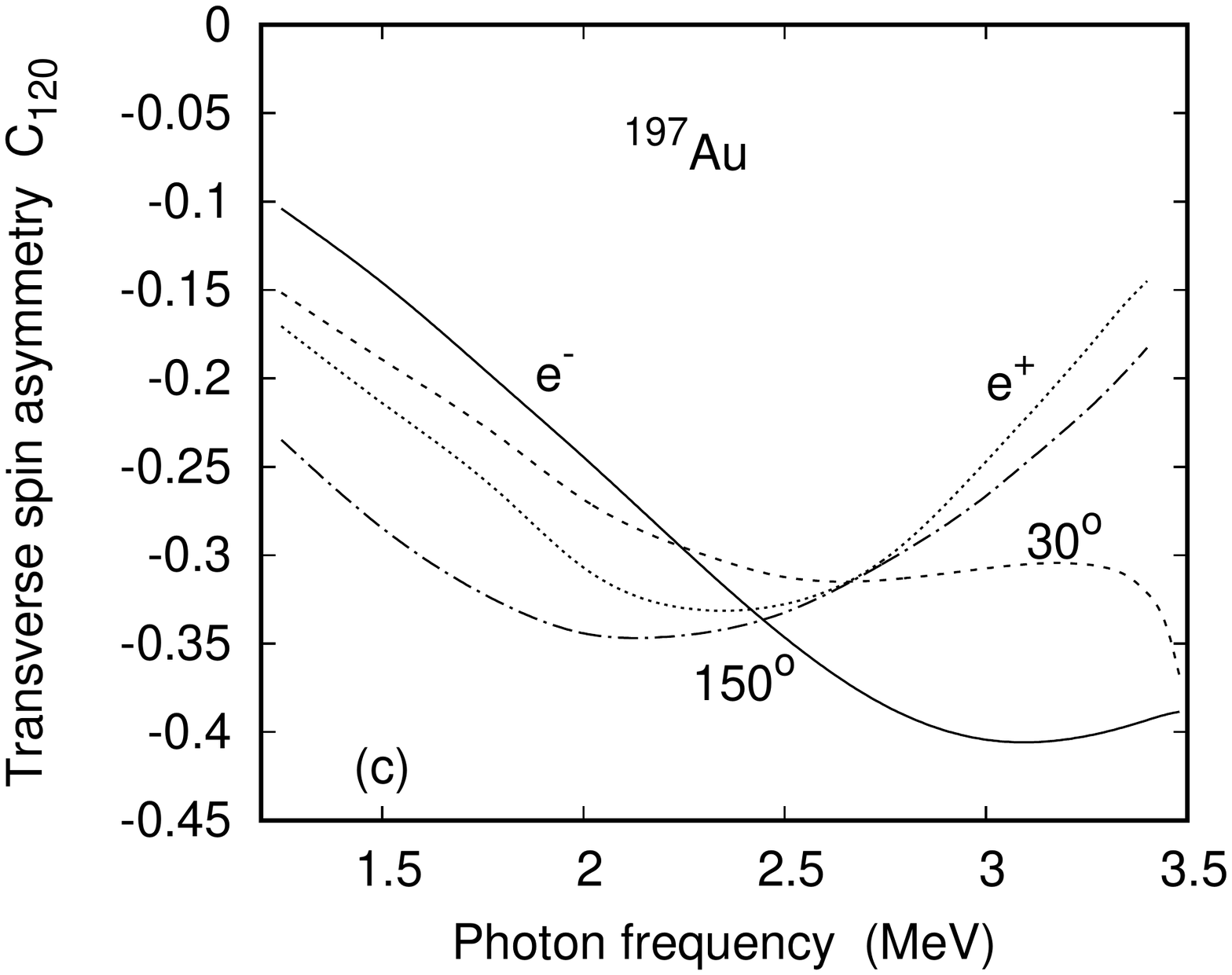}
\vspace{-1cm}
\caption{
Triply differential bremsstrahlung cross section $\frac{d^3\sigma}{d\omega d\Omega_k d\Omega_f}$ (a) and polarization correlations $C_{320}$ (b) and $C_{120}$ (c)
for 3.5 MeV electrons and positrons colliding with $^{197}$Au as a function of photon frequency $\omega$.
The photon angle is $\theta_k =20^\circ$ and $\varphi_f=0$. In (a) the scattering angle is $\vartheta_f = 30^\circ$ (upper curves) and $150^\circ$ (lower curves).
Shown is the DW for electron impact (---------------)
and for positron impact $(-\cdot - \cdot -$), as well as the PWBA $(\cdots\cdots)$.
(b) and (c) provide electron results at $30^\circ \;\,(-----$) and $150^\circ$ (-----------------) as well as positron results at $30^\circ \;\,(\cdots\cdots)$ and $150^\circ \;\,(-\cdot - \cdot -)$.}
\end{figure}

\subsection{Numerical details and results for bremsstrahlung induced by $^{197}$Au and $^{208}$Pb}

The radial integrals (\ref{3.7}) are evaluated by applying the CRM like for elastic scattering.
Nevertheless, due to the triple sum of partial waves in the transition amplitude the convergence is much poorer,
both for collision energies exceeding some tens of MeV, as well as for the emission of soft photons where a large number of final-state partial waves are required.
Since for positrons, the electron code can be used with the substitution of $Z \mapsto -Z$ in the nuclear potential and with taking care of the correct partial-wave phases according to (\ref{2.21}),
the description of the numerics given in \cite{Jaku16} is valid here too.
In some cases, a slightly higher number of partial waves may be required for the positron bremsstrahlung.

In Fig.7 we compare results for  3.5 MeV electron and positron impact on a gold nucleus, for two choices of $\omega$ and  different scattering angles.
Figs.7a  and 7b display, respectively, the triply differential cross section and the longitudinal polarization correlation $C_{320}$ for $\vartheta_f = 150^\circ$.
Here and in all subsequent figures, the azimuthal angle $\varphi_f$ of the scattered lepton 
is set to zero, such that $\theta_k =\vartheta_f$ means parallel emission of photon and lepton.
The angles $360^\circ \geq \theta_k > 180^\circ$ are equivalent to setting $\varphi_f = 180^\circ$ for $0 \leq \theta_k < 180^\circ$.

It is seen that for a final kinetic energy $E_f-c^2=1 $ MeV (corresponding to $\omega=2.5$ MeV) there is an enhancement of the cross section
near $\theta_k=\vartheta_f$, which is reduced to a shoulder at the lower energy ($E_f-c^2=0.5$ MeV and $\omega = 3$ MeV).

The positron cross sections are for fixed $\omega$ at all photon angles
below the ones for electrons due to the repulsive positron-nucleus potential
which prohibits too close positron-nucleus encounters and thus reduces the emission of hard photons \cite{FPT}.
Interestingly, for both leptons there is a crossing of the respective bremsstrahlung intensities when changing from 
 $\omega=2.5$ MeV to 3 MeV at small photon angles ($\theta_k \sim 50^\circ - 80^\circ$, for $\vartheta_f=150^\circ )$
and also at large angles ($250^\circ - 300^\circ)$.
With decreasing scattering angle, the forward crossing is shifted towards $\theta_k=0$, while the second crossing moves either closer to $360^\circ$ (e.g. for $\vartheta_f = 100^\circ)$, or it splits into several crossings (e.g. for $\vartheta_f = 30^\circ$, see Fig.7c).
This implies that for sufficiently large scattering angles, bremsstrahlung in its region of maximum intensity (close to the beam direction) may have a considerable 
fraction of photons with high frequency.
This is counterintuitive to the general assumption that hard photons are only produced in close lepton-nucleus encounters.

The angular dependence of $C_{320}$ is quite similar for electrons and positrons up to $\theta_k \sim 230^\circ$,
except for a pronounced positron peak near $150^\circ$ (for $\omega = 2.5$ MeV) where the respective cross section has a shallow minimum.
However, electron and positron spin asymmetries differ strongly above $270^\circ$, even having opposite derivatives with respect to $\theta_k$.
The effect of a change in scattering angle at a fixed frequency of 2.5 MeV is displayed in the case of the transverse spin asymmetry $C_{120}$, see Fig.7d.
Clearly, the structure in the photon angular distribution shifts from $\theta_k=150^\circ$ to $100^\circ$ when the scattering angle is decreased from $\vartheta_f=150^\circ$ to $100^\circ$.
For positrons at $\vartheta_f = 150^\circ$, the peak near $\theta_k=150^\circ$  is only small, but the maximum spin asymmetry (near $140^\circ,\;230^\circ$ and $340^\circ$) is much higher than the one
for electrons or the one for positrons at the smaller $\vartheta_f$.

The frequency dependence of the cross section for the elementary process of bremsstrahlung is shown in Fig.8a at forward photon emission and a forward and backward scattering angle for the same collision system as in Fig.7.
At $\vartheta_f=30^\circ$ the intensity drops monotonously with $\omega$, with its minimum at the SWL.
For the positrons, the decrease to zero at the SWL is exponentially, due to the normalization constant of the positron wavefunction \cite{FPT}.
In contrast, at the backward scattering angle the intensity of electron bremsstrahlung {\it increases} with $\omega$ to its  maximum at the SWL.
For the positron, the intensity peaks near $\omega = 3$ MeV, while the decrease at the higher $\omega$ is again due to the normalization constant.
This behaviour is in accord with the findings of Fig.6a where for $\vartheta_f = 150^\circ$ at $\theta_k=20^\circ$ the photons with a higher frequency are preferably emitted.

\vspace{0.2cm}
\begin{figure}[!h]
\centering
\begin{tabular}{cc}
\hspace{-1cm} \includegraphics[width=.7\textwidth]{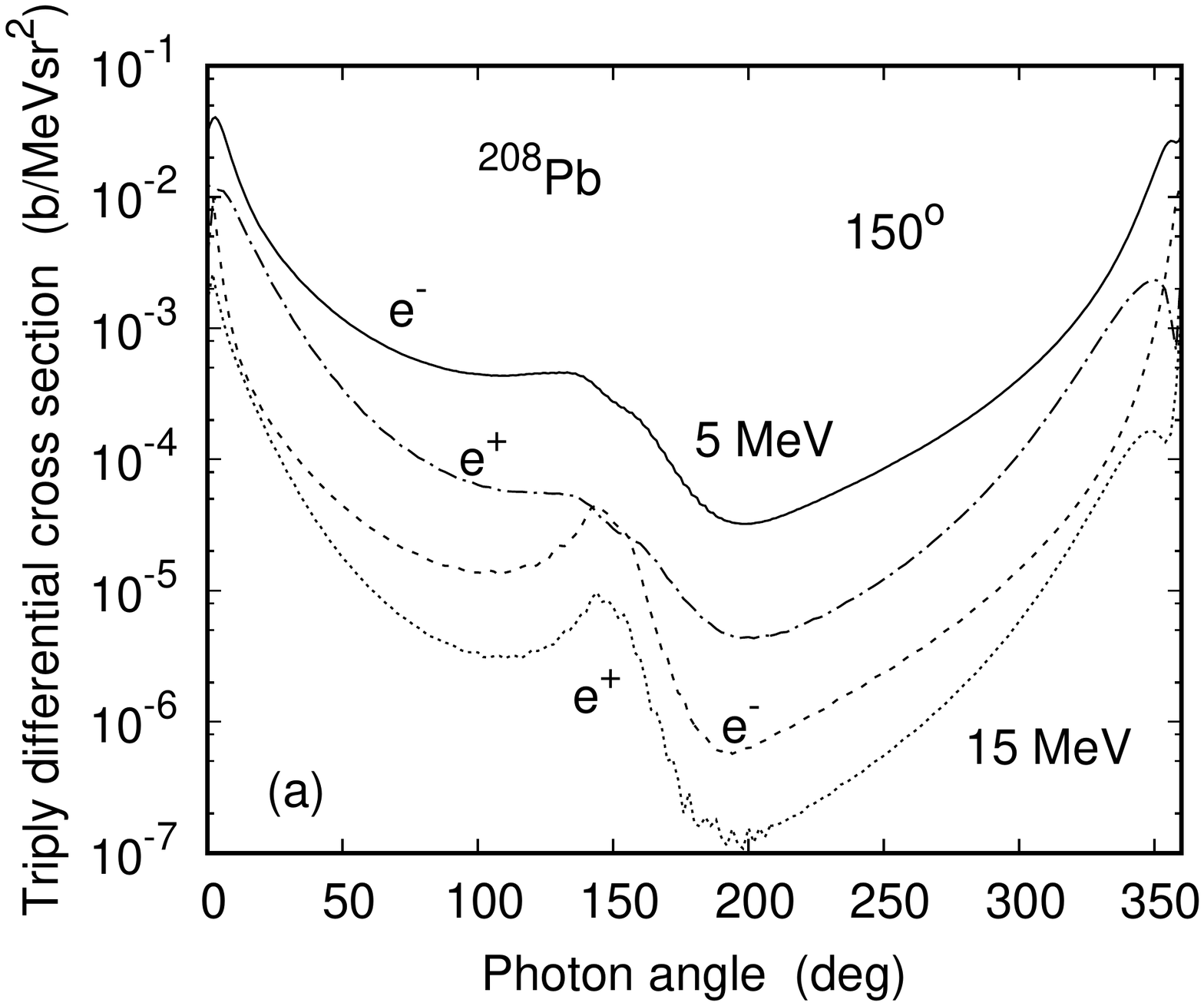}&
\hspace{-4cm} \includegraphics[width=.7\textwidth]{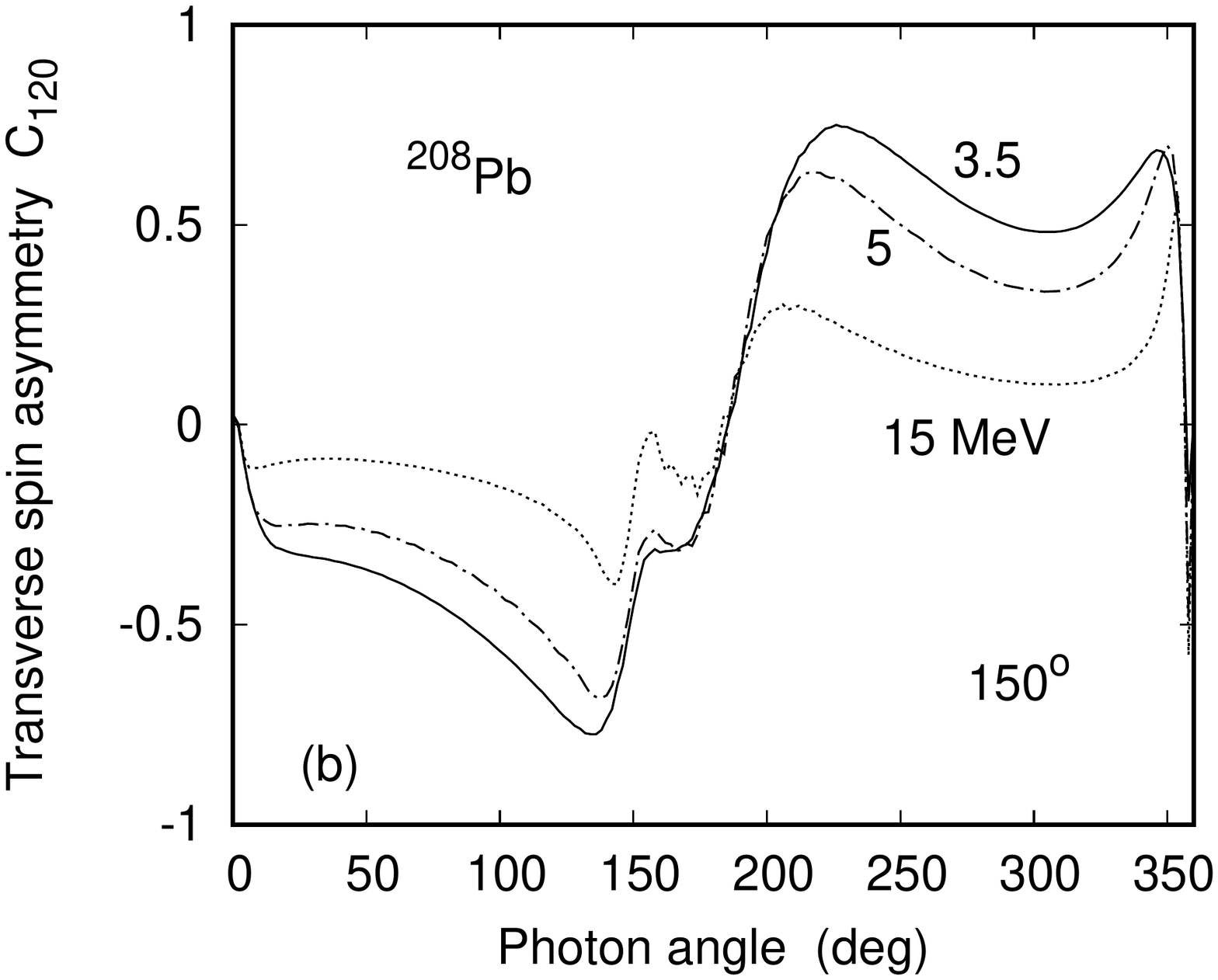}\\
\hspace{-1cm} \includegraphics[width=.7\textwidth]{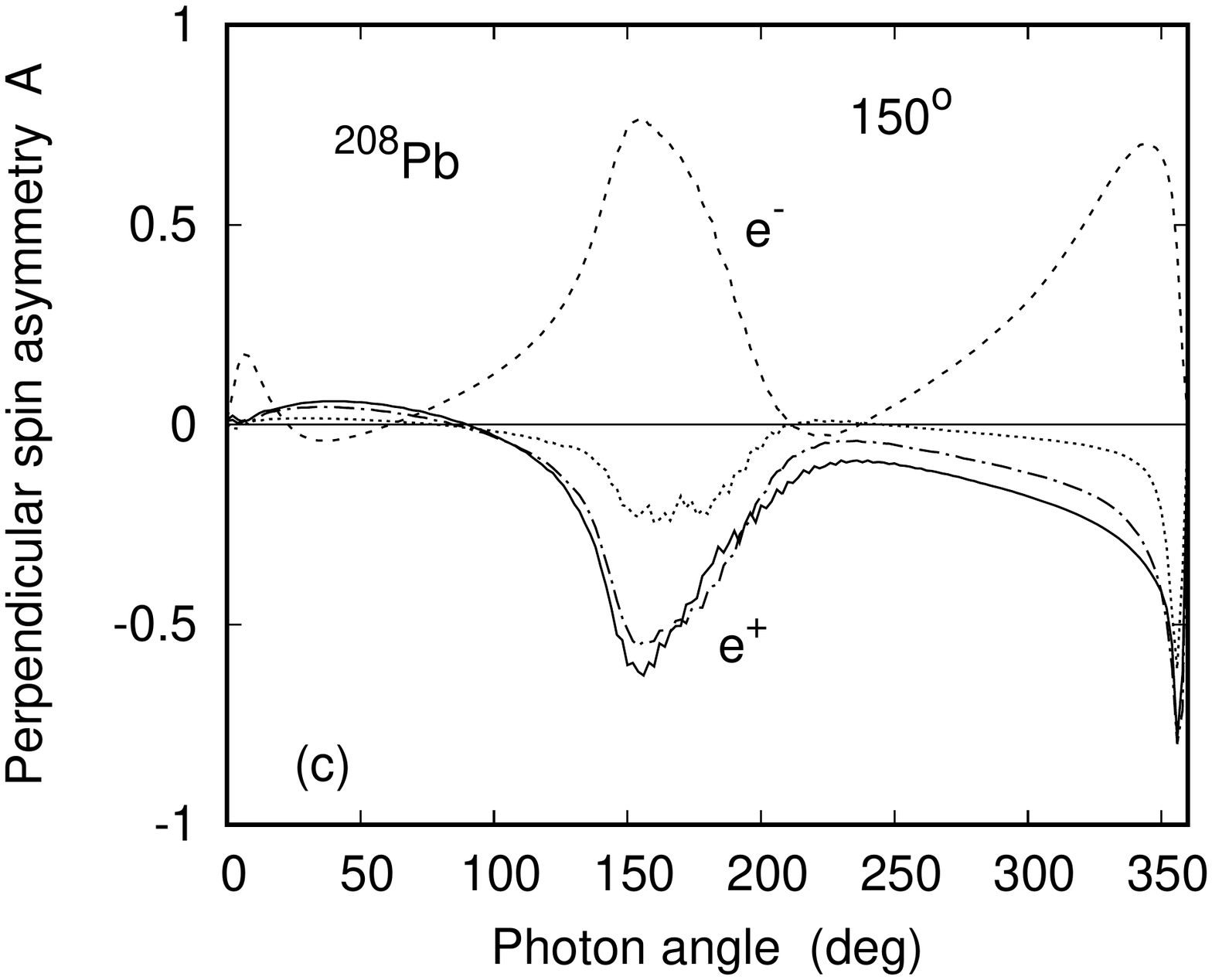}&
\hspace{-4cm} \includegraphics[width=.7\textwidth]{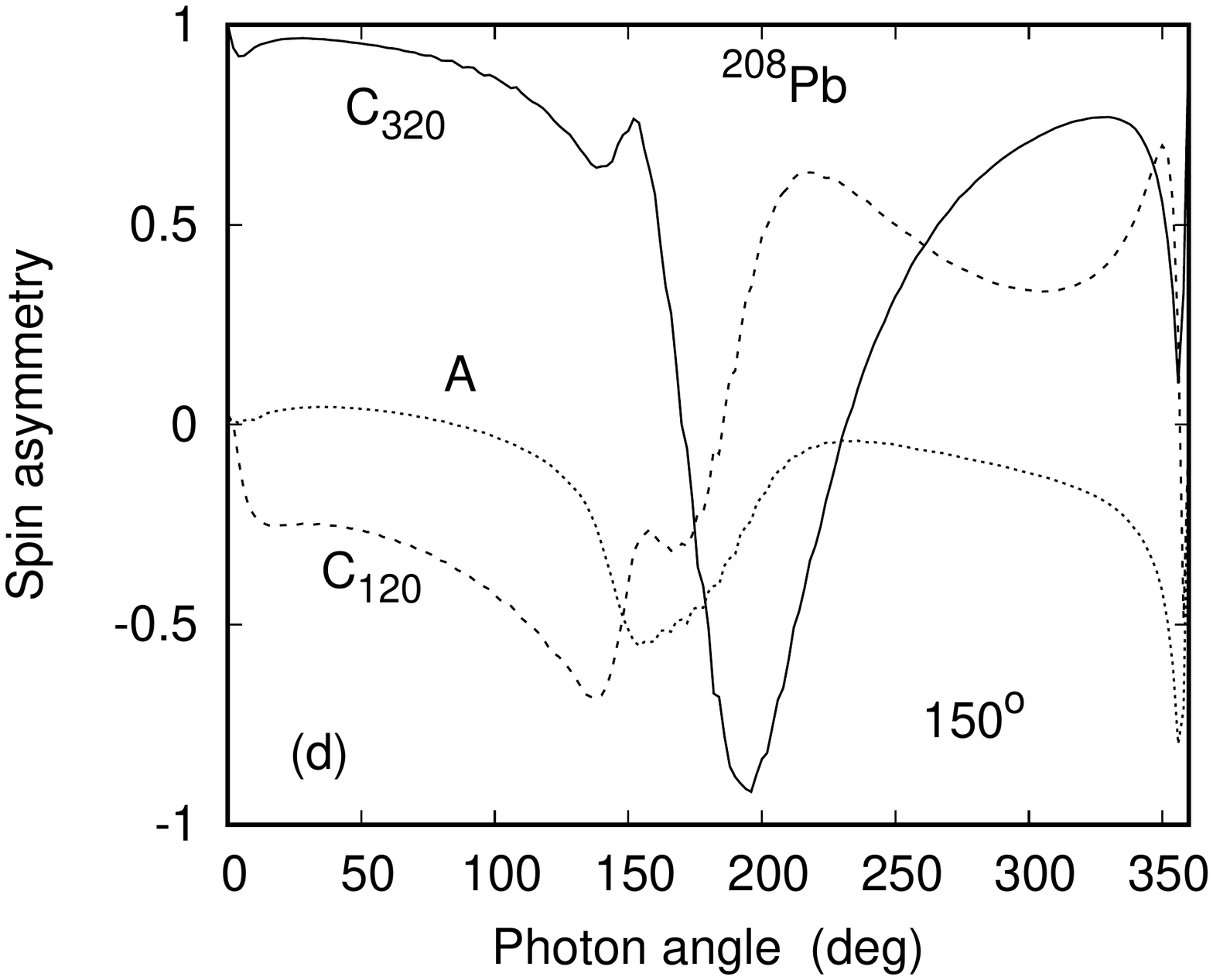}
\end{tabular}
\vspace{-1cm}
\caption{
Triply differential bremsstrahlung cross section $\frac{d^3\sigma}{d\omega d\Omega_k d\Omega_f}$ (a) and polarization correlations $C_{120}$ (b) and $A=C_{200}$ (c) for electrons and positrons colliding with $^{208}$Pb at $\vartheta_f =150^\circ,\;\varphi_f=0$ as a function of photon angle $\theta_k$.
Shown are results for  positrons at $E_e=5$ MeV $(-\cdot - \cdot -)$ and  15 MeV ($\cdots\cdots)$.
In (a), electron results at 5 MeV (-------------) and 15 MeV $(-----)$ are included.
In (b) and (c), positron results at 3.5 MeV (----------------) are shown in addition.  (c) provides also electron results at 3.5 MeV $(-----)$.\\
In (d), the three spin asymmetries $C_{320}$ (-----------------), $C_{120}\;(-----)$ and $A\;(\cdots\cdots$) for positrons are shown
at $\vartheta_f = 150^\circ,\;\varphi_f=0$ and $E_e=5$ MeV.
The ratio $\omega/E_e=3/4$ throughout.
The wiggles in $C_{120}$ and $A$ from numerics are partly smoothed by means of averaging over $\theta_k$ (over a range $\Delta \theta_k = 10^\circ$).}
\end{figure}

In order to understand the underlying physics one may study the plane-wave Born approximation (PWBA) for bremsstrahlung where the $\theta_k$-dependence is qualitatively the same as in the DW theory.
Since the prefactor of the Bethe-Heitler cross section \cite{BH,Lan} is proportional to $q^{-4}$ where $\bfq = \bfk_i-\bfk_f-\bfk$ is the momentum transfer to the nucleus, 
one expects the cross section to increase with decreasing momentum transfer.
In fact, for $\vartheta_f =150^\circ,\;\;q$ decreases monotonously with $\omega$ (for $\theta_k = 20^\circ$ and $\varphi_f=0$),
corresponding to an increase of the cross section.
However, this increase  is considerably weaker than predicted by the inverse power law $q^{-4}$.
Thus the Bethe-Heitler cross section, multiplied by $q^4$, is {\it not} slowly varying with $q$,
but counteracts the $q^{-4}$-dependence.
This fact explains also the decrease of the cross section with $\omega$ for $\vartheta_f =30^\circ$, although at this angle $q$  decreases with $\omega$ too, albeit much weaker than for $150^\circ$.
Hence, hard photons will also be produced at a low momentum transfer, corresponding to a larger lepton-nucleus distance.

The frequency dependence of the polarization correlations $C_{320}$ and $C_{120}$ is shown in Figs.8b and 8c.
Like in the case of doubly differential cross sections \cite{Jaku11,Ni}, $C_{320}$ increases monotonously with $\omega$ at forward and backward scattering angles.
The positron spin asymmetries are higher than those of electrons at $\vartheta_f=30^\circ$, but lower at $150^\circ$ if $\omega/E_e \lesssim \frac12$. 
As concerns $C_{120}$, it decreases monotonously below $\omega = 2$ MeV in all cases, but for high frequencies, $|C_{120}|$ for electrons is much larger at $\vartheta_f = 150^\circ$ than for $30^\circ$, while for positrons the increase with $\omega$ near the SWL is similar for both angles.
If, instead of $\varphi_f=0$, a coplanar geometry with $\varphi_f = 180^\circ$ is chosen (where lepton and photon emerge on different sides of the beam axis),
there appears an interference structure in the photon spectrum for small scattering  angles and photon angles (as discussed in \cite{Jaku18}), which is also visible
in the photon angular distribution  near $350^\circ$ (corresponding to $\theta_k=10^\circ$ at $\varphi_f=180^\circ$, see Fig.7c).

In Fig.9 the cross sections and polarization correlations for a $^{208}$Pb target are compared at different collision energies for a fixed scattering angle of $150^\circ$ and for a fixed ratio of $\omega/E_e$.
Like for increasing frequency, an increase of collision energy leads to a more pronounced peak in the angular dependent cross section near $\theta_k = \vartheta_f$, whereas the intensity decreases throughout (Fig.9a).
It is also seen that the intensity for positron scattering gets closer to the one for electrons when $E_e$ is growing (in concord with the case of unobserved leptons, see Fig.11a below), 
and the positron structure near $350^\circ$ is somewhat weakened.
Fig.9b shows the variation with $E_e$ of the angular dependent transverse polarization correlation $C_{120}$ for positron impact at the energies $3.5,\;5$ and 15 MeV, again for the fixed ratio $\omega/E_e = 3/4.$
Whereas the global angular variation is reduced with increasing $E_e$, the peak near $150^\circ$ is strongly enhanced.
The wiggles, particularly in the 15 MeV results, are due to numerical inaccuracies.

In Fig.9c the perpendicular spin asymmetry $A$ is displayed.
An increase of the collision energy, as shown for the positrons, leads again to a reduction of the spin asymmetry particularly in the region of its largest excursion near photon angles of $180^\circ$ and $350^\circ$.
There is kind of mirror symmetry with respect to the horizontal zero-line between the results for electrons and positrons, shown for $E_e=3.5$ MeV. 
The reason lies in the low-$Z$ (or Born) limit where $C_{320}$ and $C_{120}$ are finite (thus implying an identical spin asymmetry for electrons and positrons), while $A$ vanishes in the Born limit and is proportional to $Z$ for low $Z$.
Thus $A$ has mostly opposite signs for electrons and positrons at the moderate collision energies considered for bremsstrahlung.
As noted in \cite{Y12}, the symmetry breaking in the polarization correlations between electrons and positrons is due to the strong relativistic effects in high-$Z$ nuclei.

Finally, Fig.9d provides an overview of the three circular polarization correlations for positron impact at $E_e=5$ MeV as a function of photon angle.
Clearly, except for backward angles and near $350^\circ$, the longitudinal spin transfer $C_{320}$ is by far the dominating spin asymmetry at such a high collision energy.

\vspace{0.2cm}
\begin{figure}[!h]
\centering
\begin{tabular}{cc}
\hspace{-1cm} \includegraphics[width=.7\textwidth]{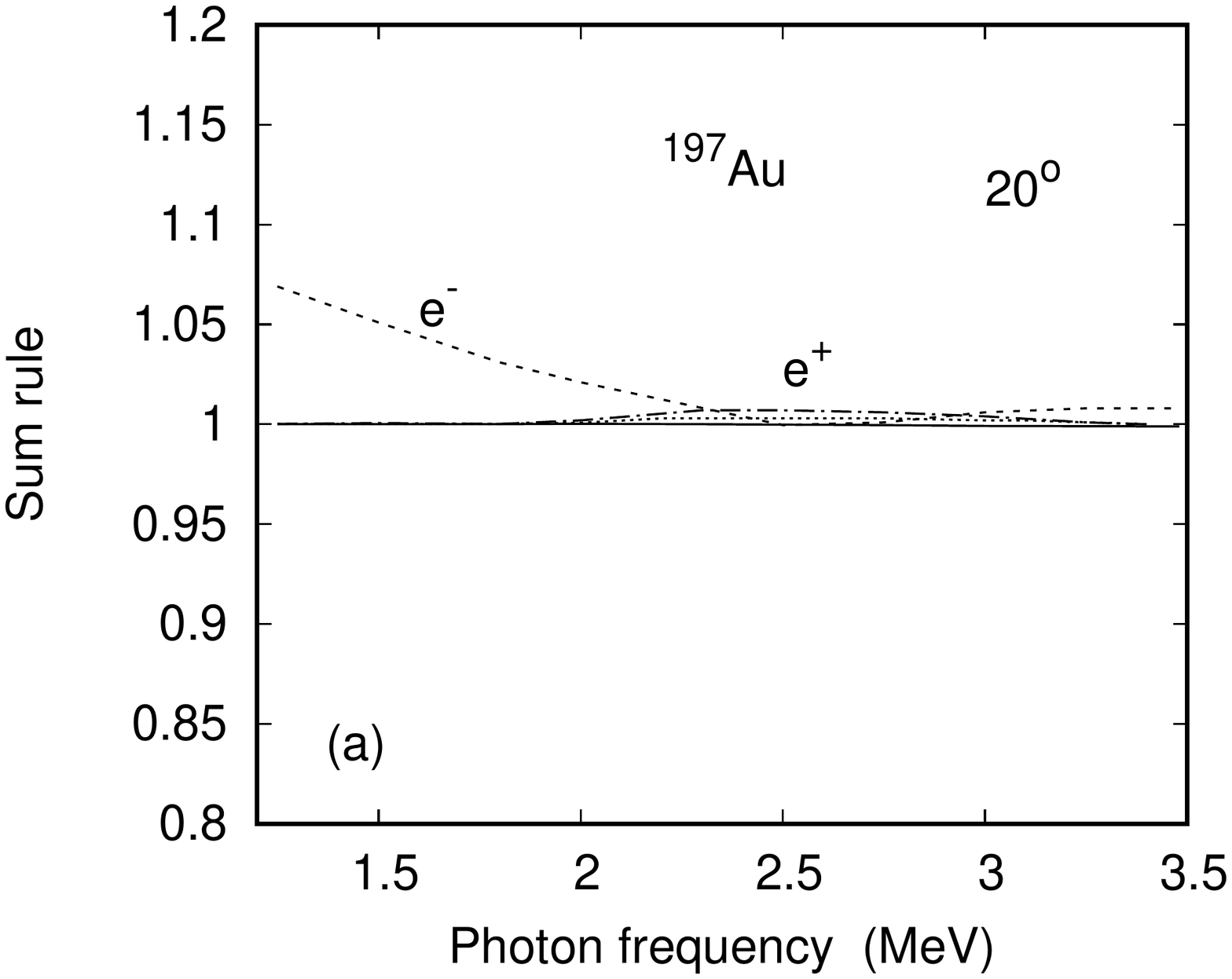}&
\hspace{-4cm} \includegraphics[width=.7\textwidth]{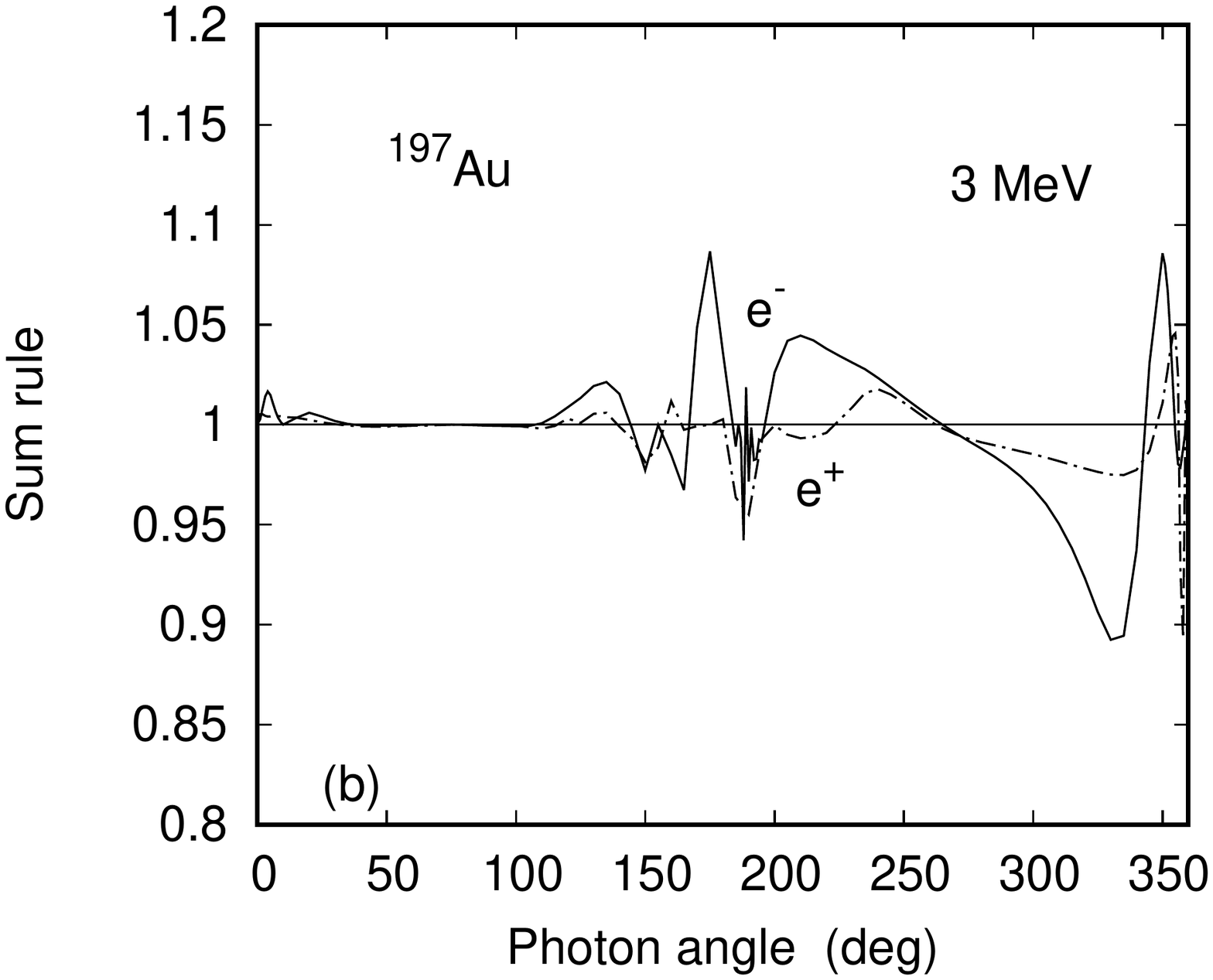}
\end{tabular}
\caption{
Lhs of the sum rule (\ref{3.18}) for bremsstrahlung from 3.5 MeV electrons and positrons colliding with $^{197}$Au (a) at $\theta_k=20^\circ$ as a function of photon frequency $\omega$.
The  electrons are scattered at  $30^\circ$  (-----------------)  and $150^\circ$  ($-----$),  and the positrons at $30^\circ\;\,(-\cdot - \cdot -$) and $150^\circ\;\,(\cdots\cdots)$. In (b), the dependence on photon angle $\theta_k$ for $\omega=3$ MeV and $\vartheta_f=150^\circ$ is given for electrons (------------------) and positrons $(-\cdot - \cdot -)$.
The azimuthal angle is $\varphi_f=0.$}
\end{figure}

\subsection{Polarization sum rule}

Based on his work on photoionization, Pratt was able to derive a sum rule for the seven polarization correlations occurring in the doubly
differential electron bremsstrahlung, and the numerical verification was provided in \cite{PMS}.
A similar sum rule is valid for the polarization correlations in the triply differential cross section (\ref{3.15}) \cite{Jaku17},
\begin{equation}\label{3.18}
C_{320}^2+C_{120}^2+C_{200}^2+C_{030}^2+C_{310}^2+C_{110}^2-C_{230}^2\;=\;1.
\end{equation}
Since the sum rule is independent of the nuclear potential, it holds also for positrons.
While it is strictly verified in the analytic plane-wave Born approximation (PWBA) where all $C_{ij0}$ vanish
except for $C_{320},\;C_{120}$ and $C_{030}$ \cite{TP73,PMS}, numerical inaccuracies occur in the partial-wave theory,
particularly for heavy targets and high collision energies, which lead to a violation of (\ref{3.18}).

As an example we show in Fig.10a the frequency-dependent results for bremsstrahlung at a collision energy of 3.5 MeV.
At the forward emission angles, the sum rule is seen to hold numerically within 1\% both for electrons and positrons.
At larger scattering angles the deviations from unity increase for electrons (but not for positrons) when the frequency
falls below 2.5 MeV.
From Fig.10b it follows that when also the photon is emitted at large angles, particularly near $\theta_k=180^\circ$ or $350^\circ$, the sum rule is violated by up to 10\%.
We note that such deviations can be used as a measure of the inaccuracies in the calculations.

\subsection{Doubly differential bremsstrahlung cross section}

When only the emitted photon is observed, but not the scattered particle, an integration over the solid angle $d\Omega_f$ has to be performed,
\begin{equation}\label{3.19}
\frac{d^2\sigma}{d\omega d\Omega_k}(\bfzeta_i,\bfeps_\lambda)\;=\;\int d\Omega_f\;\frac{d^3\sigma}{d\omega d\Omega_k d\Omega_f}(\bfzeta_i,\bfeps_\lambda).
\end{equation}
If recoil is neglected (such that $E_f=E_i-\omega$ is independent of the scattering angle $\vartheta_f$), the partial-wave representation (\ref{3.5}) of $F_{fi}$ allows for an analytical evaluation of this angular integral, turning the coherent sum over the final-state partial waves
into an incoherent one \cite{Jaku16}.

\vspace{0.2cm}
\begin{figure}[!h]
\centering
\begin{tabular}{cc}
\hspace{-1cm} \includegraphics[width=.7\textwidth]{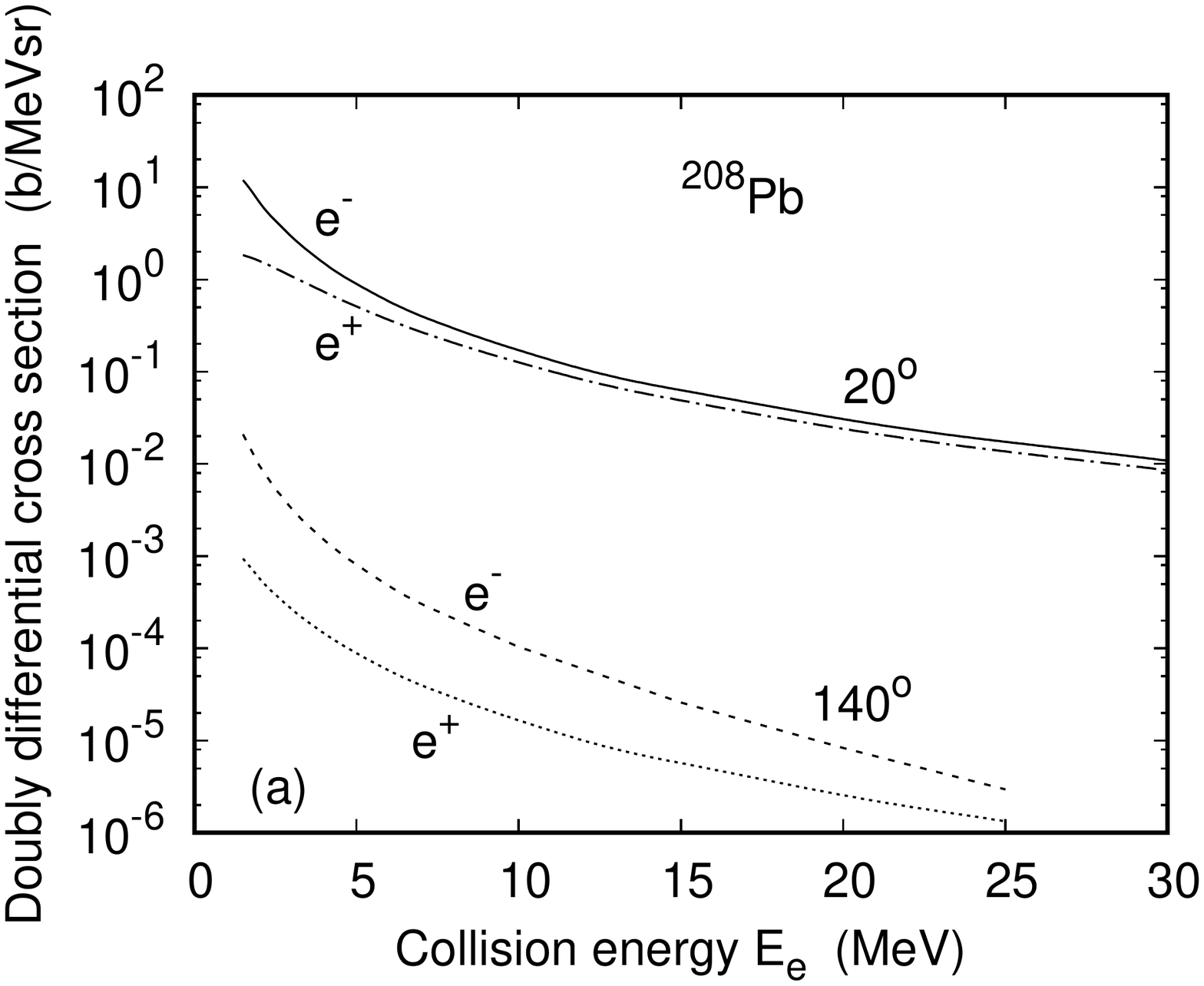}&
\hspace{-4cm} \includegraphics[width=.7\textwidth]{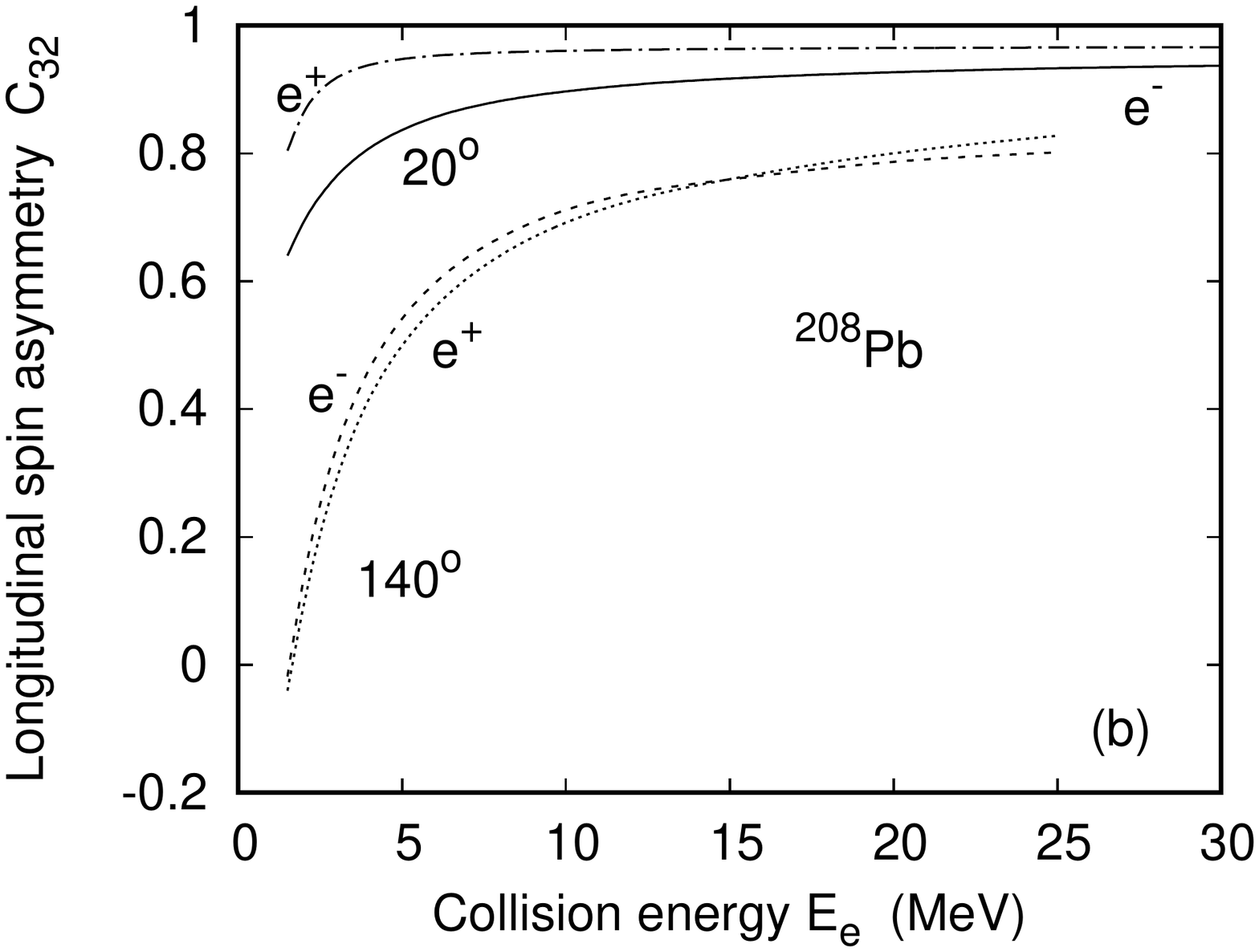}\\
\hspace{-1cm} \includegraphics[width=.7\textwidth]{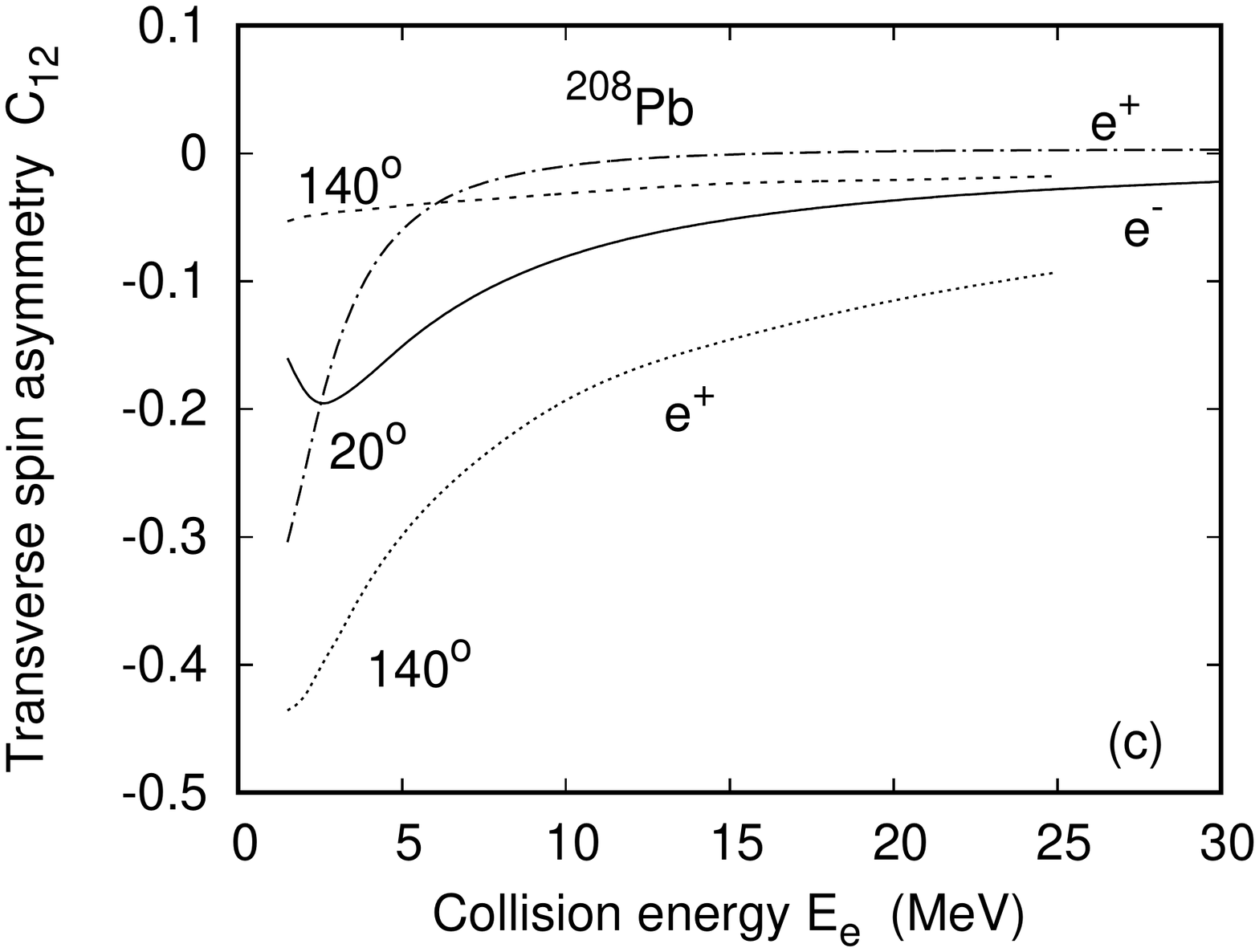}&
\hspace{-4cm} \includegraphics[width=.7\textwidth]{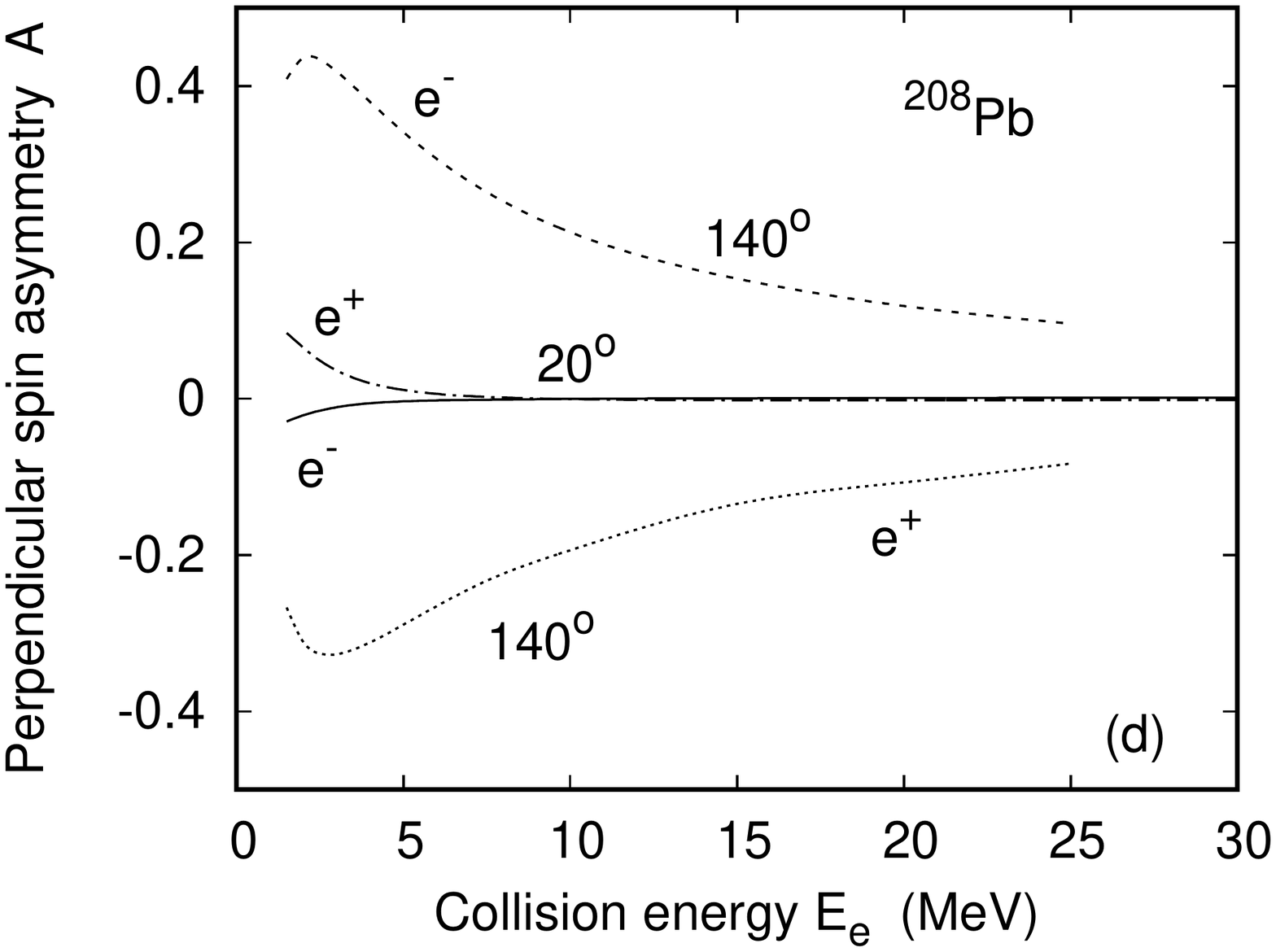}
\end{tabular}
\vspace{-1cm}
\caption{
Doubly differential bremsstrahlung cross section $\frac{d^2\sigma}{d\omega d\Omega_k}$ (a) and polarization correlations $C_{32}$ (b), $C_{12}$ (c) and $A=C_{20}$ (d) for electrons and positrons colliding with $^{208}$Pb
as a function of collision energy $E_e$. Shown are results at $\theta_k=20^\circ $ for electrons (-------------------) and positrons $(-\cdot - \cdot -)$ as
well as for $\theta_k=140^\circ$ for electrons $(-----)$ and positrons ($\cdots\cdots$).
The ratio  $\omega/E_e= 6/7 =0.857.$}
\end{figure}

Fig.11a shows the dependence of the doubly differential cross section on the collision energy for the two photon angles $20^\circ$ and $140^\circ$ at a fixed ratio of $\omega/E_e$.
It is seen that the cross section decreases monotonously with energy.
The positron intensity is again below the electron intensity throughout, and the difference between $e^+$ and $e^-$ increases with angle and decreases with energy.

Fig.11b displays the polarization correlation $C_{32}$ which tends to 1 at high energies both for electrons and positrons.  $C_{12}$ is plotted in Fig.11c.
For this polarization correlation the energy dependence of $e^+$ and $e^-$ is quite different at low collision energies.
However, this difference decreases with $E_e$, being again smaller at the forward angle.
Whereas $C_{32}$ and $C_{12}$ have the same sign for electrons and positrons, the perpendicular spin asymmetry $A$ involves a sign change as is seen in Fig.11d.
Even more, there is again some symmetry with respect to $x$-axis for both angles and all energies.
The absolute value of the electron-positron difference is also here much smaller for the forward angle than for $140^\circ$, while $|A|$ increases with angle.
The decrease of the cross section and the increase of $|A|$ with angle points to the increasing importance of close collisions where the relativistic effects
are strong. On the other hand, small angles are related to distant collisions, particularly for low photon frequencies.
Under such conditions, even the PWBA describes the intensity and the polarization correlations satisfactorily \cite{Jaku18}.
The maximum value of $|A|$ is below 1 MeV at small angles, while it is near 3 MeV
for $140^\circ$, its position increasing to about 10 MeV at the backmost angles \cite{Jaku10,Jaku13}.

\section{Correspondence between elastic scattering and bremsstrahlung}
\setcounter{equation}{0}

The photoeffect is a well-known example where the photon acts not as a wave, but as a particle. Bremsstrahlung at the short-wavelength limit (SWL) can be considered as the time-reversed process of photoionization of an electron in a high Rydberg state close to the continuum threshold.
Therefore, an SWL photon is also expected to act like a particle.
This picture has been confirmed for strong nuclear fields by comparing the polarization correlations of an  electron elastically scattered into an angle $\theta$ with those of a circularly polarized SWL photon
emitted into the same angle, $\theta_k = \theta$. For a gold target and 
sufficiently high collision energies such that the electron mass can be neglected in the cross section, it turned out that the polarization
correlations occurring in elastic scattering (i.e. $L,R,S$) are very similar to the ones originating from the doubly differential bremsstrahlung process ($C_{32},C_{12},A$) for identical spin 
polarization, and their difference decreases with increasing collision energy \cite{Jaku12}.
Even more, the polarization correlations  near the SWL were found to obey an approximate sum rule \cite{Jaku12},
\begin{equation}\label{4.1}
A^2+C_{32}^2+C_{12}^2\;\approx\;1,
\end{equation}
which corresponds to the strict sum rule for potential scattering \cite{Mo},
\begin{equation}\label{4.2}
S^2+L^2+R^2\;=\;1.
\end{equation}
In order to show that the similarity between the corresponding polarization correlations holds also for positron scattering, we have considered electron and positron scattering
from $^{208}$Pb at two collision energies, 10 MeV and 15 MeV, and a fixed final energy of 0.1 MeV.
Fig.12a provides the angular dependence of the doubly differential bremsstrahlung cross section,
showing that close to the SWL the positron cross section is about two orders of magnitude below the electron cross section, due to the
suppression of high-energy positron bremsstrahlung by the repulsive positron-nucleus potential.

\vspace{0.2cm}
\begin{figure}[!h]
\centering
\begin{tabular}{cc}
\hspace{-1cm} \includegraphics[width=.7\textwidth]{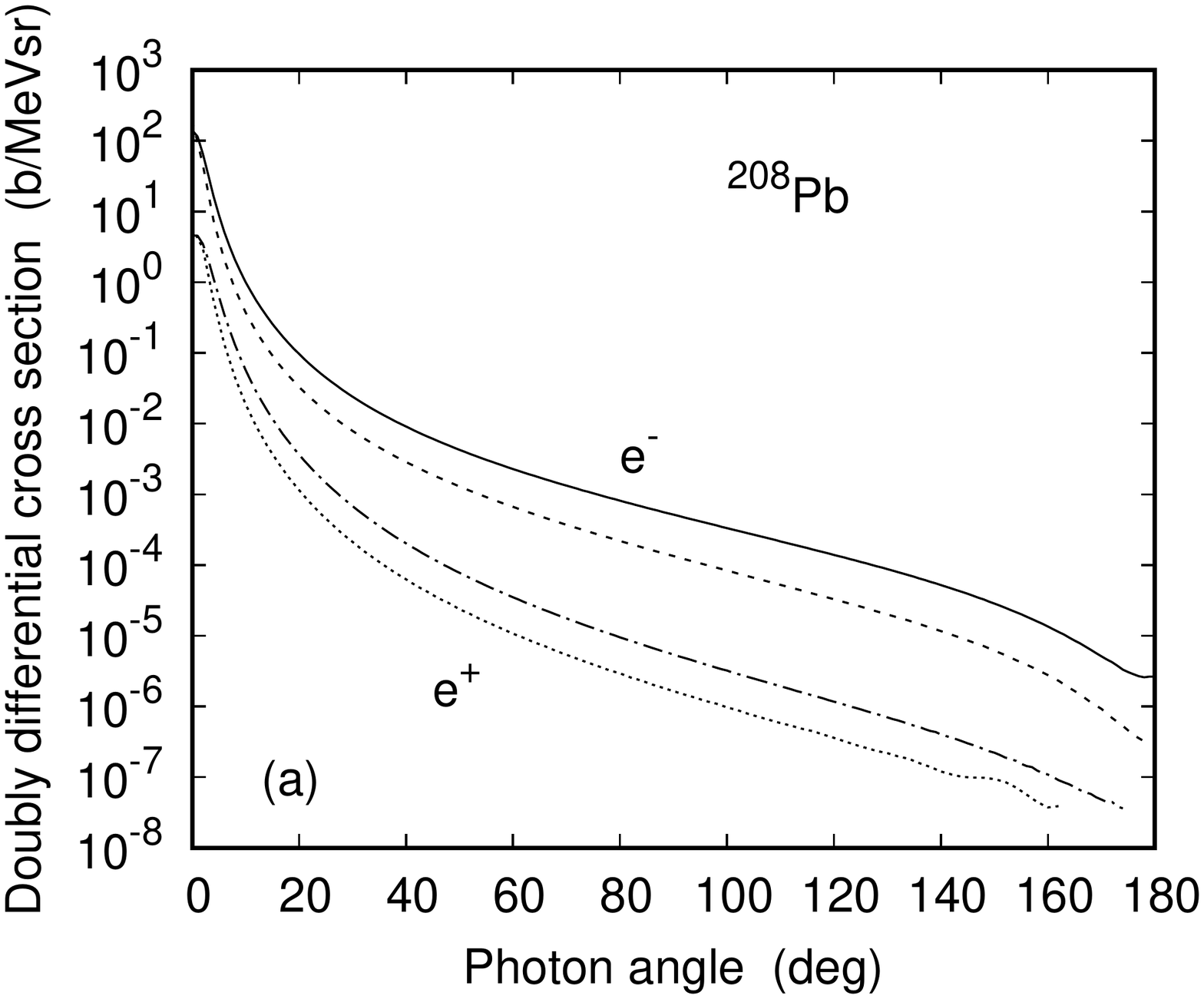}&
\hspace{-4cm} \includegraphics[width=.7\textwidth]{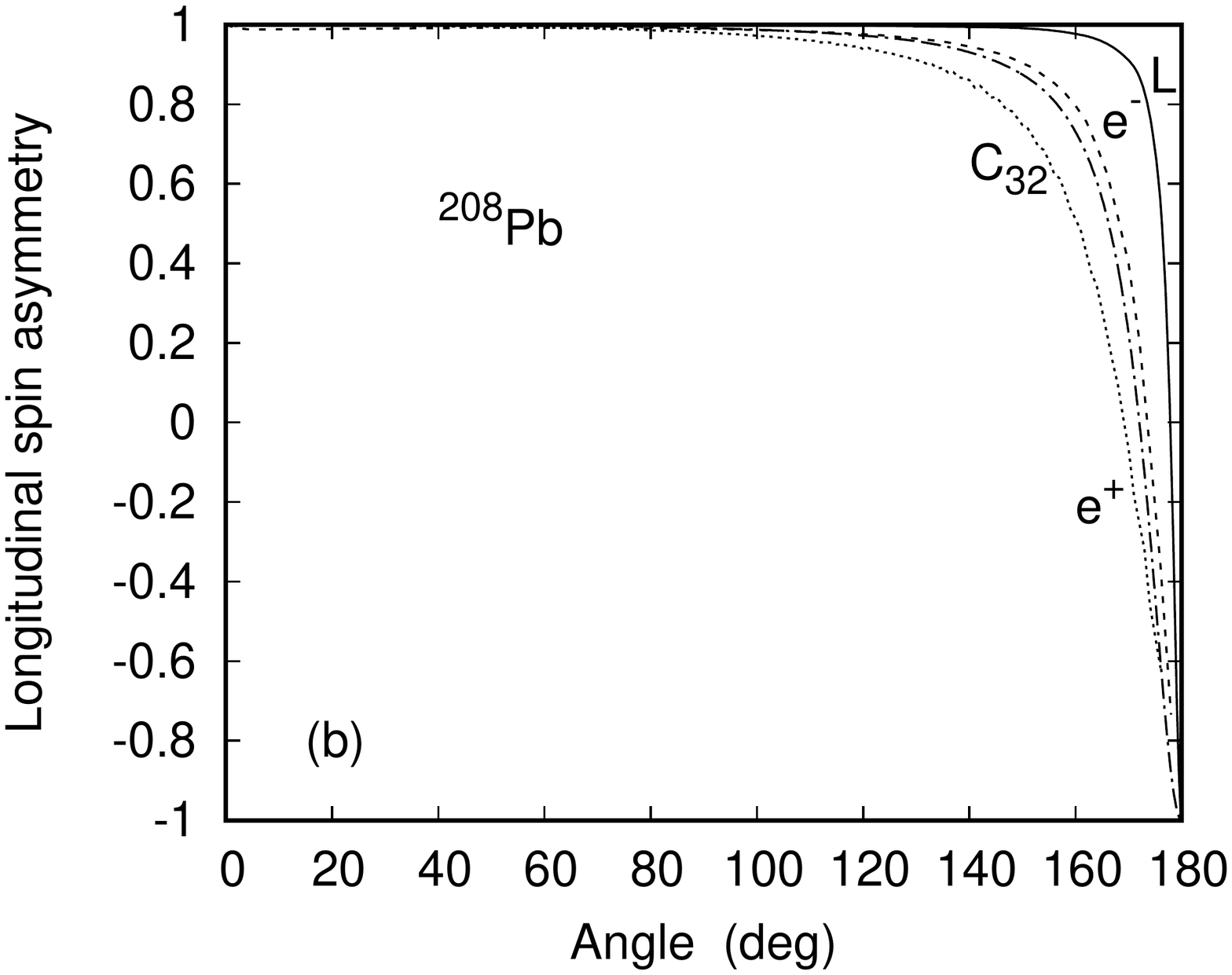}\\
\hspace{-1cm} \includegraphics[width=.7\textwidth]{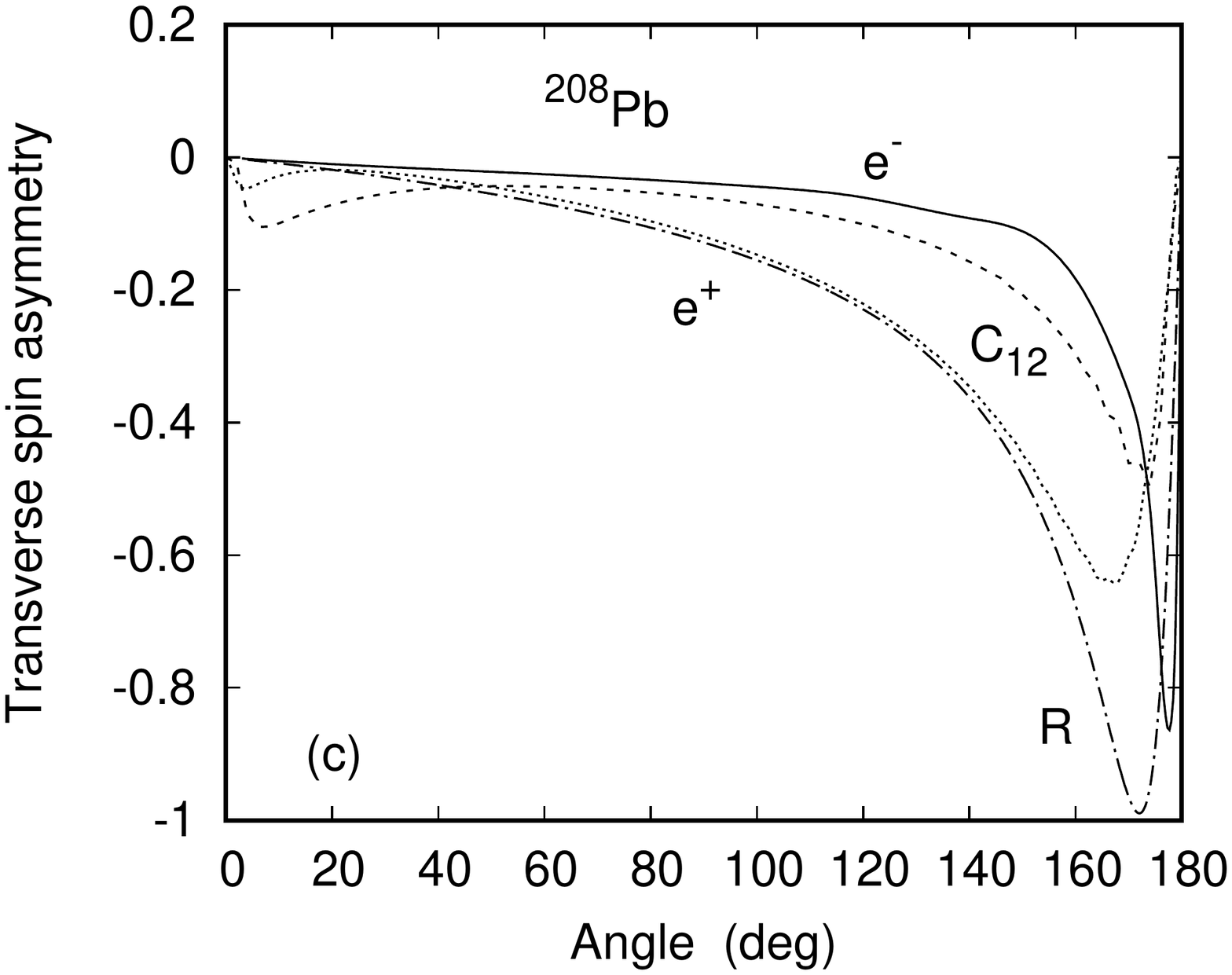}&
\hspace{-4cm} \includegraphics[width=.7\textwidth]{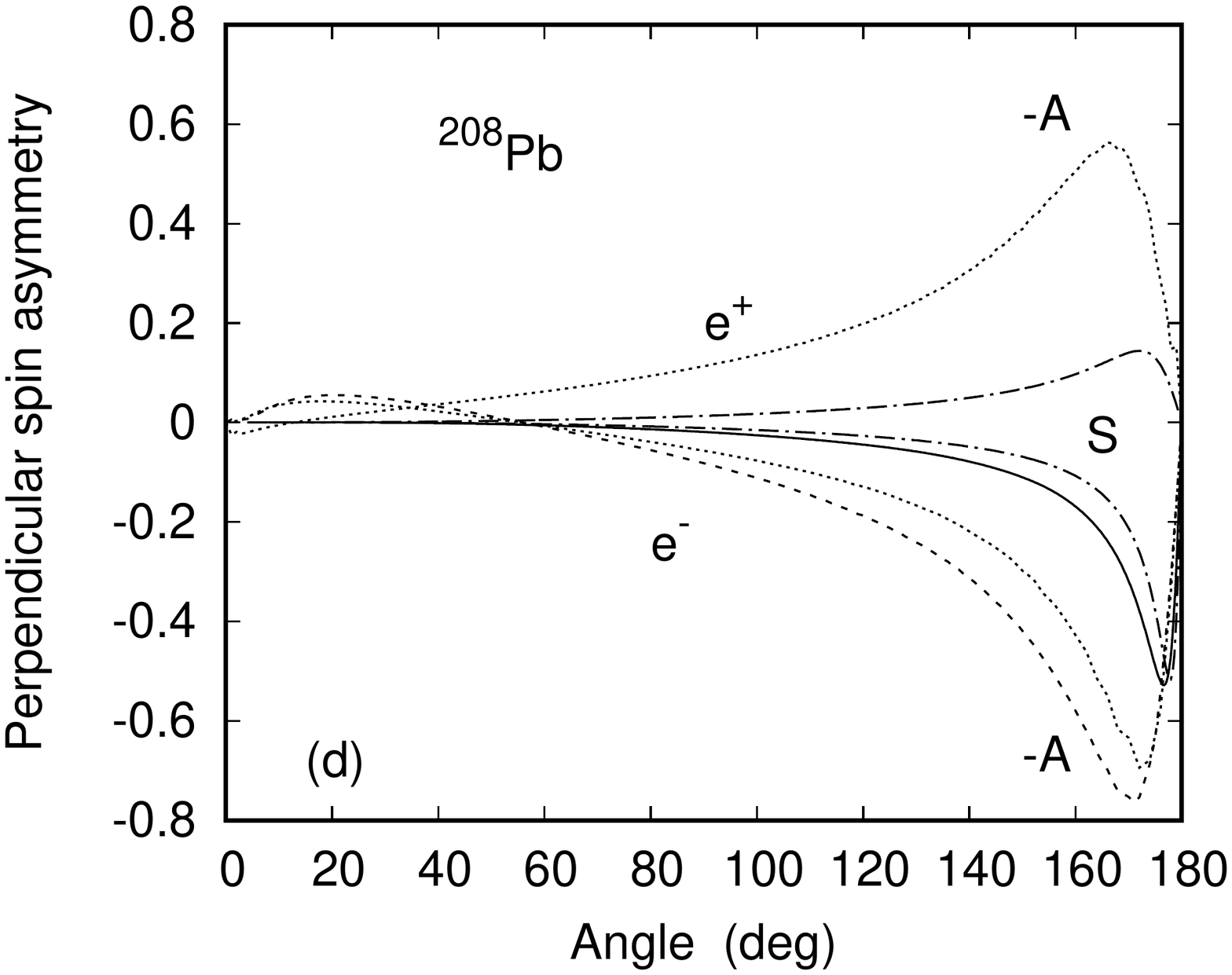}
\end{tabular}
\caption{
(a) Doubly differential bremsstrahlung cross section $\frac{d^2\sigma}{d\omega d\Omega_k}$ from electrons and positrons colliding with $^{208}$Pb as a function of photon angle $\theta_k$.
Shown are results for electron impact at 10 MeV (----------------)and 15 MeV ($-----$), and for positron impact at 10 MeV $(-\cdot - \cdot -$)  and 15 MeV  $(\cdots\cdots$).
The final kinetic lepton energy is 0.1 MeV.\\
(b) Longitudinal spin transfer $C_{32}$ as a function
of photon angle $\theta_k$ at $E_f-c^2=0.1$ MeV and $L$ as a function of scattering angle $\theta$ in collision with $^{208}$Pb.
Shown are results for $C_{32}$ from electrons at 15 MeV $(-----)$ and from positrons at 10 MeV $(\cdots\cdots)$, and for $L$ from electrons at 15 MeV (--------------)  and from positrons at 10 MeV $(-\cdot - \cdot -$).\\
(c) Angular dependence of the transverse polarization correlations $C_{12}$ and $R$ in collision with $^{208}$Pb. Shown are results for $C_{12}$ at $E_f-c^2=0.1$ MeV from electrons at 15 MeV ($-----)$ and from positrons at 10 MeV $(\cdots\cdots)$, and for $R$ from electrons at 15 MeV (-------------------) and from positrons at 10 MeV $(-\cdot - \cdot -)$.\\
(d) Angular dependence of the perpendicular spin asymmetries $A=C_{20}$ and $S$ in collision with $^{208}$Pb.
Shown are results for $-A$ at $E_f-c^2=0.1$ MeV from
electrons at 10 MeV $(-----)$ and at 15 MeV $(\cdots\cdots,$ lower curve) and from positrons at 10 MeV ($\cdots\cdots$, upper curve).
Results for $S$ are from electrons at 10 MeV (----------------) and at 15 MeV $(-\cdot - \cdot -$, lower curve) 
and from positrons at 10 MeV ($-\cdot - \cdot -$, upper curve).} 
\end{figure}

Fig.12b compares $C_{32}$ and $L$ for 15 MeV electrons and for 10 MeV positrons, and it is seen that for either particle, both $C_{32}$ and $L$ are close to unity up to an angle of $100^\circ$,
decreasing sharply near $180^\circ$.
There is also a marked similarity between $C_{12}$ and $R$ (Fig.12c).

Fig.12d shows the respective comparison between $A$ and $S$ (due to a different choice of the coordinate system when originally defining $A$ and $S$ there appears a minus sign in $A$, such that in fact the shapes of $S$ and $-A$ have to be compared).
We have considered electron results for the two collision energies, showing that the minima of both $S$ and $-A$ are located at larger angles and become slightly shallower when $E_e$ is increased.
While the positron differences for $C_{32}$ with respect to $L$ and for $C_{12}$ with respect to $R$ are considerably smaller than those for electrons (despite  the lower collision energy for positrons which had to be taken due to reasons of convergence), this behaviour is reversed for $S$ with respect to $-A$ even at the same collision energy of 10 MeV.
It is also clearly  seen that both $S$ and $A$ switch sign when changing from electrons to positrons except at very small angles,
while the other polarization correlations do not switch sign. 
Thus, at high energy, this particular Born-type behaviour remains true for the strong nuclear fields considered here.

\section{Conclusion}

Two basic processes were considered which occur during the scattering of high-energy leptons with heavy nuclei,
elastic scattering and bremsstrahlung emission.
Both processes were described within the state-of-the-art theories, the relativistic phase shift analysis for elastic scattering supplemented by the DWBA theory
to account for magnetic scattering in the case of nuclei with spin, and the relativistic Dirac partial-wave theory for bremsstrahlung.

For elastic scattering, the experimental cross section data for electron and positron impact on  lead are well reproduced.
We have confirmed the phase shift between electron and positron intensities in the region of the diffraction structures.
We have shown that this phase shift increases with potential strength and changes with scattering angle.
While the shape of the nuclear charge distribution influences the location of the diffraction minima, it does, however, not alter the electron-positron phase shift.

Comparing the polarization correlations for electrons and positrons, their energy- and angular dependence is much alike in the case of $L$ and $R$.
Existing electron-positron deviations decrease with collision energy but increase with scattering angle.
The Sherman function $S$, on the other hand, differs considerably for the two lepton species.
$S$ is not only of opposite sign at low energies and backward scattering angles, but the damping of the diffraction structures in the presence
of the magnetic interaction is much stronger for positrons than it is for electrons.

For positron bremsstrahlung no experimental data are yet available, so only predictions could be presented.
Due to the repulsive potential, positron bremsstrahlung is much weaker than electron bremsstrahlung, in particular close to the short-wavelength limit.
For the triply differential  cross section we have found a decrease with frequency $\omega$ at forward photon and positron angles as for  electron impact  or for 
the doubly diferential cross section.
However, at backward scattering angles the positron bremsstrahlung intensity increases with $\omega$ to a maximum
and eventually falls off towards the SWL.

For  the circular polarization correlation $C_{320}$ (respectively $C_{32}$ for unobserved leptons) there is a striking similarity between electrons and positrons (like for $L$ in the case of elastic scattering),
but for $C_{120}$ there is a large difference between the results for electrons and positrons, particularly at high frequencies
where this spin asymmetry is strongly suppressed for the positrons.
The behaviour of the spin asymmetry $A$ resembles  the Sherman function $S$ for elastic scattering.
In particular, it has opposite signs for electrons and positrons in  a large parameter region, which lead to a kind of mirror symmetry with respect to the zero-line in the energy or angular dependence.
This behaviour, which is true in the Born approximation valid for low nuclear charge or low frequencies in forward emission, is somewhat unexpected in the case of strong nuclear fields or hard photons.

For electrons, the above-mentioned similarity between the polarization correlations of elastic scattering and the non-coincident bremsstrahlung, 
for a fixed spin polarization of projectiles and ejectiles, is well-known in  case of a complete energy transfer in strong nuclear fields.
We have investigated the respective behaviour for positron scattering at a high collision energy $E_e$ and a photon frequency near the SWL, $\omega/E_e = 0.99.$
We have found that there is a visible similarity for the pairs $L/C_{32}$ and $R/C_{12}$, but the differences
in the pair $S/-A$ are considerably larger for positrons than they are for electrons.

For estimating the precision of the numerical calculations, strict sum rules for the polarization correlations can be applied, both for potential scattering and for the elementary process of bremsstrahlung.
While the  accuracy is high for elastic scattering (well below one permille for potential scattering, and in the percent region if magnetic scattering is dominant),
even at collision energies of 300 MeV or beyond, it is insufficient for bremsstrahlung at backward angles
for collision energies as low as $10-20$ MeV.
An accurate bremsstrahlung  theory for high-energy projectiles and heavy targets, which goes beyond the PWBA, has still to be awaited.

\vspace{0.5cm}

{\large\bf Acknowledgment}

\vspace{0.2cm}

I would like to thank K.Aulenbacher for initiating the
positron studies and V.A.Yerokhin for providing additional positron
bremsstrahlung results to test my code. I am also indepted to H.Hergert and R.Roth for the nuclear structure calculation of the sodium charge density.



\vspace{1cm}

\end{document}